\documentclass[reprint,superscriptaddress,showpacs,amsmath,amssymb,aps,prb]{revtex4-2}

\usepackage{soul}
\usepackage{dsfont}
\usepackage[utf8]{inputenc}
\usepackage{physics}
\usepackage{float}
\usepackage{dsfont}
\usepackage{appendix}
\usepackage{xcolor}
\usepackage{graphicx,subfigure,bbm}
\usepackage{tabularray}
\usepackage[hidelinks]{hyperref}
\hypersetup{
	colorlinks,
	linkcolor={red!50!black},
	citecolor={blue!50!black},
	urlcolor={blue!80!black}
}
\usepackage{mathtools}
\usepackage{wasysym}

\usepackage{chngcntr}
\counterwithout{table}{section}

\newcommand{\he}{\text{HE}}
\newcommand{\me}{\text{ME}}
\newcommand{\lb}{\text{LB}}

\begin{document}
\title{Exact finite-time correlation functions for multi-terminal setups: \\ Connecting theoretical frameworks for quantum transport and thermodynamics}

\author{Gianmichele Blasi}
	\affiliation{Department of Applied Physics, University of Geneva, 1211 Geneva, Switzerland}%\email{gianmichele.blasi@unige.ch}
	
	\author{Shishir Khandelwal} 
 \affiliation{Department of Applied Physics, University of Geneva, 1211 Geneva, Switzerland}
 \affiliation{Physics Department, Lund University, Box 118, 22100 Lund, Sweden}
  \affiliation{NanoLund, Lund University, Box 118, 22100 Lund, Sweden}
	\author{G\'eraldine Haack}
	\affiliation{Department of Applied Physics, University of Geneva, 1211 Geneva, Switzerland}
	%\email{shishir.khandelwal@unige.ch}
	
\begin{abstract}

Transport in open quantum systems can be explored through various theoretical frameworks, including the quantum master equation, scattering matrix, and Heisenberg equation of motion. The choice of framework depends on factors such as the presence of interactions, the coupling strength between the system and environment, and whether the focus is on steady-state or transient regimes. Existing literature mainly treats these frameworks independently. Our work establishes connections between them by clarifying the role and status of these approaches using two paradigmatic models for single and multipartite quantum systems in a two-terminal setup under voltage and temperature biases. 
We derive analytical solutions in both steady-state and transient regimes for the populations, currents and current correlation functions. Exact results from the Heisenberg equation are shown to align with scattering matrix and master equation approaches within their respective validity regimes. Crucially, we establish a protocol for the weak-coupling limit, bridging the applicability of master equations at weak-coupling with Heisenberg or scattering matrix approaches at arbitrary coupling strength. 
\end{abstract}

% \date{\today}
\maketitle

\section{Introduction}

Recent advancements in quantum technologies have revitalized interest in the field of open quantum systems, captivating diverse physics communities, each with distinct methodologies and goals. A focal point within this domain is quantum transport \cite{Beenakker1991,Datta1997,Nazarov2009}, where the emphasis lies on assessing out-of-equilibrium average currents and corresponding correlations for particles, charges and energy. These quantities hold significance in thermodynamics from both theoretical \cite{Vinjanampathy2016,Mitchison2019,Bhattacharjee2021} and experimental perspectives \cite{Giazotto2006,Pekola2015,Chien2015,Pekola2021,Myers2022}, playing a crucial role in addressing fundamental topics such as fluctuation-dissipation relations~\cite{Altaner2016,Barker2022} and thermodynamic uncertainty relations~\cite{Guarnieri2019,Miller2021,Potts2019,Lopez2023}, and practical ones such as quantum metrology~\cite{Cavina2018,Salvia2023,Rodriguez2023}.

Transport in open quantum systems can be explored through various theoretical frameworks, each relying on distinct central concepts and deriving average currents and current correlations from different perspectives. Conventional quantum transport approaches, such as the Landauer-B\"uttiker theory \cite{Landauer1957,Buttiker1986}, Green's functions and Keldysh formalism \cite{Konig96, Economou2006, Splettstoesser06, Arrachea2006},  are centered around the number of particles in a reservoir without direct reference to the quantum state of the system. Conversely, the knowledge of the state of a quantum system is fundamentally important in the field of quantum information. The typical approach towards this problem is the quantum master equation \cite{Wiseman,Breuer2007,Rivas2012,Schaller2014,Soret2022}, focusing on the density operator describing the state of the system and incorporating environmental effects perturbatively under the assumption of weak system-bath coupling. Given that significant progress in quantum thermodynamics has been driven by quantum information \cite{Goold2016}, the master equation approach has naturally emerged as the primary choice for computations in this area.

\begin{figure*}
    [!htb]
	\centering
 \includegraphics[width=0.75\paperwidth]{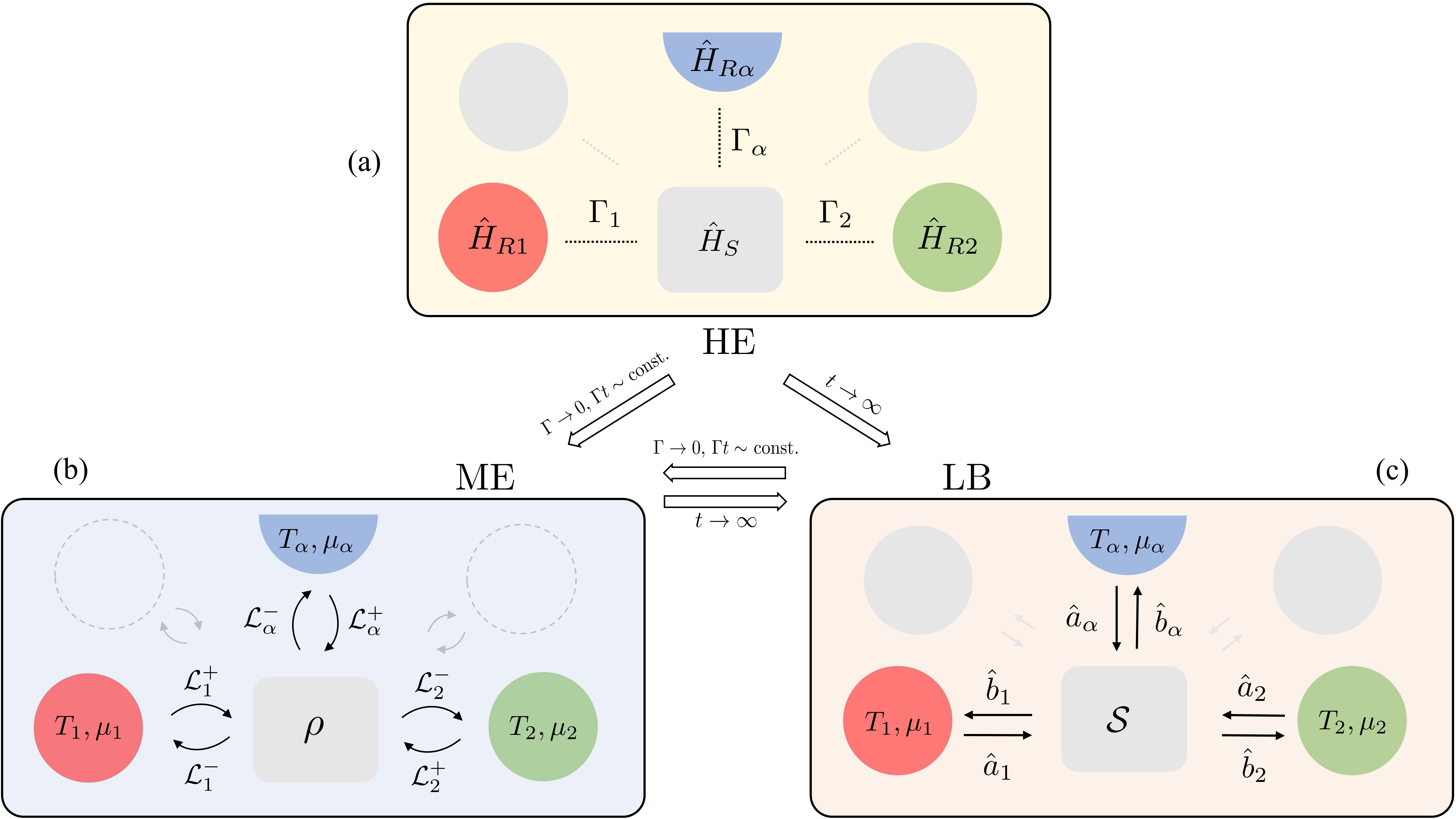}
 	\caption{
Schematic representation of the three approaches considered in this work: Heisenberg equation of motion (HE), Master equation (ME) and Landauer-Büttiker (LB). The  details of these approaches to assess the dynamics of an open quantum system coupled to multiple reservoirs can be found in Sec.~\ref{sec:modframe}. The thick arrows represent the limits necessary to move between the results of each approach.
% Each Hamiltonian and operators appearing here are defined in Sec. II.
% In particular, this work develops a weak-coupling procedure allowing for deriving observables in the ME framework, from HE and LB results.
  % An open quantum system coupled to different thermal reservoirs. (a) The HE framework leads to an exact transient solution within the wide-band approximation. (b) The transient ME solution can be obtained from the HE and LB solutions by taking the weak-coupling limit, $\Gamma\to 0$, under the constraint $\Gamma t\sim\text{constant}$. (c) Both the HE and ME solutions reduce to LB predictions at long times, $t\to\infty$.
  }
	\label{fig:setups}
\end{figure*}

Irrespective of the approach, existing literature predominantly explores steady-state observables due to the inherent complexity in deriving exact analytical finite-time solutions. 
For example, recent efforts have introduced techniques combining quantum stochastic Hamiltonians with Keldysh formalism~\cite{Jin2022,Ferreira2023}, however, at present, their complexity limits their applicability to the steady-state. 
Notably, exact results for current correlations are known only for a single resonant level with non-equilibrium Green's functions approach \cite{Zhang2014}. Furthermore, in the master equation approach, currents and fluctuations are typically derived either in the steady state with full counting statistics (see, for example, Ref. \cite{Schallernotes}), 
%\gh{other refs missing, check works by Flindt and the Danish guy he made his PhD with, have a look at schaller's biblio}
or under a stochastic framework \cite{Wiseman1993,Zoller1997,Chantasri2015,Jin2022,Landi2024}, while the Landauer-B\"uttiker approach predominantly focuses on the zero-frequency component of currents correlations, limited to the steady state \cite{Blanter2000}.

%\gh{given the importance of the work by the Zhang et al, it would be important to give them some credit in the intro, and find a good way to discuss their work with respect to this one. We can not just ignore it when setting the landscape. Please give a try.}\sk{[SK: I am adding a sentence in the above paragraph. The exact expressions we present are  not the main result of this paper, it is the connections between approaches that are aided with GFCS. We have not claimed anywhere that we are the first to give exact solutions and have cited Zhang where necessary. Zhang himself is not the first to do this stuff, one can find similar expressions in Schaller's book.]}

Despite shared motivations, the techniques employed in quantum transport and thermodynamics often differ, posing a challenge for researchers transitioning between these backgrounds. Connecting approaches from quantum transport and thermodynamics is therefore a natural and timely pursuit. Practically, this entails the possibility to exactly recover results from any approach by elucidating connections across the different theoretical frameworks.
As discussed in %noted in earlier works
\cite{Purkayastha2016, Purkayastha2022, Bhandari2021}, this is not a trivial task, as each framework has its own domain of validity and unique assumptions. Therefore, it is typically challenging to determine how the results of one framework and its corresponding implications can be understood in the context of another. Specifically, establishing a connection between approaches at the level of current correlations is an open problem.
% A unified framework that encapsulates approaches from both quantum transport and thermodynamics is a natural and timely pursuit. This necessitates the precise recovery of results from any approach within this unifying framework, a non-trivial task given the distinct domains of validity, unique assumptions, and varied understandings of quantities such as currents and fluctuations in different frameworks. 
% \gh{are they really understood differently? Do you believe so? I imagine what you mean, but I do not think this is the right formulation. Also, to discuss: do we want to sell a "unifying framework" or rather solid connections between frameworks? I think I would go for the second one, as the unifying framework can be considered to be HE -> exact solution! By the way, if I am correct, Heisenberg equations as exact approach is not mentioned. We should,no? }

In this work, we focus on three main approaches to open quantum systems and quantum transport: the Heisenberg equation of motion ($\he$), the quantum master equation ($\me$), and the Landauer-Büttiker approach ($\lb$). 
In the $\me$ framework, we further generalize the full counting statistics (FCS) approach, to account for multiple terminals, multiple times, and non-instantaneous jumps, as compared to existing literature~\cite{Schallernotes,Marcos2007,Marcos2010}.
Our work pursues two primary objectives: (1) establishing connections between different frameworks for assessing open quantum system dynamics, developing a protocol for the weak-coupling limit from exact solutions, as depicted in Fig.~\ref{fig:setups}; and (2) providing explicit and complete analytical results for two paradigmatic models for single and multipartite quantum systems.
We first treat the case of a single-level quantum dot model, for which we analytically compute the average particle and heat currents, as well as the auto- and cross-correlation functions of current. This minimal model serves as a benchmark, facilitating comparisons across approaches and validating the protocol for deriving weak-coupling and stationary limits. We then go beyond this minimal model, and compute exact solution of the density operator at all times for two interacting quantum dots. This bipartite system allows us to discuss the case of different interaction regimes set by inter-dot versus system-bath coupling strengths, providing a generalization of our results and weak-coupling protocol.
%for auto- and cross-correlation functions of current considering a minimal single-level quantum dot model, using the Heisenberg equation of motion and the master equation, as summarized in Table~\ref{Table1}. This paradigmatic model serves as a benchmark, facilitating comparisons across approaches and validating the protocol for deriving weak-coupling and stationary limits. 
%\sk{[How about this?]Crucially, we go beyond this model to investigate the transient dynamics of a two-terminal double-quantum-dot setup, further demonstrating how to go to a many-dot setup. }
%Importantly, our results go beyond the simple example discussed and can serve as building blocks to investigate higher-dimensional and many-body open quantum systems. %\gb{Here we have to say that in order to prove the generality of our approach, we consider also the case of a double quantum dot.}
The techniques introduced in this work can therefore be applied to understand the operation of a large class of quantum devices for any couplings and for finite-times, for example, to investigate the build up of quantum correlations beyond minimal setups \cite{BohrBrask2015}.

The article is organized as follows. We start in Sec. \ref{sec:modframe} by delineating a general model of an open quantum system alongside the $\he$, $\me$, and $\lb$ frameworks within the context of quantum transport. 
In Sec.~\ref{GFCS}, we introduce a generalized full counting statistics approach (FCS) for the calculation of current and current correlations accounting for multiple times, multiple terminals and extended to arbitrary quanta of particles, charges, or energies exchange between system and reservoirs.
After providing a brief description of the single-level quantum dot model, we present exact analytical solutions for currents and their fluctuations using the $\he$ approach in Sec.~\ref{sec_HE_for_QD}, followed by the corresponding solutions using the $\me$ approach in Sec.~\ref{sec_ME_for_QD}. We benchmark both approaches against the LB approach in the steady state. In Sec.~\ref{sec:connection}, we present a weak-coupling protocol to derive observables calculated within the $\me$ framework from the results obtained within the $\he$ and $\lb$ frameworks. To underscore the generality of our results, we then treat the case of a double quantum dot system in Sec.~\ref{sec:DQD}. We discuss two interacting regimes and configurations for the dots, coupled in series or in parallel with respect to the two terminals. We obtain exact solutions at all times for the populations of the density operator from the $\he$ approach, and demonstrate the validity of our weak coupling protocol by recovering results obtained previously with the so-called local and global master equations \cite{Hofer2017, Gonzalez2017, Mitchison2018, DeChiara2018, Dann2018, Cattaneo2019, Potts2021, Dann2023}. We conclude in Sec. VIII, offering perspectives for further exploration.
%\gh{In Sec.~\ref{sec:conc}, we summarize our main results, and list few research directions that could benefit from our connections and exact results for currents and current correlation functions, offering perspectives for further exploration.}

\section{Model and frameworks for quantum transport}
\label{sec:modframe}
We consider a time-independent system of $D$ fermionic quantum dots, coupled to $N$ reservoirs through tunneling interactions as depicted in Fig.~\ref{fig:setups}~(a). Furthermore, we consider tunnel-like interaction terms between the dots. The total Hamiltonian $\hat{H}$ is then composed of the following terms, 
\begin{eqnarray}
\label{eq:totalH}
   \hat{H} =  \hat{H}_S + \sum_{\alpha=1} ^N\hat{H}_{R_\alpha} + \sum_{\alpha=1}^N \hat{H}_{SR_{\alpha}}\,,
\end{eqnarray}
with the following system Hamiltonian, 
\begin{eqnarray}
\label{eq:system}
\hat{H}_S = \sum_{j=1}^D \epsilon_j \hat{d}_j^\dagger \hat{d}_j +\sum_{i\neq j}^{D}g_{ij}\hat{d}_i^{\dagger}\hat{d}_j\,.
\end{eqnarray}
The bare energies of each dot $j$ are labelled $\epsilon_j$, while the inter-dot tunneling strength is set by $g_{ij}$. The fermionic creation and annihilation operators $\hat{d}_j^\dagger, \hat{d}_j$ obey the anti-commutation relations $\{ d_i, d_{j}^\dagger\} = \delta_{ij}$. The free fermionic Hamiltonian for each reservoir $\alpha$ is given by,
\begin{eqnarray}
\hat{H}_{R_\alpha} = \sum_k \epsilon_{k\alpha} \hat{c}_{k\alpha}^\dagger \hat{c}_{k\alpha} \,,
\end{eqnarray}
where $\hat c_{k\alpha}^\dagger$ and $\hat c_{k\alpha}$ are the creation and annihilation operators for the $k$-th mode in reservoir $\alpha$, corresponding to energy $\epsilon_{k\alpha}$ and obeying the anti-commutation relations $\{ c_{k \alpha}, c^\dagger_{k' \alpha'}\} = \delta_{kk'} \delta_{\alpha \alpha'}$. Finally, the system-reservoir interaction Hamiltonian takes the form,
\begin{eqnarray}
\label{eq:H_SR}
   \hat{H}_{SR_{\alpha}} =  \sum_{j k} t_{jk\alpha}^{*}\hat{c}_{k\alpha}^{\dagger}\hat{d}_j+t_{jk\alpha} \hat{d}^{\dagger}_j\hat{c}_{k\alpha} \,,
\end{eqnarray}
where $t_{jk\alpha}$ represents the tunneling amplitude between the $j$-th level and the $k$-th mode of reservoir $\alpha$. The total Hamiltonian $\hat{H}$ 
describes a family of resonant-level systems \cite{Caroli1971,Meir1992,Mitchison2018}, and represents a paradigmatic model for investigating quantum transport. 
% In this way, the interacting problem can be treated in a manner similar to the non-interacting case, with the energies of the system fermions shifted due the interaction strengths. \gh{gh: this is not correct. Depending on the type of interactions, if you write everything in the eigenbasis of the system, the reservoirs will not anymore be in a diagonal state.}

The main goal of this work is to explore the transient and steady-state transport properties of the open quantum system governed by the Hamiltonian $\hat{H}$ as defined in Eq.~\eqref{eq:totalH}. Our investigation focuses on three distinct approaches. The first approach employs the Heisenberg equation of motion ($\he$), providing exact solutions for all times and system-environment coupling strengths. The second approach involves the quantum master equation ($\me$), which is applicable in the weak-coupling regime. The third approach follows the Landauer-Büttiker formalism ($\lb$), utilizing the scattering matrix in the steady-state regime. 
For each approach, we outline the key assumptions and detail the computation of fundamental transport quantities. In particular we will focus on particle current $I_\alpha$, energy current $J_\alpha$, and current correlation function $S_{\alpha\alpha^\prime}$. We specifically concentrate on particle and energy currents within the reservoirs due to their importance in quantum thermodynamics and their utility in deriving other thermodynamically relevant quantities. 
% For instance, the charge current ($I^c_\alpha$) can be expressed as $I^c_\alpha=eI_\alpha$ (where $e$ represents the electron charge), and the heat current ($J^h_\alpha$) can be derived from the first law of thermodynamics: \gb{$J^h_\alpha=J_\alpha+J_{c\alpha}/2-\mu_\alpha I_\alpha$}, where $\mu_\alpha$ denotes the chemical potential at reservoir $\alpha$\gb{, and $J_{c\alpha}$ represents the energy current in the contact region between the system and the reservoir $\alpha$}.
For instance, the charge current ($I^c_\alpha$) can be expressed as $I^c_\alpha(t) = eI_\alpha(t)$, where $e$ represents the elementary charge. Similarly, the heat current ($J^h_\alpha$) can be written as $J^h_\alpha(t) = J_\alpha(t) + J_{c\alpha}(t)/2 - \mu_\alpha I_\alpha(t)$, with $\mu_\alpha$ denoting the chemical potential at reservoir $\alpha$, and $J_{c\alpha}$ representing the energy current in the contact region between the system and reservoir $\alpha$.
As highlighted in Refs.~\cite{Ludovico2014,Ludovico2016}, including the contact energy current in the heat current definition is crucial to uphold the second law of thermodynamics in the transient regime. In our investigation, we do not focus on this term, as we have verified its negligibility in both the stationary and weak-coupling regimes. Specifically, in Appendix~\ref{appendix:contact_current}, we provide evidence for the single-level quantum dot case, demonstrating that this term precisely vanishes in the steady-state and 
only contributes in second order to the coupling strength in the weak-coupling regime.
% remains of order $\mathcal{O}(\Gamma^2)$ in the weak-coupling regime.

\subsection{Heisenberg equation approach}

In the Heisenberg equation ($\he$) picture, the evolution of the operators $\hat d_j$ and $\hat c_{k\alpha}$ of the system and reservoir $\alpha$ is given by the Heisenberg equations of motion (we assume %hereafter 
$\hbar,k_B = 1$),
\begin{align}
&\frac{d}{dt} \hat{d}_j = i [ \hat H, \hat{d}_j]\, \label{eq:Heisenberg_d} \\
& \frac{d}{dt} \hat{c}_{k\alpha} = i [ \hat H, \hat{c}_{k\alpha}] \label{eq:Heisenberg_c}
\end{align} 
where $\hat H$ is the total Hamiltonian of Eq.~\eqref{eq:totalH}. 
In the solution to Eqs.~\eqref{eq:Heisenberg_d} and \eqref{eq:Heisenberg_c}, a crucial quantity arises: the bare tunneling rate. In full generality, starting from Eq.~\eqref{eq:H_SR}, this quantity is represented by a matrix with elements $\Gamma^\alpha_{ij}$ % which sets the strength of local describes the tunneling coupling of the quantum system to the reservoir $\alpha$~
\cite{Datta1997, Guevara2003},
%which we will further examine in the specific case of a single-level quantum dot system in Sec.~\ref{sec_HE_for_QD} \gb{and for a double quantum dot scenario in Sec.~\ref{sec:DQD}}. 
%\st{This quantity is represented by the bare tunneling rate} 
%\gb{In the general case of a system composed by multiple quantum dots, this quantity is represented the the bare tunneling rate matrix $\Gamma^\alpha$, which describes the tunneling coupling of the quantum system to the reservoir $\alpha$~\cite{Datta1997,Guevara2003}:
\begin{equation}
\label{eq:tunnel_rate}
\Gamma_{ij}^{\alpha}(\epsilon) = 2\pi \sum_{k}  t_{i k\alpha}^*t_{j k\alpha}  \delta\left(\epsilon - \epsilon_{k\alpha}\right).
\end{equation}
While the diagonal terms ($i=j$) set the strength of local tunneling processes between reservoir $\alpha$ and dot $i$, the non-diagonal terms ($i\neq j$) correspond to nonlocal co-tunneling processes 
% \gb{gh: At the beginning, I tent to also disagree, but then Gianmichele explained to me what he had in mind, and I think it's ok. It's clearly written in the most general case. }
% \sk{[I don't agree with your distinction between local and nonlocal or cotunnelling effects - If you see eq. 4, we have equivalent direct interactions between all dots and all reservoirs. This means that $\alpha=j$ and $\alpha\neq j$ are treated on an equal footing.]} 
involving transitions between reservoir $\alpha$ and different quantum dots. In the case of local tunneling processes involving a reservoir and a single dot, we use a simplified notation for convenience, $\Gamma_{j \alpha}(\epsilon)\equiv\Gamma_{jj}^{\alpha}(\epsilon) = 2\pi \sum_{k}  \abs{t_{jk\alpha}}^2  \delta\left(\epsilon - \epsilon_{k\alpha}\right)$. It is a function of energy in general.
%When Hereafter we will consider only the case in which the non-diagonal terms are equal to zero, namely $\Gamma_{\alpha j}(\epsilon)\equiv\Gamma_{jj}^{\alpha}(\epsilon) = 2\pi \sum_{k}  \abs{t_{jk\alpha}}^2  \delta\left(\epsilon - \epsilon_{k\alpha}\right)$.
%In this case (when $i=j$) the bare tunneling rate only depends on the tunneling probabilities $\abs{t_{jk\alpha}}^2$, and in general is a function of the energy.}
%\st{$\Gamma_{j \alpha}(\epsilon)$, which describes the strength of the system-bath coupling between the energy level $j$ of the system and the reservoir $\alpha$,}
%\textcolor{red}{TO BE REMOVED
%\begin{equation}
%\label{eq:tunnel_rate}
%\Gamma_{j \alpha}(\epsilon) = 2\pi \sum_{k} \vert t_{j k\alpha} \vert^2 \delta\left(\epsilon - \epsilon_{k\alpha}\right).
%\end{equation}
%}
%\st{This function encapsulates the impact of the reservoirs on the system and generally depends on the energy.}
Further assumptions on the sum $\sum_{k} \vert t_{jk\alpha} \vert^2$ yield well-known profiles for this quantity, such as the Ohmic or Lorentzian shapes as functions of energy~\cite{Weiss2012}. An important limit under consideration in this work is the wide-band limit (WBL), where the energy width of $\Gamma_{j\alpha}(\epsilon)$ is larger than all other energy scales of the system. The WBL assumes an energy-independent spectral density, expressed as~\cite{Brako1989,Baldea2016,Covito2018}, 
\begin{equation}
\label{eq:WBL}
\Gamma_{j\alpha}(\epsilon)\equiv \Gamma_{j\alpha}\,.
\end{equation}
In essence, in the WBL, the tunneling probabilities of the particles become insensitive to the energy structure of the bath; the system cannot resolve it. In the specific case of a single-level quantum dot discussed subsequently, the WBL approximation enables the derivation of analytical results.
In Fig.~\ref{fig:setups} (a), we indicate with $\Gamma_\alpha=\sum_j \Gamma_{j\alpha}$ the total tunneling rate associated with reservoir $\alpha$.

Using the solutions to the above Eqs.~\eqref{eq:Heisenberg_d} and \eqref{eq:Heisenberg_c}, observable quantities can be calculated for all times. Specifically, quantum transport observables, i.e., currents and current-correlation function can be defined using the time-dependent occupation number operator in a given reservoir $\alpha$, 
\begin{eqnarray}
\label{eq:Nalpha}
    \hat{N}_\alpha (t) = \sum_k \hat c_{k\alpha}^\dagger(t) \hat c_{k\alpha}(t),
\end{eqnarray}
the operators $\hat c_{k\alpha}^\dagger(t), \hat c_{k\alpha}(t)$ being solutions of Eq.~\eqref{eq:Heisenberg_c} for the creation and annihilation operators. The average particle current, $I_\alpha^{\he}$, and energy current, $J_\alpha^\he$,  in lead $\alpha$, are defined as,
\begin{eqnarray}
\label{eq:def_HE_particle_current}
I_\alpha^\he(t) &= \expval{\hat I_\alpha(t)}&\equiv  -\frac{d}{dt} \langle \hat{N}_\alpha \rangle = - i \left\langle [ \hat H, \hat{N}_\alpha ]  \right\rangle\,, \label{def_current} \\
\label{eq:def_HE_energy_current}
J_{\alpha}^\he(t)&= \expval{\hat J_\alpha(t)}&\equiv -\frac{d}{dt} \langle \hat{H}_{R_\alpha} \rangle=-i\left\langle [\hat{H},\hat  H_{R_\alpha}] \right\rangle\,,
\label{eq:def_HE_current_correlation}
\end{eqnarray}
where $\langle \ldots \rangle$ denotes the quantum statistical average. 
The solutions depend on initial conditions at time $t=t_0$,
\begin{eqnarray}
\label{eq_ini_lead}
&& \expval{\hat{c}_{k\alpha}^{\dagger}(t_0)\hat{c}_{k^{\prime}\alpha^{\prime}}(t_0)} = \delta_{k\alpha,k^{\prime}\alpha^{\prime}} f_{\alpha}(\epsilon_{k\alpha}) \,, \\
\label{eq:ini_dot}
&& \expval{ \hat{d}^{\dagger}_i(t_0)\hat{d}_j(t_0)}=\delta_{ij}n_j\,.
\end{eqnarray}
The first condition assumes that the initial mean occupation of the $\alpha$ reservoir is set by the Fermi-Dirac distribution at temperature $T_\alpha$ and chemical potential $\mu_\alpha$,
\begin{equation}
f_{\alpha}( \epsilon)=\{e^{(\epsilon-\mu_{\alpha})/ T_{\alpha}}+1\}^{-1}\,.
\end{equation}
The second condition assumes that the initial occupation population of the $j$-th energy level is $n_j$, while initial coherences among different levels are assumed to be zero.
Fluctuations of the particle current at the reservoir $\alpha$ at time $t$ are accessed through the operator $\Delta\hat I_\alpha(t)\equiv \hat I_\alpha(t) - \expval{\hat I_\alpha}$, from which the current correlation function depending on times $t,t'$ and leads $\alpha, \alpha'$ is defined,
\begin{eqnarray}
    S_{\alpha\alpha'}^\he(t,t') &=& \expval{\Delta \hat{I}_{\alpha}(t) \Delta \hat{I}_{\alpha^{\prime}}(t^{\prime})} \nonumber\\
&=& \expval{\hat{I}_{\alpha}(t)\hat{I}_{\alpha^{\prime}}(t^{\prime})}- I_\alpha^\he(t)  I_{\alpha'}^\he(t')\,. \label{def_noise}
\end{eqnarray}
The case $\alpha=\alpha^\prime$ corresponds to the auto-correlation function, while $\alpha\neq\alpha^\prime$ gives the cross-correlations.\par
%We emphasize that solving the Heisenberg equations provides solutions that are entirely quantum coherent.
The $\he$ approach exactly addresses the evolution of operators for both the quantum system and the reservoirs, stemming from the unitary evolution dictated by total Hamiltonian $\hat{H}$, giving solutions that are entirely quantum coherent. In the subsequent discussion (see Sec.~\ref{sec_HE_for_QD}), we use this framework to derive exact expressions for currents and correlations for a single-level quantum dot model.
%we focus on the single-level quantum dot case to illustrate features derived from this comprehensive quantum coherent treatment of dynamics. 
% \sk{\st{Notably, this approach differs from the quantum master equation method discussed in the section below. The latter considers the impact of environments by offering an effective solution solely for the evolution of the quantum system, assuming that the reservoirs remain in their equilibrium state.}[I think this has been mentioned sufficiently before.]}

%%%%%%%%%%%%%%%%%%%%%%%%%%%%%%%%%%%%%%%%%%%%%%%%%%%%%%%%%%%%%%%%%%%%%%%

\subsection{Master equation approach}

The master equation (\me) framework serves as a paradigmatic approach for analyzing the dynamics of quantum systems weakly coupled with one or multiple reservoirs. In contrast to the Heisenberg equation approach, it is crucial to highlight that the central quantity of interest within the master equation framework is the reduced density operator of the quantum system, denoted as $\rho$, which satisfies the non-unitary evolution equation~\cite{Breuer2007},

\begin{eqnarray}
\label{eq:not_closed}
    \dot{\rho} = - i~\text{Tr}_R\left\{ [\hat{H}, \rho_{\text{tot}}] \right\}\,.
\end{eqnarray}

Here, $\hat{H}$ represents the total Hamiltonian from Eq.~\eqref{eq:totalH}, and $\rho_{\text{tot}}$ is the total density operator describing the system and all reservoirs. In contrast, $\rho$ is the reduced density operator of the quantum system, obtained by tracing out the degrees of freedom of the reservoirs ($\text{Tr}_R\left\{\ldots\right\}$). Under the assumption of weak-coupling between the system and reservoirs, the evolution equation for $\rho$ is governed by the Liouvillian superoperator $\mathcal{L}$. This superoperator corresponds to a CPTP (Completely Positive and Trace Preserving) map in Lindblad form \cite{Gorini1975, Lindblad1976, Breuer2007}, encompassing the unitary evolution of the quantum system according to its Hamiltonian $\hat{H}_S$, along with a dissipative part. Unlike Eq.~\eqref{eq:not_closed}, it exclusively describes the evolution of the reduced density operator $\rho$ of the system, and in general, it takes the form,
% \gb{GB: HERE we should consider the general ME like Eq.~\eqref{eq:local_global_Lindblad}!!! At the end , for this section what is important is Eq. 21.}
\begin{align}
\label{eq:Lindblad}
     \dot\rho(t)
=\mathcal{L} \rho(t)= &-i[\hat H_S,\rho(t)]\nonumber \\
&+ \sum_{j \alpha}\left(\Gamma^{+}_{j \alpha}  \mathcal D\left[\hat L_{j \alpha}^{\dagger} \right] + \Gamma^{-}_{j \alpha}  \mathcal D\left[\hat L_{j \alpha} \right] \right)\rho(t).  
 \end{align}\noindent where $\hat L_{j \alpha}^\dagger$ and $\hat L_{j \alpha}$ are jump operators which add and remove excitations between the system at energy $\epsilon_j$ and the reservoir $\alpha$ respectively.
It can also be derived using the Feynman-Vernon influence functional theory~\cite{Feynman2000, Aurell2020, Aurell2021, Muratore2023}, or using  Green’s function formalism~\cite{Bhandari2021}, by performing a perturbative expansion in the system-reservoir coupling.
In the above equation, the in- and out-tunneling rates are the product of the bare tunneling rates introduced in Eq.~\eqref{eq:tunnel_rate} and the occupation probability of the reservoir $\alpha$ given by the Fermi-Dirac distribution evaluated at the energy level $\epsilon_j$ \cite{Breuer2007},
\begin{eqnarray}
    \Gamma_{j \alpha}^+ &=& \Gamma_{j \alpha}(\epsilon_j) f_\alpha(\epsilon_j) \\
    \Gamma_{j \alpha}^- &=& \Gamma_{j \alpha}(\epsilon_j) \big( 1- f_\alpha(\epsilon_j) \big)\,.
\end{eqnarray}
% \begin{eqnarray}
%     \Gamma_{j\alpha}^+ &=& \Gamma_{j\alpha}(\epsilon_j) f_\alpha(\epsilon_j) \\
%     \Gamma_{j\alpha}^- &=& \Gamma_{j\alpha}(\epsilon_j) \big( 1- f_\alpha(\epsilon_j) \big)\,.
% \end{eqnarray}
% Let us recall that, under WBL approxiamtion, the bare rates are energy-independent, $\Gamma_{j,\alpha} = \Gamma_\alpha$. The energy dependence of $\Gamma^\pm_{j\alpha}$ then only enters through the Fermi distribution.
The dissipators $\mathcal{D}$ are superoperators acting on $\rho$ and are defined as $\mathcal{D}[X] \rho \coloneqq X \rho X^\dagger - 1/2 (X^\dagger X \rho + \rho X^\dagger X)$ with $X=\hat L_{j \alpha},\hat L_{j \alpha}^{\dagger}$. The first term, $X \rho X^\dagger$, corresponds to \textit{quantum jumps}, while the last two terms ensure trace conservation of the density operator $\rho$ for $\mathcal{L}$ to be a valid CPTP map.

It is convenient, for the remaining part of the section, to single out the role of the quantum jumps, which will directly enter the definitions of the currents and current correlation functions, and rewrite the Liouvillian superoperator $\mathcal{L}$ of Eq.~\eqref{eq:Lindblad} in the following form,
\begin{eqnarray}
\label{eq:lind}
\dot{\rho}(t)=\mathcal{L} \, \rho(t) = \left( \mathcal L_0 + \sum_{j \alpha } (\mathcal L^+_{j\alpha} +\mathcal L^-_{j \alpha}) \right)\rho(t) .
\end{eqnarray}
Here, $\mathcal L_0$ represents the coherent non-unitary evolution of the reduced state of the system and can be expressed in terms of a non-hermitian commutator, $\left[A,B\right]_\dagger = AB- B^\dagger A^\dagger$,
\begin{equation}
\label{eq:L0}
    \mathcal L_0 \rho = -i\left[\hat H_S - \frac{i}{2} \sum_{j \alpha } (\Gamma_{j \alpha}^-\hat L_{j \alpha}^\dagger \hat L_{j \alpha} + \Gamma_{j \alpha}^+\hat L_{j \alpha} \hat L_{j \alpha}^\dagger), \rho \right]_\dagger.
\end{equation}

\noindent The jump superoperator $\mathcal L^+_{j\alpha}$ ($\mathcal L^-_{j\alpha}$) captures the tunneling in (out) of particles from (to) reservoir $\alpha$ to (from) the energy level $\epsilon_j$ of the quantum system with corresponding rate $\Gamma_{j\alpha}^+$ ($\Gamma_{j\alpha}^-$),
\begin{eqnarray}
    \mathcal L^+_{j\alpha} \rho &=& \Gamma_{j\alpha}^+ \, \hat L_{j \alpha}^\dagger \, \rho \, \hat L_{j \alpha} \label{eq:jump_lind_op1}\\
    \mathcal L^-_{j\alpha} \rho &=& \Gamma_{j\alpha}^- \, \hat L_{j \alpha} \,  \rho  \, \hat L_{j \alpha}^\dagger \,. \label{eq:jump_lind_op2}
\end{eqnarray}

\subsubsection{Generalized Full Counting Statistics}
\label{GFCS}
%A convenient method for solving Eq.~\eqref{eq:lind} and calculating currents and correlation functions is to employ Full Counting Statistics (FCS).
A convenient method for calculating currents and correlation functions by solving Eq.~\eqref{eq:lind} is Full Counting Statistics (FCS). In the standard FCS approach, the emphasis is on the steady state \cite{Schallernotes,Braggio2006,Flindt2008,Altaner2016}, and quantum jumps are assumed to be instantaneous \cite{Landi2024, Zoller1997}. Here, we develop a generalized approach to FCS that not only accommodates multiple terminals and times but also considers non-instantaneous jumps. To achieve this, we initially discretize Eq.~\eqref{eq:lind} in terms of the net amount of particles, charges, or energy quanta exchanged between the reservoirs and the system. For this purpose, we introduce the $\vec x$-resolved state of the system, denoted as $\rho^{(\vec x)}\left(t\right)$, which satisfies \cite{Schaller2014},
\begin{align}
\label{eq:lindnresolved}
\dot{\rho}^{(\vec x)}(t) &=\mathcal L_0 \rho^{(\vec x)}(t)\nonumber\\ &+\sum_{j \alpha } \left(\mathcal L_{j\alpha}^+\rho^{(x_\alpha -\nu_{j\alpha })}(t) +\mathcal L_{j\alpha}^-\rho^{(x_\alpha+ \nu_{j\alpha })}(t)\right).
\end{align}
Here, the vector $\vec x = \left\{x_1,\ldots,x_\alpha,\ldots,x_N \right\}$ contains the counting variables describing the net amount of particles, charges, or energy quanta exchanged between the system and reservoir $\alpha=1, 2, \ldots, N$, and $\nu_{j\alpha }$ corresponds to the change in $x_{\alpha}$ via the jump occurred between the $\alpha$-th reservoir and the $j$-th energy level of the system,
\begin{eqnarray}
\label{nu}
    \nu_{j \alpha} = \left\{ \begin{array}{ccl}
    1 & \text{for} & \text{particles} \\
    e & \text{for} & \text{charges} \\
    \epsilon_{j} & \text{for} & \text{energy quanta}\,. \end{array} \right.
\end{eqnarray}
The $\vec x$-resolved density operator $\rho^{(\vec x)}$ and the reduced state of the system $\rho$ are related via $\rho(t) = \sum_{\vec x} \rho^{(\vec x)}\left( t\right)$.
%%%%%%%%%%%%%%%%%%%%%%%%%%%%%%%%%%%%%%%%%%%%%%%%%%%
Eq.~\eqref{eq:lindnresolved} can be solved by taking the Fourier transform,
\begin{equation}
     \rho(\vec\chi,t) = \sum_{\vec{x}} \rho^{(\vec x)}(t)e^{i\vec x\cdot\vec\chi},
\end{equation}
where $\vec{\chi} = \{\chi_1, \ldots, \chi_\alpha, \ldots,\chi_N \}$ is the counting field vector.
Substituting the above expression in Eq. \eqref{eq:lindnresolved}, we obtain the following evolution equation,
\begin{eqnarray}
\label{eq:lindchi}
    \dot{\rho}(\vec{\chi},t) &= \mathcal{L}(\vec{\chi})\rho(\vec{\chi},t),
\end{eqnarray}
where the Lindbladian is now a function of the counting fields, 
\begin{eqnarray}
\label{eq:Lindbladian_fourier_transformed}
    \mathcal{L}(\vec{\chi}) = \mathcal L_0 +\sum_{j\alpha} \left(e^{+i\chi_\alpha \nu_{j\alpha}}\mathcal L_{j\alpha}^+ +e^{-i\chi_\alpha \nu_{j\alpha}}\mathcal L_{j\alpha}^-\right).
\end{eqnarray}
\begin{figure}[!htb]%[h!]
\centering
\includegraphics[width=0.9\columnwidth]{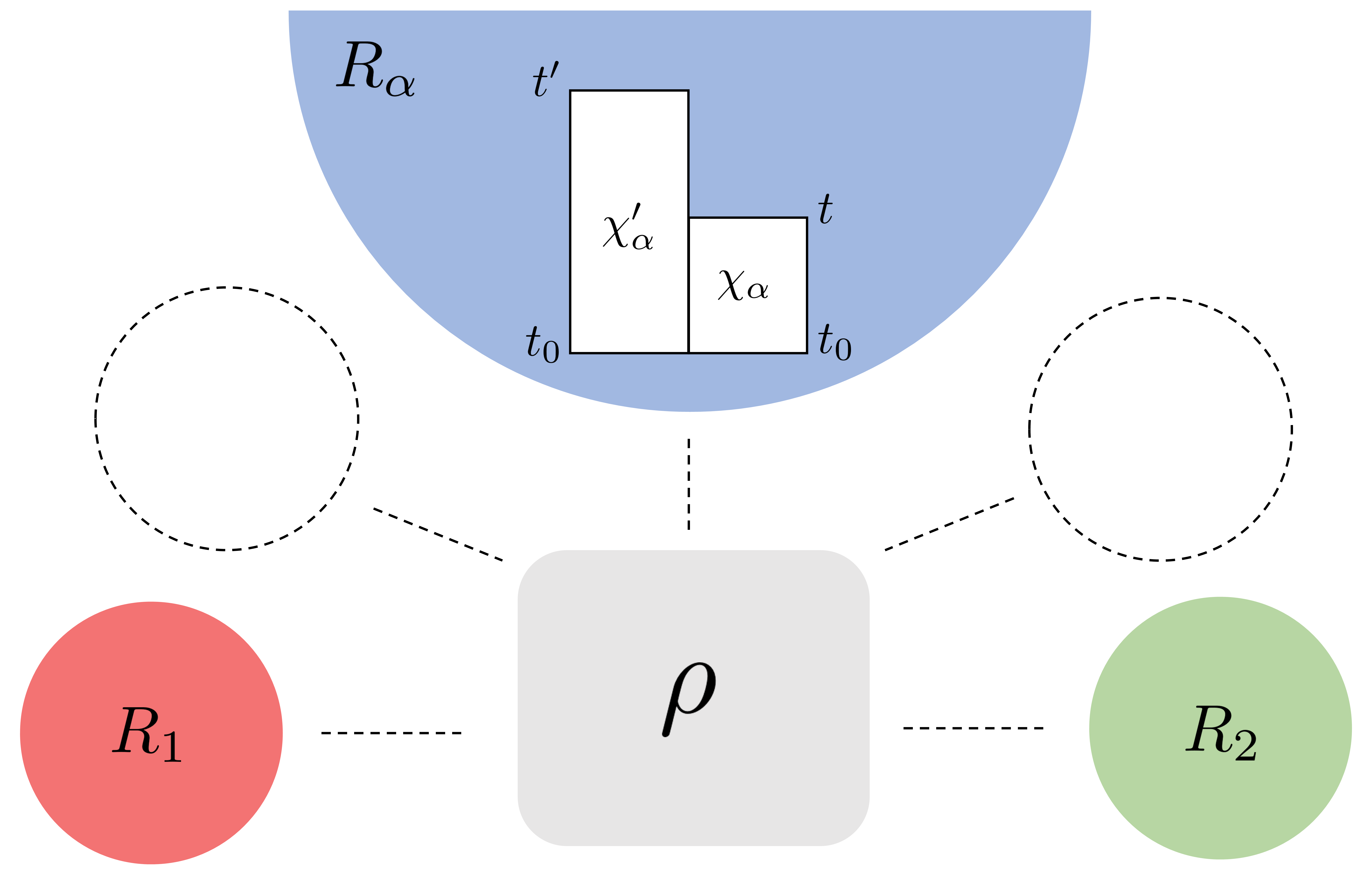}
\caption{Scheme outlining the conceptual framework for counting statistics for multiple terminals and multiple times. In reservoir $R_\alpha$, distinct boxes represent different time bins, each associated with a specific counting field ($\chi_\alpha$ for $t - t_0$ and $\chi'_\alpha$ for $t' - t_0$). The joint probability of observing a net transfer of $x'_\alpha$ quanta during the time $t' - t_0$ is obtained by counting both fields, $\chi_\alpha + \chi_\alpha^\prime$, for the interval $t - t_0$, and solely $\chi_\alpha^\prime$ for $t' - t$. Generalizing this approach to multiple terminals involves analogous operations with the vectors $\vec{\chi}$ and $\vec{\chi}^\prime$.} 
\label{fig:FCS}
\end{figure}

Notice that with $\vec{\chi}=0$, the above equation reduces to the standard form of Eq.~\eqref{eq:lind}, with $\mathcal L(\vec{\chi}=0)=\mathcal L$.
The formal solution of Eq.~\eqref{eq:lindchi} is given by,
\begin{equation}
\rho(\vec{\chi},t)=\Omega(\vec{\chi},t-t_0)\rho(\vec{\chi},t_0)=\Omega(\vec{\chi},t-t_0)\rho_0,
\end{equation}
with $\rho_0=\rho(t_0)$ the initial state of the system, and where we assumed no tunneling at time $t_0$, i.~e.~$\rho(\vec{\chi},t_0)=\rho_0$.

In the above equation we introduced the propagator in $\vec{\chi}$-space $\Omega(\vec{\chi},t)=e^{\mathcal L(\vec{\chi})t}$. The propagator in the $\vec{x}$-space is simply given by the inverse Fourier transform $\Omega(\vec{x},t)=\int d\vec{\chi}/(2\pi)^N e^{-i\vec{x}\cdot\vec{\chi}}~\Omega(\vec{\chi},t)$, such that $\rho^{(\vec{x})}(t)=\Omega(\vec{x},t-t_0)\rho_0$.
The propagator in the $\vec{x}$-space is essential for computing the joint probability $P^>(\vec{x},t;\vec{x}',t')$ of having a net amount $\vec{x}$ of particles, charges, or energy quanta transferred within time $t-t_0$ and $\vec{x}'$ within time $t'-t_0$, where the superscript ``$>$" indicates $t' > t$. This joint probability is expressed as,
\begin{align}
\label{eq.x-joint_probability}
P^>(\vec{x},t;\vec{x}',t')=\Tr{\Omega(\vec{x}'-\vec{x},t'-t)\Omega(\vec{x},t-t_0)\rho_0}.\nonumber\\
\end{align}
The above expression can be derived using the
Chapman–Kolmogorov property for Markovian evolution, as detailed in Appendix~\ref{appendix:multi-time_joint_probability}, where we also provide a generalized multi-time expression. We note that the presented expression generalizes the two-time joint probability found in Refs.~\cite{Marcos2007,Marcos2010}, which was obtained under the conditions of a single terminal.

The joint probability of Eq.~\eqref{eq.x-joint_probability} is crucial for calculating average quantities within the master equation formalism, denoted as $\expval{\ldots}_{\me}$. For instance, the average number of tunneled particles, charges, or energy quanta exchanged with reservoir $\alpha$ within time $t'-t_0$, is defined from the marginal $P(\vec{x}',s')$ of the joint probability given in Eq.~\eqref{eq.x-joint_probability} as detailed in App.~\ref{appendix:observables_GFCS}. The average number is then simply  expressed through the derivative of the joint probability with respect to the counting field, $\chi'_\alpha$,
\begin{align}
\label{eq:general_number}
\expval{x^\prime_\alpha(t')}_{\me} &= \int_{t_0}^{t^{\prime}}ds^\prime F(t^{\prime}-s^\prime) \sum_{\vec{x}'} x'_\alpha  P({\vec x}',s') \nonumber\\
& = \int_{t_0}^{t^{\prime}}ds' F(t^{\prime}-s^\prime)\partial_{i\chi'_\alpha} P^>(\vec{\chi},s;\vec{\chi}',s')\big\rvert_{\vec{\chi},\vec{\chi}'=0}.
\end{align}

Here, the function $F$ serves as a detection response function, similar to ones introduced in the context of quantum optics~\cite{Carmichael1989,Ueda1990,Barchielli1990}, with boundary conditions $F(0)=F(t-t_0)=0$ ~\footnote{It worth noticing that in the context of quantum optics, the detector response function is a property of the detector. In contrast, in our case, the reservoirs themselves play the role of detectors of quanta exchanged between the environment and the system.}. In typical resources on FCS, the tunneling of quanta is assumed to be instantaneous, corresponding to the detection response function being a delta function, i.e. $F(t-s)=\delta(s-t)$~\footnote{Technically the instantaneous jump condition for the detection response function reads as $F(t-s) = \delta (s-(t-0^-))$, where $0^-$ is needed such that Dirac delta is centered inside the time window of detection and $\int_{t_0}^tds~F(t-s)=\int_{t_0}^tds~\delta(s-(t-0^-))=1$}.
In Eq.\eqref{eq:general_number} we have introduced the joint probability in the $\vec{\chi}$-space,
\begin{align}
\label{eq:CHI-joint_probability}
P^>(\vec{\chi},t;\vec{\chi}',t')=\Tr{\Omega(\vec{\chi}',t'-t)\Omega(\vec{\chi}+\vec{\chi}',t-t_0)\rho_0}.
\end{align}
The expression can be intuitively understood using Fig. \ref{fig:FCS}. The counting procedure involves first counting with both counting-field vectors $\vec\chi$ and $\vec\chi^\prime$ in the time-bin $t-t_0$ (i.e., evolving with $\Omega\left(\vec\chi + \vec\chi^\prime, t-t_0 \right)$) and then with just $\vec\chi^\prime$ in the time bin $t^\prime-t$ (i.e., evolving with $\Omega\left(\vec\chi^\prime, t'-t \right)$). 

The average current associated to $x_\alpha(t)$ can now be calculated by taking the time derivative of Eq.~\eqref{eq:general_number}. Utilizing the property of the derivative of a convolution product, the boundary conditions of $F$, and employing the trace-preserving property of the Lindbladian $\Tr{\mathcal L(0)\sigma}=0$ (valid for any operator $\sigma$), we obtain (see Appendix \ref{appendix:observables_GFCS} for more details), 
\begin{align}
\label{eq:general_current}
\partial_{t}\expval{x_\alpha(t)}_{\me}
\hspace*{-0.05cm}=\hspace*{-0.1cm} \int_{t_0}^{t}\hspace*{-0.2cm}ds\,F\left(t\!-\!s\right)\Tr\{\, \partial_{i\chi_\alpha}\mathcal L(\vec{\chi})\rvert_{\vec\chi=0} e^{\mathcal L \left(s-t_0\right)}\rho_0 \},
\end{align}
where we have now changed primed labels to unprimed ones, for clarity. The particle and energy currents can be obtained from the above equation by considering $x_\alpha\equiv n_\alpha$ (with $\nu_{j \alpha} = 1$) for the former, and $x_\alpha\equiv \epsilon_\alpha$ (with $\nu_{j \alpha} = \epsilon_{j}$) for the latter, see Eq.~\eqref{nu}
\begin{eqnarray}
\left\langle I_\alpha(t)\right\rangle_{\me} &=& \int_{t_0}^{t}ds\,F\left(t-s\right)\Tr\{\mathcal I_\alpha e^{\mathcal L \left(s-t_0\right)}\rho_0\}, \label{eq:av_current_I}\\
\left\langle J_\alpha(t)\right\rangle_{\me} &=& \hspace*{-0.02cm}\int_{t_0}^{t}ds\,F\left(t-s\right)\Tr\{\mathcal J_\alpha e^{\mathcal L \left(s-t_0\right)}\rho_0\}, \label{eq:av_current_J}
\end{eqnarray}
where we have introduced the particle current and energy current superoperators, 
\begin{align}
\label{eq:super_current_I}
\mathcal I_{\alpha} &\coloneqq \partial_{i\chi_\alpha}\mathcal L(\vec{\chi})\big\rvert_{\substack{\vec\chi=0\\\nu_{j\alpha}=1}} \,\,= \sum_j \left( \mathcal L_{j \alpha}^+ - \mathcal L_{j \alpha}^- \right), \\
\mathcal J_{\alpha} &\coloneqq \partial_{i\chi_\alpha} \mathcal L(\vec{\chi})\big\rvert_{\substack{\vec\chi=0\\\nu_{j\alpha}=\epsilon_j}} = \sum_j \epsilon_{j} \left(\mathcal L_{j \alpha}^+ - \mathcal L_{j \alpha}^-\right).\label{eq:super_current_J}
\end{align}
The intuition behind the forms of these equations is self-evident; the superoperators count the net number of particles or the net amount of energy exchanged between the system and reservoir $\alpha$. %\begin{equation}
It is important to note that Eqs.~\eqref{eq:av_current_I} and \eqref{eq:av_current_J} are valid at all times, for any detection response function. 
To make a comparison with the usual master equation results, one may consider the case in which jumps happen instantaneously. In this case, $F(t-s) = \delta (s-t)$, and familiar expressions for the currents can be recovered, which we will refer to later in this work as $I^{\me}_\alpha$ and $J^{\me}_\alpha$, respectively,
\begin{align}
\label{eq:Currents_ME}
I^{\me}_\alpha(t) &= \Tr\left\{\mathcal I_\alpha\rho\left(t\right)\right\}=\Tr\left\{\mathcal I_\alpha e^{\mathcal L \left(t-t_0\right)}\rho_0\right\},\\
J^{\me}_\alpha(t) &= \Tr\left\{\mathcal J_\alpha\rho\left(t\right)\right\}=\Tr\left\{\mathcal J_\alpha e^{\mathcal L \left(t-t_0\right)}\rho_0\right\}.
\end{align}
The above expressions coincide with those recently introduced in Ref.~\cite{Landi2024}, derived within the framework of stochastic master equation.
% \gb{but have not been derived from first principles as we have proceeded above (I DON'T LIKE IT)}. 
% We believe the details provided in this section are of interest to a wide community of researchers interested in open quantum systems.\\

Similarly to Eqs.~\eqref{eq:general_number}, the above framework can be extended to calculate the two-time average, 
\begin{align}
\label{eq:general_number_correlation}
\expval{x_\alpha(t)x'_{\alpha'}(t')}_{\me} &=\int_{t_0}^{t}ds\int_{t_0}^{t^{\prime}}ds^\prime F(t-s)F(t^{\prime}-s^\prime)\nonumber\\
&\times\partial_{i\chi_\alpha}\partial_{i\chi'_{\alpha'}} P(\vec{\chi},s;\vec{\chi}',s')\big\rvert_{\vec{\chi},\vec{\chi}'=0}
\end{align}
where we have introduced the time-ordered joint probability, 
\begin{align}
    P(\vec{\chi},t;\vec{\chi}',t')=&\Theta (t'-t)P^>\left(\vec{\chi},t;\vec{\chi}',t'\right)\nonumber\\+&\Theta (t-t')P^<\left(\vec{\chi},t;\vec{\chi}',t'\right)
\end{align}
with $P^<\left(\vec{\chi},t;\vec{\chi}',t'\right)=P^>\left(\vec{\chi}',t';\vec{\chi},t\right)$.
Taking the double derivative of Eq.~\eqref{eq:general_number_correlation}  with respect to times $t$ and $t'$, and using $x_\alpha,x'_{\alpha'}=n_{\alpha},n'_{\alpha'}$, we can compute the two-time current correlation, which takes the following form,
\begin{widetext}
\begin{align}
\label{eq:two_time_current_correlation_main}
\left\langle I_\alpha(t)I_{\alpha^\prime}(t^\prime)\right\rangle_{\me} &=\partial_t\partial_{t'}\int_{t_0}^{t}ds\int_{t_0}^{t^{\prime}}ds^\prime F(t-s)F(t^{\prime}-s^\prime)\times\partial_{i\chi_\alpha}\partial_{i\chi'_{\alpha'}} P(\vec{\chi},s;\vec{\chi}',s')\big\rvert_{\vec{\chi},\vec{\chi}'=0}\nonumber\\
&=\delta_{\alpha\alpha^\prime}\int_{t_0}^{t}ds\,F\left(t-s \right)F\left(t^{\prime}-s\right)\text{Tr}\{\mathcal A_{\alpha} \, e^{\mathcal L\left(t^\prime-t_0 \right)} \rho_0\} \nonumber \\
& + \int_{t_0}^{t}ds\int_{t_0}^{t^{\prime}}ds^\prime F(t-s)F(t^{\prime}-s^\prime) \Theta\left(s^\prime-s\right)\text{Tr}\{\mathcal I_{\alpha'} e^{\mathcal L \left(s'-s\right)}\mathcal I_{\alpha}\rho(s)\}\nonumber\\
& + \int_{t_0}^{t}ds\int_{t_0}^{t^{\prime}}ds^\prime F(t-s)F(t^{\prime}-s^\prime) \Theta\left(s-s'\right)\text{Tr}\{\mathcal I_\alpha e^{\mathcal L \left(s-s'\right)}\mathcal I_{\alpha'}\rho(s')\}.
\end{align}
\end{widetext}
The detailed derivation of the expression above can be found in Appendix~\ref{appendix:observables_GFCS}. This expression, which agrees with the results of Ref.~\cite{Barchielli1990} obtained in the context of quantum optics, is composed of three terms.
The last two terms contain the particle current superoperators $\mathcal I_{\alpha},\mathcal I_{\alpha^{\prime}}$, defined in Eq.~\eqref{eq:super_current_I}. In the first term, we have identified the dynamical activity superoperator $\mathcal A_\alpha$ defined as \cite{Landi2024},
\begin{align}
\label{eq:activity_superop}
   \mathcal A_\alpha= -\partial^{2}_{\chi_\alpha}\mathcal L(\vec\chi)\big\rvert_{\substack{\vec\chi=0\\\nu_{j\alpha}=1}}=\sum_{j}\left(\mathcal L_{j\alpha}^+ + \mathcal L_{j\alpha}^-\right),
\end{align}
which is a measure of how frequently jumps occur between the system and the reservoir $\alpha$.

In the case of instantaneous jumps, the current correlation function within the master equation approach, denoted as $S^{\me}_{\alpha \alpha'}(t,t')$, is derived from Eq.~\eqref{eq:two_time_current_correlation_main} by substituting $F(t-s)=\delta (s-t)$ and subtracting the product of currents, $I_\alpha(t)^\me I_{\alpha^\prime}^\me(t^\prime)$. The resulting expression simplifies to the following form,
\begin{align}
\label{eq:S_ME}
S^{\me}_{\alpha \alpha'}(t,t')
&= \delta_{\alpha\alpha^{\prime}} \delta(t-t') \Tr{\mathcal A_{\alpha}e^{\mathcal{L} (t-t_0)}\rho_0}\nonumber\\
&+\Theta(t'-t)\Tr{\mathcal{I}_{\alpha'} e^{\mathcal{L} (t'-t)} \mathcal{I}_{\alpha} e^{\mathcal{L} (t-t_0)}\rho_0}\nonumber\\
&+\Theta(t-t')\Tr{\mathcal{I}_{\alpha} e^{\mathcal{L} (t-t')} \mathcal{I}_{\alpha'} e^{\mathcal{L} (t'-t_0)}\rho_0}\nonumber\\
&- \Tr{\mathcal I_{\alpha}e^{\mathcal{L} (t-t_0)}\rho_0}  \Tr{\mathcal I_{\alpha'}e^{\mathcal{L} (t'-t_0)}\rho_0}.
\end{align}
The last line simply corresponds to the product of the average currents at leads $\alpha$ and $\alpha'$ at times $t$ and $t'$, respectively. 
The first line is proportional to a Dirac delta function which, in the Fourier domain will correspond to white noise: it is present at all frequencies with equal strength \cite{Landi2024}. Due to the Kronecker delta, this term is present only for auto-correlations ($\alpha=\alpha^\prime$) and is zero for cross-correlations.
The second and third lines can be obtained alternatively by applying the quantum regression theorem \cite{Carmichael1999}, while the first cannot.
% As we will see in the next sections, the first term is $\mathcal O\left( \Gamma\right)$ in the coupling, while all the other terms are $\mathcal O\left( \Gamma^2\right)$.
Overall, the above expression agrees with previous results obtained in different forms in the context of quantum stochastic master equation \cite{Barchielli1990,Wiseman1993,Korotkov1994,Zoller1997, Landi2024}, all derived under the assumption of instantaneous jumps. This result therefore contributes to one of the main goal of this work, to connect theoretical frameworks for quantum transport and thermodynamics. In the following section, we will return to this important point, recovering the above expressions with a concrete example of a single-level quantum dot using exact principles. 

\SetTblrInner{rowsep=5pt,colsep =5pt}
\begin{table*}
	\centering
		\begin{equation*}
			\begin{tblr}{|c|ccccc|}
				\hline  & \text{Master Equation} & \xleftarrow[\Gamma t \sim \text{const.}]{\Gamma\rightarrow 0 } & \text{Heisenberg Equation} &\xrightarrow[]{t\rightarrow\infty} & \text{Landauer B\"uttiker}   \\
				\hline I_{\alpha}(t) & \Tr\left\{\mathcal I_\alpha e^{\mathcal L \left(t-t_0\right)}\rho_0\right\} & & -i\expval{[\hat{ H},\hat{N}_\alpha]} & & \int \frac{d\epsilon}{2\pi}\mathcal{T}\left(f_{\alpha}-f_{\overline{\alpha}}\right) \\
				\hline J_{\alpha}(t) & \Tr\left\{\mathcal J_\alpha e^{\mathcal L \left(t-t_0\right)}\rho_0\right\} & & -i\expval{[\hat{ H},\hat{H}_{R_\alpha}]} & & \int \frac{d\epsilon}{2\pi}\epsilon \mathcal{T}\left(f_{\alpha}-f_{\overline{\alpha}}\right) \\
				\hline S_{\alpha\alpha^{\prime}}(t,t^{\prime}) & \begin{array}{c} \delta_{\alpha\alpha^{\prime}} \delta(t-t') \Tr{\mathcal A_{\alpha}e^{\mathcal{L} (t-t_0)}\rho_0}\\+\Theta(t'-t)\Tr{\mathcal{I}_{\alpha'} e^{\mathcal{L} (t'-t)} \mathcal{I}_{\alpha} e^{\mathcal{L} (t-t_0)}\rho_0}\\+\Theta(t-t')\Tr{\mathcal{I}_{\alpha} e^{\mathcal{L} (t-t')} \mathcal{I}_{\alpha'} e^{\mathcal{L} (t'-t_0)}\rho_0}\\- \Tr{\mathcal I_{\alpha}e^{\mathcal{L} (t-t_0)}\rho_0}  \Tr{\mathcal I_{\alpha'}e^{\mathcal{L} (t'-t_0)}\rho_0}\end{array} & & \begin{array}{cc}\expval{\hat{I}_{\alpha}(t)\hat{I}_{\alpha^{\prime}}(t^{\prime})}\\-\expval{\hat{I}_{\alpha}(t)}\expval{\hat{I}_{\alpha^{\prime}}(t^{\prime})}\end{array} & &  \begin{array}{c}\int \frac{d\epsilon}{2\pi} \mathcal{T}\sum_{\beta}f_{\beta}\left(1-f_{\beta}\right)\\+\mathcal{T}(1-\mathcal{T})\left(f_L-f_R\right)^2\end{array} \\
				\hline
			\end{tblr}
		\end{equation*}
  \caption{Different rows correspond to different observables: particle current ($I_\alpha$), energy current ($J_{\alpha}$), and  current correlation fuction ($S_{\alpha\alpha^{\prime}}$). Different columns correspond to different approaches: master equation ($\me$), Heisenberg equation of motion (HE), and Landauer B\"uttiker (LB). Each entry in the table provides an expression for computing the corresponding quantity within the relevant framework. In the text, we explain how to obtain results in the $\me$ and LB frameworks from the exact results obtained in the HE framework. This involves applying the weak-coupling protocol for the $\me$ framework and taking the stationary limit $t\to\infty$ for the LB framework. Note that the expression for the correlation function in the $\lb$ regime corresponds to the zero-frequency component of the noise power, i.e. the shot noise given in Eq. \eqref{eq:shot_noise_lb}.} 
\label{Table1}
\end{table*}
\vspace{0.5cm}

\subsection{Landauer-B\"uttiker approach}
\label{sub_sec_LB}

The Landauer-B\"uttiker ($\lb$)~\cite{Landauer1957,Buttiker1986,Datta1997,Blanter2000,Moskalets2011} or scattering matrix (s-matrix) theory of transport  provides a simple and powerful theoretical framework for the description of currents and current correlations in mesoscopic conductors, when coherent and elastic scattering processes are present in the active region. In contrast to the Heisenberg equation and master equation approaches, the Landauer-Büttiker theory is restricted to steady-state observables and non-interacting (quadratic) quantum systems. Nevertheless, this framework accounts for arbitrary system-environment coupling strengths and allows for periodically-driven Hamiltonians~\cite{Moskalets2011}.
Below, we briefly recall the central concepts of this approach, to fix the terms and notations.
This preparatory step is essential as we subsequently delve into a specific example, utilizing the $\lb$ formalism to benchmark results against the $\he$ and $\me$ approaches.

As illustrated in Fig.~\ref{fig:setups} (c), the central object is the s-matrix $\mathcal{S}$, which relates the incoming and outgoing annihilation fermionic operators in the reservoirs, $\hat{b}_\alpha$ and $\hat{a}_\alpha$, respectively, with $\alpha = 1, \ldots, N$. For a setup with $N$ reservoirs, and one transport channel per lead, it is therefore a $N \times N$ matrix with elements denoted $\mathcal{S}_{\alpha \beta}$, defined as,
\begin{equation}
\label{scattering_eq}
\hat{b}_\alpha(\epsilon)=\sum_{\beta}\mathcal{S}_{\alpha \beta}(\epsilon)\hat{a}_{\beta}(\epsilon).
\end{equation}
A corresponding equation establishes the connection between the creation operators $\hat{b}^\dagger_\alpha$ and $\hat{a}^\dagger_\alpha$ via $\mathcal{S}^*(\epsilon)$.
Particle conservation is ensured through the unitarity of the s-matrix, $\mathcal{S} \mathcal{S}^\dagger = \mathcal{S}^\dagger \mathcal{S} = \mathds{1}$.
The particle current operator can be expressed in terms of the s-matrix, and incoming and outgoing operators in the following way \citep{Blanter2000},
\begin{equation}
\label{current_1}
\hat I_{\alpha}^{LB}(t)=\!\!\int_{-\infty}^{\infty} \!\!\frac{d\epsilon~d\epsilon^{\prime}}{2\pi}e^{-i(\epsilon-\epsilon^{\prime})t}\left[\hat{a}_{\alpha}^{\dagger}(\epsilon)\hat{a}_{\alpha}(\epsilon^{\prime})-\hat{b}_{\alpha}^{\dagger}(\epsilon)\hat{b}_{\alpha}(\epsilon^{\prime})\right].
\end{equation}
The average particle current is calculated by substituting Eq.~\eqref{scattering_eq} in the above expression, taking the statistical average, and using the relation $\expval{\hat a_{\alpha}^{\dagger}\left(\epsilon\right)\hat a_{\beta}\left(\epsilon^{\prime}\right)}=\delta_{\alpha\beta}\delta{\left(\epsilon-\epsilon^{\prime}\right)}f_{\alpha}\left(\epsilon\right)$, where $f_{\alpha}\left(\epsilon\right)$ is the Fermi-Dirac distribution function. 
The method explained so far is valid for an arbitrary number of reservoirs; however, simpler expressions for currents and current correlations can be obtained in the case of a two-terminal scenario, i.e.,~$\alpha=L,R$ (for left and right respectively). In this case, the average particle current in the steady state takes the form, 
\begin{align}
\label{eq:LB_particle}
I^\lb_\alpha &= \sum_{\beta} \int \frac{d\epsilon}{2\pi} \mathcal{T}(\epsilon)  \left(f_{\alpha}(\epsilon) -f_{\bar{\alpha}}(\epsilon)\right),
\end{align}
where $\bar{\alpha}$ denotes the opposite lead of $\alpha=L,R$, and where we introduced the so-called transmission function $\mathcal{T}(\epsilon)=\abs{\mathcal{S}_{\alpha\bar{\alpha}}}^2$. Similarly we can obtain the energy current,
\begin{align}
\label{eq:LB_energy}
J^\lb_\alpha &=  \int_{-\infty}^\infty \frac{d\epsilon}{2\pi} \, \epsilon  \, \mathcal{T}(\epsilon)   \left(f_{\alpha}(\epsilon) -f_{\bar{\alpha}}(\epsilon)\right),
\end{align}
which simply contains an extra energy term $\epsilon$ under the integral. For quadratic Hamiltonians like Eq.~\eqref{eq:totalH}, the transmission function $\mathcal{T}$ can be computed from the Green functions \cite{Datta1997, Nazarov2009}, providing a general method to access it,
\begin{eqnarray}
\label{eq:T_from_Green}
    \mathcal{T}(\epsilon) = \Tr\{ G^{a}(\epsilon) \Gamma^\alpha G^{r}(\epsilon) \Gamma^{\bar{\alpha}}\}\,,
\end{eqnarray}
with $G^{a,r}(\epsilon)$ being the Fourier transforms of the retarded $G^r(t) = -i \Theta(t)  \expval{\{ d_i (t), d_j^\dagger(0)}$ and advanced $G^a(t) = i \Theta(-t)  \expval{\{ d_i (t), d_j^\dagger(0)}$ Green's functions, that can be expressed from the lesser and greater Green's functions given in App.~\ref{Green_functions}. The matrices $\Gamma^{\alpha, \bar{\alpha}}$ have elements $\Gamma^{\alpha, \bar{\alpha}}_{ij}$ defined in Eq.~\eqref{eq:tunnel_rate}.

These average currents correspond to the steady-state limit of the currents obtained from the $\he$ approach, valid for all system-bath coupling strengths in the limit of $t\to \infty$.
% ,
% \begin{eqnarray}
%     I^\lb_\alpha = \lim_{t\rightarrow \infty} I^\he_\alpha (t)\,.
% \end{eqnarray}
In the example of a single-level quantum dot treated in the following section, we will show how to take this limit explicitly.\\ 
The Landauer-B\"uttiker approach also enables the computation of the steady-state current correlation function. In the stationary regime, due to time-translation invariance, the current correlation function depends solely on the time difference, denoted as $S^\lb_{\alpha \alpha'}(\tau)$ with $\tau = t - t'$.
Furthermore, within the $\lb$ framework, it is customary to analyze the spectrum of $S^\lb_{\alpha \alpha'}(\omega)$ (also known as noise power), which corresponds to the Fourier transform~\cite{Blanter2000},
\begin{eqnarray}
    S^\lb_{\alpha \alpha'}(\omega) = \int_{-\infty}^{\infty}d\tau e^{-i\omega \tau} S^\lb_{\alpha \alpha'}(\tau).
\end{eqnarray}It can be shown that the finite-frequency noise satisfies $S_{\alpha \alpha'}(\omega) = S_{\alpha' \alpha}(-\omega)$. At zero frequency, the so-called shot noise takes the following simple form for the auto-correlation functions,
\begin{align}
\label{eq:shot_noise_lb}
S_{LL}^\lb(\omega=0) &= S_{RR}^\lb(\omega=0) \nonumber \\
&=  \int \frac{d\epsilon}{2\pi} \mathcal{T}(\epsilon)\sum_{\beta=L,R}f_{\beta}(\epsilon)\left(1-f_{\beta}(\epsilon)\right)\nonumber\\
&+\mathcal{T}(\epsilon)(1-\mathcal{T}(\epsilon))\left(f_L(\epsilon)-f_R(\epsilon)\right)^2.
\end{align}
The auto- and cross-correlation functions are then connected via the following relation,
\begin{eqnarray}
\label{shot_noise_lb_relations}
S_{\alpha \alpha}^\lb(\omega=0) &= -S^\lb_{\alpha \beta}(\omega =0)\,, \quad \beta \neq \alpha\,.
\end{eqnarray}
Zero-frequency exact expressions for the multi-terminal case can be found in Refs.~\cite{Buttiker1992, Blanter2000}.

In the subsequent sections, we will focus on the specific case of a single quantum dot coupled to two reservoirs. We will use the shot noise $S_{\alpha\alpha'}^\lb(\omega=0)$ derived from the $\lb$ approach to benchmark our steady-state expressions obtained from the exact $\he$ approach for all system-reservoir coupling strengths, and from the $\me$ approach for weak coupling. 

In Table~I, we summarize the main formal expressions of average currents and current correlation functions discussed so far.

\section{Application: The single-level quantum dot}
\label{sec:HE_QD}
\vspace{0.3cm}
\begin{figure}[!htb]%[h!]
\centering
\includegraphics[width=0.8\columnwidth]{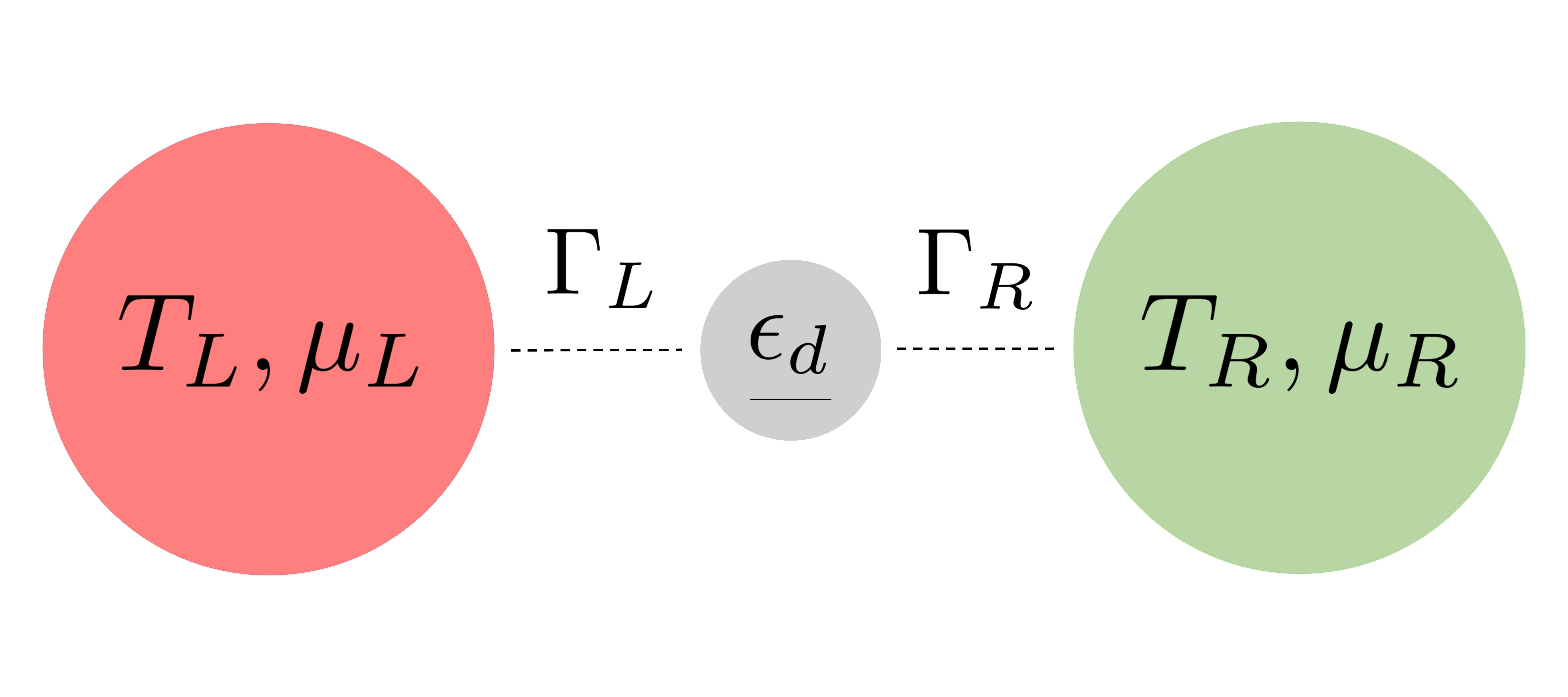}
\caption{Scheme of the single-level quantum dot with energy $\epsilon_d$, coupled to two reservoirs at temperature $T_\alpha$ and chemical potential $\mu_\alpha$, with $\alpha=L,R$. In the WBL, $\Gamma_L$ and $\Gamma_R$ represent the energy-independent bare tunneling rates describing the coupling of the dot with the left and right reservoir respectively.} 
\label{fig:single_quantum_dot}
\end{figure}
\vspace{0.3cm}
In this section, we analyze the specific case of a single-energy level quantum dot (QD) coupled with two reservoirs as depicted in Fig.~\ref{fig:single_quantum_dot}.  
The Hamiltonian describing this system corresponds to the model introduced in Eq.~\eqref{eq:totalH} with $D=1$ and $N=2$. Specifically, by indicating with $\epsilon_d$ the energy of the dot, the Hamiltonian takes the following form,
\begin{equation}
\label{eq:totalH_QD}
		\hat{H}=\epsilon_{d} \hat{d}^{\dagger}\hat{d}+\sum_{k \alpha}\epsilon_{k\alpha}\hat{c}_{k\alpha}^{\dagger}\hat{c}_{k\alpha}+\sum_{k \alpha}\left[t_{k\alpha}^{*}\hat{c}_{k\alpha}^{\dagger}\hat{d}+t_{k\alpha} \hat{d}^{\dagger}\hat{c}_{k\alpha}\right].
\end{equation}
Each fermionic reservoir $\alpha=L,R$ is characterized by its Fermi-Dirac distribution, $f_{\alpha}(\epsilon)=\{e^{(\epsilon-\mu_{\alpha})/ T_{\alpha}}+1\}^{-1}$, with $\mu_\alpha$ the chemical potential and $T_\alpha$ the temperature.

In the following, we start first by computing the exact expressions of currents and current correlation function obtained from the $\he$ approach (Sec. \ref{sec_HE_for_QD}), and benchmark them with the $\lb$ results in the steady-state regime (Sec. \ref{sec:QD_benchmark_LB}). Moving forward, in Sec. \ref{sec_ME_for_QD}, we present the results obtained from the $\me$ approach. Finally, in Sec.\ref{sec:connection}, we delve into the discussion of the interconnections between these three frameworks, as outlined in Fig.\ref{fig:setups}.

%%%%%%%%%%%%%%%%%%%%%%%%%%%%%%%%%%%%%%%%%%%%%%%%%%%%%%%%%%%%%%%%%%%%%%
	
\section{Heisenberg equation for the single-level QD}
\label{sec_HE_for_QD}

Inserting Eq.~\eqref{eq:totalH_QD} into the Heisenberg equations Eqs.~\eqref{eq:Heisenberg_d} and \eqref{eq:Heisenberg_c}, we obtain the following evolution equations for the operators $\hat d$ and $\hat c_{k\alpha}$ in the case of the single-level QD~\cite{Mitchison2018,Schaller2014},
\begin{eqnarray}
\label{diff_Eq_d_c}
\frac{d}{d t} \hat{d} &=& i[\hat{ H}, \hat{d}]=-i\epsilon_d \hat{d} -i \sum_{k \alpha} t_{k \alpha} \hat{c}_{k \alpha}\,,\nonumber \\
\frac{d}{d t} \hat{c}_{k \alpha} &=& i[\hat{ H}, \hat{c}_{k \alpha}]=-i \epsilon_{k \alpha} \hat{c}_{k \alpha} -i\, t_{k \alpha}^* \hat{d} \,.
\end{eqnarray}
% \begin{eqnarray}
% \label{diff_Eq_d_c}
% \frac{d}{d t} \hat{d}^{\dagger} &=& i[\hat{ H}, \hat{d}^{\dagger}]=i\epsilon_d \hat{d}^{\dagger} +i \sum_{k, \alpha} t_{k \alpha}^{*} \hat{c}_{k \alpha}^{\dagger} \,,\nonumber \\
% \frac{d}{d t} \hat{c}_{k \alpha}^{\dagger} &=& i[\hat{ H}, \hat{c}_{k \alpha}^{\dagger}]=i \epsilon_{k \alpha} \hat{c}_{k \alpha}^{\dagger}+i t_{k \alpha} \hat{d}^{\dagger}\,.
% \end{eqnarray}
The Heisenberg equations for $\hat d^{\dagger}$ and $\hat c_{k\alpha}^{\dagger}$ 
are simply the hermitian conjugate of the above. 
Similarly, by inserting Eq.~\eqref{eq:totalH_QD} into Eqs.~\eqref{eq:def_HE_particle_current} and \eqref{eq:def_HE_energy_current}, we can compute the particle and energy currents,
\begin{eqnarray}
I_\alpha^{\he}(t) &=&  -2 \text{Im}\left\{\sum_{k}t_{k\alpha}^{*}\expval{\hat{c}_{k\alpha}^{\dagger}(t)\hat{d}(t)}\right\} ,\label{eq:IHEstarting} \\
J_{\alpha}^\he(t) &=& -2 \text{Im}\left\{\sum_{k}\epsilon_{k\alpha}t_{k\alpha}^{*}\expval{\hat{c}_{k\alpha}^{\dagger}(t)\hat{d}(t)}\right\}. \label{eq:JHEstarting}
\end{eqnarray}

The current correlation function is derived by substituting the particle current operator expression into the definition given in Eq.~\eqref{def_noise}. Employing Wick's theorem, it can be expressed as a sum of products of two-point correlators, as explained in Appendix \ref{app_Noise}. The resulting form is as follows,
\begin{align}
\label{eq:noiseHE}
S^\he_{\alpha\alpha^{\prime}}(t,t^{\prime}) &=-\sum_{k k^{\prime}}\left\{t_{k\alpha}^{*}t_{k^{\prime}\alpha^{\prime}}^{*}\expval{\hat{c}_{k\alpha}^{\dagger}(t)\hat{d}(t^{\prime})}\expval{\hat{d}(t)\hat{c}_{k^{\prime}\alpha^{\prime}}^{\dagger}(t^{\prime})}\right.\nonumber\\
& \left.-t_{k\alpha}^{*}t_{k^{\prime}\alpha^{\prime}}\expval{\hat{c}_{k\alpha}^{\dagger}(t)\hat{c}_{k^{\prime}\alpha^{\prime}}(t^{\prime})}\expval{\hat{d}(t) \hat{d}^{\dagger}(t^{\prime})}\right.\nonumber\\
& -\left.t_{k\alpha}t_{k^{\prime}\alpha^{\prime}}^{*}\expval{ \hat{d}^{\dagger}(t)\hat{d}(t^{\prime})}\expval{\hat{c}_{k\alpha}(t)\hat{c}_{k^{\prime}\alpha^{\prime}}^{\dagger}(t^{\prime})}\right.\nonumber\\
&\left.+t_{k\alpha}t_{k^{\prime}\alpha^{\prime}}\expval{ \hat{d}^{\dagger}(t)\hat{c}_{k^{\prime}\alpha^{\prime}}(t^{\prime})}\expval{\hat{c}_{k\alpha}(t) \hat{d}^{\dagger}(t^{\prime})}\right\} .\nonumber \\
&
\end{align}
It is worth noting that while auto- ($\alpha=\alpha'$) and cross- ($\alpha\neq \alpha'$) correlations are related in the stationary regime according to Eq.\eqref{shot_noise_lb_relations}, this relationship does not hold in general in the transient regime. Notably, as elucidated in Appendix~\ref{app_Noise}, the two-point correlators in the above expression correspond to lesser Green functions for the dot, reservoirs, or between the dot and reservoirs. 
% These correlators intuitively capture the propagator dynamics of the dot and the reservoirs independently, as well as the propagator associated with tunneling \sk{via or through the dot[What does this mean?]}.

\vspace{-.2cm}
\subsection{HE exact solutions for the QD}

In the rest of the work, analytical solutions are derived by assuming the wide-band limit (WBL) as discussed along with Eq.~\eqref{eq:WBL}, which entails a broad spectral density of the reservoirs with respect to the energy scales of the open quantum system, $\Gamma_{\alpha}(\epsilon) \equiv \Gamma_\alpha$, with $\alpha=L,R$.
This assumption is justified, for example, in case of flat or wide-Lorentzian spectra. While closed-form solutions are possible in certain cases for a Lorentzian spectrum (by going in the Laplace space~\cite{Zedler2009}) beyond the WBL, they are typically far more cumbersome. We therefore focus exclusively on the WBL which leads to simpler expressions, while capturing essential physics.
% \gh{This assumption is verified for a flat or a wide Lorentzian spectrum for instance. Let us note that beyond the WBL, analytical results can be obtained if a Lorentzian spectrum is assumed, by going into the Laplace space \cite{Zedler2009}. For clarity, we restrict to WBL which leads to simpler analytical expressions.} 
For convenience, we also define the total tunneling rate,
\begin{equation}
\label{eq:Gamma}
    \Gamma = \Gamma_L + \Gamma_R\,.
\end{equation}
Within the WBL, we obtain the analytical formal solution of the coupled differential equations~\eqref{diff_Eq_d_c}, for the operator $\hat{d}(t)$,
\begin{align}
\label{d_op_sol}
\hat{d}(t) &= e^{\left(-\frac{\Gamma}{2}-i\epsilon_d\right)(t-t_0)} \hat{d}(t_0)+\!\int_{t_0}^{t} \!\!ds~  e^{\left(-\frac{\Gamma}{2}-i\epsilon_d\right)\left(t-s\right)}\hat{\xi}(s), \nonumber\\
\end{align}
% \begin{align}
% \label{d_op_sol}
% \hat{d}^{\dagger}(t) &= e^{\left(-\frac{\Gamma}{2}+i\epsilon_d\right)(t-t_0)} \hat{d}^{\dagger}(t_0)+\!\int_{t_0}^{t} \!\!ds~  e^{\left(-\frac{\Gamma}{2}+i\epsilon_d\right)\left(t-s\right)}\hat{\xi}^{\dagger}(s), \nonumber\\
% \end{align}
and for $\hat{c}_{k \alpha}(t)$,
\begin{align}
\hat{c}_{k \alpha}(t) &= e^{-i \epsilon_{k \alpha} (t-t_0)} \hat{c}_{k\alpha}(t_0)- i \int_{t_0}^{t} ds~ e^{-i \epsilon_{k \alpha}(t-s)} t_{k \alpha}^* \hat{d}(s)\nonumber\\
	&= e^{-i \epsilon_{k \alpha} (t-t_0)} \hat{c}_{k\alpha}(t_0)\nonumber \\
	&-i \int_{t_0}^{t} ds~ e^{\left(-\frac{\Gamma}{2}-i\epsilon_d\right)(s-t_0)} e^{-i \epsilon_{k \alpha}(t-s)} t_{k \alpha}^* \hat{d}(t_0)\nonumber\\
	&-i \int_{t_0}^{t}\!\! ds \int_{t_0}^{s}\!\! ds^{\prime}~ e^{-i \epsilon_{k \alpha}(t-s)}e^{\left(-\frac{\Gamma}{2}-i\epsilon_d\right)(s-s^{\prime})}t_{k \alpha}^* \hat{\xi} \left(s^{\prime}\right). \label{c_op_sol} \nonumber \\
\end{align}
where we have introduced the function, 
% \vspace{-0.25cm}
% $\label{xi}
% \hat{\xi}(t)=-i\sum_{k\alpha}t_{k\alpha} e^{-i \epsilon_{k \alpha} (t-t_0)} \hat{c}_{k\alpha}(t_0).$
\begin{equation}
\label{xi}
\hat{\xi}(t)=-i\sum_{k\alpha}t_{k\alpha} e^{-i \epsilon_{k \alpha} (t-t_0)} \hat{c}_{k\alpha}(t_0).
\end{equation}

Additional details for the derivation of the above solutions can be found in Appendix.~\ref{appendix_A}.
Let us note that similar expressions for this paradigmatic model were already discussed in Refs.~\cite{Schaller2014,Rolandi2023}.
The solutions for $\hat{d}^{\dagger}(t)$ and $\hat{c}_{k\alpha}^{\dagger}(t)$ can be obtained by taking the hermitian conjugate of Eqs.~\eqref{d_op_sol} and \eqref{c_op_sol}, respectively. We now substitute the above solutions into Eqs.~\eqref{eq:IHEstarting} and \eqref{eq:JHEstarting}, and apply the initial conditions at time $t_0$: $\expval{\hat{c}_{k\alpha}^{\dagger}(t_0)\hat{c}_{k^{\prime}\alpha^{\prime}}(t_0)} = \delta_{k\alpha,k^{\prime}\alpha^{\prime}} f_{\alpha}(\epsilon_{k\alpha})$ and $\expval{\hat{d}^{\dagger}(t_0)\hat{d}(t_0)} = n_d$.
We finally obtain the time-dependent expressions for the particle and energy currents,
\begin{widetext}
\begin{eqnarray}
\label{eq:IHE}
I_\alpha^\he(t) &=& \Gamma_\alpha\sum_{\beta} \frac{\Gamma_\beta}{\Gamma}\left\{-n_d\,e^{-\Gamma\left(t-t_0\right)} +  \left(A_\alpha^0-A_\beta^0\right) -  e^{-\frac{\Gamma}{2}\left(t-t_0\right)}\left[A_\beta^0 +B_\alpha^0- 2 B_\beta^0 +  C_\alpha^0 \right]\right\} \\
\label{eq:JHE}
J_\alpha^\he(t) &=&\Gamma_\alpha\sum_{\beta} \frac{\Gamma_\beta}{\Gamma}\left\{ -\epsilon_d n_d\,e^{-\Gamma\left(t-t_0\right)} +  \left(A_\alpha^1 -A_\beta^1\right) - e^{-\frac{\Gamma}{2}\left(t-t_0\right)} \left[B_\alpha^1 - B_\beta^1 +C_\alpha^1\right]\right\},
\end{eqnarray}
\end{widetext}
with the definitions of the integrals,
\begin{align}
\label{eq:A_coeff}
A_{\gamma}^m &\coloneqq \Gamma\int\frac{d\epsilon}{2\pi}\frac{( \epsilon+\epsilon_d)^m}{\frac{\Gamma^2}{4}+\epsilon^2} f_{\gamma}( \epsilon+\epsilon_d), \quad m=0,1\,,\\
\label{eq:B_coeff}
B_{\gamma}^m &\coloneqq \Gamma\int\frac{d\epsilon}{2\pi}\frac{( \epsilon+\epsilon_d)^m}{\frac{\Gamma^2}{4}+\epsilon^2} f_{\gamma}( \epsilon+\epsilon_d) \cos\left( \epsilon\left(t-t_0 \right)\right),\\
\label{eq:C_coeff}
 C_{\gamma}^m &\coloneqq 2\int\frac{d\epsilon}{2\pi}\frac{( \epsilon+\epsilon_d)^m}{\frac{\Gamma^2}{4}+\epsilon^2} f_{\gamma}( \epsilon+\epsilon_d) \epsilon \sin\left( \epsilon\left(t-t_0 \right)\right).
\end{align}
In Eqs.~\eqref{eq:IHE} and \eqref{eq:JHE}, the first term that depends on initial occupation of the dot $n_d$ corresponds to a shift in the current that decays exponentially in time. Since the steady state cannot depend on the initial state, the contribution of $n_d$ naturally has to be exponentially decaying. The second term, being independent of time, corresponds to the steady-state current. The third term is an oscillating, exponentially decaying contribution, which does not depend on the initial state of the system.

\begin{figure}[htb!]
	\centering
\includegraphics[width=1\columnwidth]{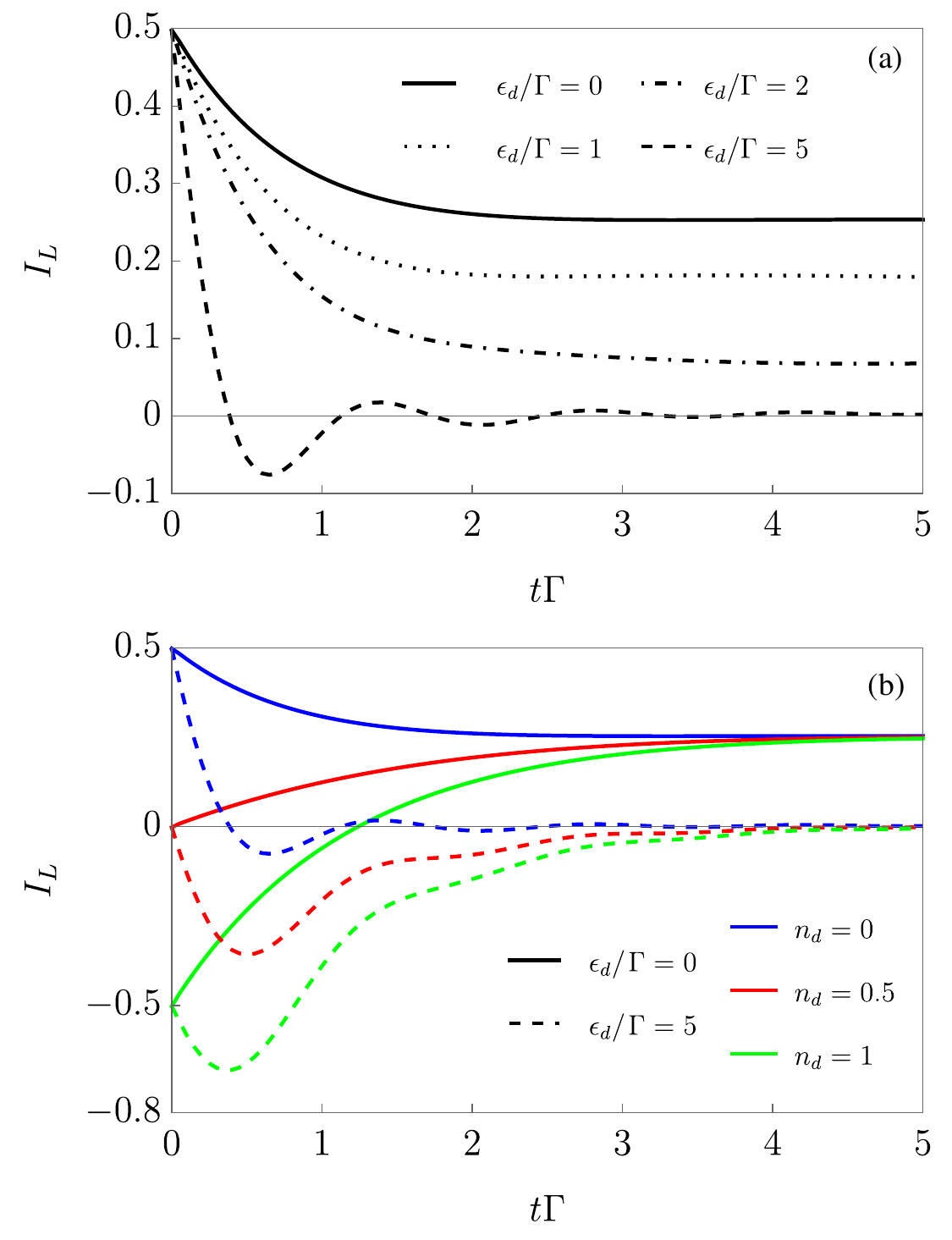}
\caption{Particle current as a function of $t\Gamma$ with (a) constant $n_d=0$ and different values of $\epsilon_d/\Gamma=0,\,1,\,2,\,5$, and (b) with different values of $n_d=0,\,0.5,\,1$ and two values of $\epsilon_d/\Gamma=0,\,5$. The other parameters are $\Gamma_L=\Gamma_R=1$, $\mu_L=-\mu_R=\Gamma/2$ and $T_L=T_R=\Gamma/100$ ($\hbar,k_B=1$). }
\label{fig:currents1}
\end{figure}

In Fig. \ref{fig:currents1}, we illustrate the results of Eqs.~\eqref{eq:IHE} and \eqref{eq:JHE} by plotting the particle and energy currents as functions of time, for various energies of the dot in panel (a), and fixed initial dot occupation $n_d=0$, and then for different $n_d$, for two values of dot energies in panel (b). 
We have considered the two reservoirs to be at the same temperature ($T_L=T_R$) with opposite chemical potentials ($\mu_L=-\mu_R$), such that the transport energy window is symmetric around the zero energy.
When the energy of the dot lies at the centre of the transport window ($\epsilon_d/\Gamma=0$), we find classical behaviour of the current, marked by exponential decay towards the steady state. However, as the energy of the dot moves away from the transport window, we start probing quantum coherent effects such as quantum tunnelling, seen clearly in the form of oscillations in the current at large $\epsilon_d$. Interestingly, such behaviour cannot be seen with the master equation approach, which only involves simple exponential decay toward the steady state, as we will see in the next section on ME. 

For the current correlation function, we proceed in a similar way as for the currents. We insert Eqs.~\eqref{d_op_sol} and \eqref{c_op_sol} into Eq.~\eqref{eq:noiseHE}. A long calculation leads to the following exact analytical expression of $S^\he_{\alpha\alpha^{\prime}}(t,t^{\prime})$, valid at all times,
\begin{widetext}
\begin{align}
\label{SHE}
S^\he_{\alpha\alpha^{\prime}}(t,t^{\prime})&= \Gamma_{\alpha}\Gamma_{\alpha^{\prime}} \sum_{\beta,\beta^{\prime}} \frac{\Gamma_{\beta}\Gamma_{\beta^{\prime}}}{\Gamma^2}\left[\Lambda_0(t,t^{\prime})-\Lambda^{(1)}_{\alpha}(t,t^{\prime})+\Lambda^{(2)}_{\beta}(t,t^{\prime})\right]\left[\overline{\Lambda}_0(t^{\prime},t)-\overline{\Lambda}^{(1)}_{\alpha^{\prime}}(t^{\prime},t)+\overline{\Lambda}^{(2)}_{\beta^{\prime}}(t^{\prime},t)\right]\nonumber\\
&+\Gamma_{\alpha}\Gamma_{\alpha^{\prime}}\sum_{\beta,\beta^{\prime}} \frac{\Gamma_{\beta}\Gamma_{\beta^{\prime}}}{\Gamma^2}\left[\Lambda_0(t,t^{\prime})-\Lambda^{(1)}_{\alpha}(t,t^{\prime})-\Lambda^{(1)}_{\alpha^{\prime}}(t^{\prime},t)^*+\Lambda^{(2)}_{\beta}(t,t^{\prime})\right]\left[\overline{\Lambda}_0(t^{\prime},t)+\overline{\Lambda}^{(2)}_{\beta^{\prime}}(t^{\prime},t)\right]\nonumber\\
&+\Gamma_{\alpha}\delta_{\alpha\alpha^{\prime}}\sum_{\beta}\Gamma_{\beta}\Lambda_{\alpha}(t,t')\left[\overline{\Lambda}_0(t^{\prime},t)+\overline{\Lambda}^{(2)}_{\beta}(t^{\prime},t)\right]+\left\{c.c.\quad\text{with} \quad \Lambda \leftrightarrow \overline{\Lambda}\right\},
\end{align}
\end{widetext} 
with the definitions,
\begin{equation}
\label{Lambda_HE}
\begin{aligned}
& \Lambda_0(t,t^{\prime}) = \frac{1}{2} e^{-\frac{\Gamma}{2}(t+t^{\prime})}e^{\Gamma t_0}e^{i \epsilon_0 (t-t^{\prime})}n_d\\
&\Lambda_{\gamma}(t,t^{\prime}) = \frac{2}{\Gamma}\int \!\frac{d\epsilon}{2\pi} e^{i\epsilon\left(t-t^\prime\right)} f_\gamma\left(\epsilon\right)\\
&\Lambda^{(1)}_{\gamma}(t,t^{\prime}) = 2\int \!\frac{d\epsilon}{2\pi}  e^{i\epsilon\left(t-t'\right)}g_{-}(\epsilon-\epsilon_d,t)f_\gamma\left(\epsilon\right)\\
&\Lambda^{(2)}_{\gamma}(t,t^{\prime}) = 2\Gamma\!\!\int\!\! \frac{d\epsilon}{2\pi} e^{i\epsilon\left(t-t'\right)}g_{-}(\epsilon-\epsilon_d,t)g_{+}(\epsilon-\epsilon_d,t^{\prime}) f_\gamma(\epsilon),
\end{aligned}
\end{equation}

% \begin{align}
% \label{Lambda_HE}
% & \Lambda_0(t,t^{\prime}) = \frac{1}{2} e^{-\frac{\Gamma}{2}(t+t^{\prime})}e^{\Gamma t_0}e^{i \epsilon_0 (t-t^{\prime})}n_d\nonumber\\
% &\Lambda_{\gamma}(t,t^{\prime}) = \frac{2}{\Gamma}\int \!\frac{d\epsilon}{2\pi} e^{i\epsilon\left(t-t^\prime\right)} f_\gamma\left(\epsilon\right)\nonumber\\
% &\Lambda^{(1)}_{\gamma}(t,t^{\prime}) = 2\int \!\frac{d\epsilon}{2\pi}  e^{i\epsilon\left(t-t'\right)}g_{-}(\epsilon-\epsilon_d,t)f_\gamma\left(\epsilon\right)\nonumber\\
% &\Lambda^{(2)}_{\gamma}(t,t^{\prime}) = 2\Gamma\!\!\int\!\! \frac{d\epsilon}{2\pi} e^{i\epsilon\left(t-t'\right)}g_{-}(\epsilon-\epsilon_d,t)g_{+}(\epsilon-\epsilon_d,t^{\prime}) f_\gamma(\epsilon),
% \end{align}
and,
\begin{equation}
    g_{\pm}(\epsilon,t)=\frac{e^{-\left(\frac{\Gamma}{2}\pm i\epsilon\right)\frac{t-t_0}{2}}}{\frac{\Gamma}{2}\pm i\epsilon}\sinh{\left[\left(\frac{\Gamma}{2}\pm i\epsilon\right)\frac{t-t_0}{2}\right]},
\end{equation}
where the $\overline{\Lambda}$ functions are obtained from Eqs.~\eqref{Lambda_HE} substituting $n_d\rightarrow (1-n_d)$ and $f_{\gamma}\left(\epsilon \right)\rightarrow \left(1-f_{\gamma}\left(\epsilon \right)\right)$. Let us remark that this general exact expression of the correlation function is complex and satisfies the relation $S_{\alpha\alpha^{\prime}}(t,t')=S^*_{\alpha^{\prime}\alpha}(t',t)$.
% \begin{figure*}[htb]
%  	\centering
%  \includegraphics[width=0.7\paperwidth]{Particle_Current_HE_horizontal.pdf.pdf}
%  \caption{Particle current as a function of $t\Gamma$ with (a) constant $n_d=0$ and different values of $\epsilon_d/\Gamma=0,\,1,\,2,\,5$, and (b) with different values of $n_d=0,\,0.5,\,1$ and two values of $\epsilon_d/\Gamma=0,\,5$. The other parameters are $\Gamma_L=\Gamma_R=1$, $\mu_L=-\mu_R=\Gamma/2$ and $T_L=T_R=\Gamma/100$ ($\hbar,k_B=1$). }
%  \label{fig:currents1}
% \end{figure*}

We note that the above expression of the correlation function can equivalently be obtained using non-equilibrium Green’s function techniques in the Schwinger-Keldysh theory~\cite{Schwinger1961}, as explained in Ref.~\cite{Zhang2014}. Our expression \eqref{SHE} singles out the explicit time-dependence of each term through the $\Lambda$-functions defined in Eq.~\eqref{Lambda_HE}, as well as the dependence on the coupling rates. This considerably simplifies the calculations of the long-time limit of the current correlation functions as discussed below, and simplifies applying the weak-coupling protocol to connect HE and ME frameworks, as presented in Sec.~\ref{sec:connection}.

\subsection{Steady-state limit: Benchmark with LB}
\label{sec:QD_benchmark_LB}

In this section, we recall known expressions from the LB approach for the steady-state particle and energy currents, as well as for the shot noise, for the single-level QD in a two-terminal setup. These expressions will then be used to benchmark the exact results in the steady state obtained from the HE in the previous section.
As discussed in Sec.~\ref{sub_sec_LB}, the key quantity of interest in the LB approach is the transmission probability, which has to be computed explicitly, typically from Green functions as discussed before, see Eq.~\eqref{eq:T_from_Green}. For general interest, we also provide in Appendix~\ref{app:trans} an alternative and intuitive derivation for this simple model. It takes the following form~\cite{Stone1985,Buttiker1986,Buttiker1988,Datta1997},
\begin{eqnarray}
\label{eq:Transmission_function}
\mathcal T\left(\epsilon \right) = \frac{\Gamma_L\Gamma_R}{\frac{\Gamma^2}{4}+(\epsilon-\epsilon_d)^2},
\end{eqnarray}
and corresponds to a Lorentzian centered around the energy of the dot $\epsilon_d$, with a width set by the total tunneling rate $\Gamma = \Gamma_L + \Gamma_R$. Inserting $\mathcal{T}(\epsilon)$ into Eqs.~\eqref{eq:LB_particle} and \eqref{eq:LB_energy}, we obtain the well-known expressions,
\begin{align}
\label{eq:ILB}
I_L^{\lb} &=\!\!-I_R^{\lb}=\!\! \int \frac{d \epsilon}{2 \pi} \frac{\Gamma_L \Gamma_R}{\frac{\Gamma^2}{4}+(\epsilon-\epsilon_d)^2}  \big( f_L(\epsilon)\!\! - f_R(\epsilon) \big),\\
J_L^{\lb} &=\!\!-J_R^{\lb} =\!\!   \int \frac{d \epsilon}{2 \pi} \frac{\epsilon\,\Gamma_L \Gamma_R}{\frac{\Gamma^2}{4}+(\epsilon-\epsilon_d)^2}  \big( f_L(\epsilon)\!\! - f_R(\epsilon) \big).\label{eq:JLB}
\end{align}

\subsubsection{Benchmarking with $\lb$}
% \subsubsection{Benchmarking average currents}

To benchmark the exact results we obtained within the $\he$ approach, we take the limit $t \rightarrow \infty$ (or equivalently taking $t_0\rightarrow -\infty$) in Eqs.~\eqref{eq:IHE} and \eqref{eq:JHE}. In this limit, all terms proportional to exponential time-decaying functions vanish, which lead us to the compact expressions,
\begin{align}
\label{IHE_ss}
I_{\alpha,ss}^\he \equiv \lim_{t \to \infty} I_\alpha^\he (t ) &= \Gamma_\alpha \sum_{\beta} \frac{\Gamma_\beta}{\Gamma}\left[A_\alpha^0-A_\beta^0 \right], \\
\label{JHE_ss}
J_{\alpha,ss}^\he \equiv \lim_{t \to \infty} J_\alpha^\he(t ) &= \Gamma_\alpha \sum_{\beta}\frac{\Gamma_\beta}{\Gamma} \left(A_\alpha^1 -A_\beta^1\right).
\end{align}
It is simple to show that, by substituting the integrals $A_{\gamma}^0,A_{\gamma}^1$ defined in Eq. \eqref{eq:A_coeff}, in the above equations and by performing the change of variable $\epsilon\to(\epsilon-\epsilon_d)$, we recover the well-known expressions for the steady-state particle and energy currents in the reservoir $\alpha=L,R$ for a single-level QD in a two-terminal setup, coinciding exactly with Eqs.~\eqref{eq:ILB}-\eqref{eq:JLB},
\begin{align}
\label{eq:Iss_Jss_MEvs_LB}
I_{\alpha,ss}^\he &=  I_\alpha^{\lb} \quad \text{and} \quad J_{\alpha,ss}^\he = J_\alpha^{\lb}.
\end{align}

% \subsubsection{Benchmarking current correlation functions}

  Similarly to the average currents, the long-time limit of the current fluctuation functions $S^{\he}_{\alpha \alpha',ss}$ is obtained by taking the limit $t\rightarrow \infty$ in Eqs.~\eqref{Lambda_HE},
\begin{equation}
\label{Lambda_long_time_1}
\begin{aligned}
&\lim_{t \to \infty}	\Lambda_0(t,t') = 0,\\
&\lim_{t \to \infty}	\Lambda_{\gamma}(t,t') =\frac{2}{\Gamma}\int \frac{d\epsilon}{2\pi}  f_\gamma\left(\epsilon\right)e^{i\epsilon(t-t^\prime)}\, ,\\
&\lim_{t \to \infty}	\Lambda^{(1)}_{\gamma}(t,t') = \int \frac{d\epsilon}{2\pi}\frac{e^{i \epsilon(t-t^\prime)}}{\frac{\Gamma}{2}-i(\epsilon-\epsilon_d)}f_{\gamma}(\epsilon),\\
&\lim_{t \to \infty}	\Lambda^{(2)}_{\gamma}(t,t') =\frac{\Gamma}{2}\int \frac{d\epsilon}{2\pi} \frac{e^{i \epsilon(t-t^\prime)}}{\frac{\Gamma^2}{4}+(\epsilon-\epsilon_d)^2}f_{\gamma}(\epsilon).
\end{aligned}
\end{equation}
The above functions only depend on the time difference $\tau=t-t^{\prime}$, as expected in the steady-state regime. We then substitute the above expressions into Eq.~\eqref{SHE}, and take the Fourier transform with respect to the time delay $\tau$. In the next step, one has to exploit the convolution theorem to get an exact expression for the finite-frequency auto- and cross-current correlations,
\begin{widetext}
\begin{align}
\label{eq:noise_HE_frequency}
\small
S^{\he}_{\alpha\alpha^{\prime},ss}(\omega) &= \frac{\Gamma _{\alpha} \Gamma _{\alpha^{\prime}} }{4\Gamma^2} \sum _{\beta,\beta^{\prime}}  \Gamma _{\beta }\Gamma _{\beta^{\prime}} \int \frac{d\epsilon^{\prime}}{2\pi} \left[  \frac{\Gamma f_{\beta }(\epsilon^{\prime})}{\frac{\Gamma^2}{4}+(\epsilon^{\prime}-\epsilon_d)^2} - \frac{2f_{\alpha }(\epsilon^{\prime})}{\frac{\Gamma}{2}-i(\epsilon^{\prime}-\epsilon_d)} \right]\left[\frac{\Gamma (1-f_{\beta^{\prime}}(\epsilon^{\prime}-\omega))}{\frac{\Gamma^2}{4}+(\epsilon^{\prime}-\omega-\epsilon_d)^2}-\frac{2(1-f_{\alpha^{\prime}}(\epsilon^{\prime}-\omega))}{\frac{\Gamma}{2}-i(\epsilon^{\prime}-\omega-\epsilon_d)}\right]\nonumber\\
& + \int \frac{d\epsilon^{\prime}}{2\pi}\sum _{\beta\beta^{\prime} } \frac{\Gamma _{\alpha} \Gamma _{\alpha^{\prime}} \Gamma _{\beta }\Gamma _{\beta^{\prime}}}{4\Gamma^2} 
\left[\frac{\Gamma f_{\beta^{\prime}}(\epsilon^{\prime})}{\frac{\Gamma^2}{4}+(\epsilon^{\prime}-\epsilon_d)^2}-\frac{2f_{\alpha^{\prime}}(\epsilon^{\prime})}{\frac{\Gamma}{2}+i(\epsilon^{\prime}-\epsilon_d)}-\frac{2f_{\alpha}(\epsilon^{\prime})}{\frac{\Gamma}{2}-i(\epsilon^{\prime}-\epsilon_d)}\right]\frac{\Gamma (1-f_{\beta^{\prime}}(\epsilon^{\prime}-\omega))}{\frac{\Gamma^2}{4}+(\epsilon^{\prime}-\omega-\epsilon_d)^2}\nonumber\\
& + \int \frac{d\epsilon^{\prime}}{2\pi}\sum _{\beta} \frac{\Gamma _{\alpha}  \Gamma _{\beta }}{2\Gamma}\frac{\Gamma f_{\alpha}(\epsilon^{\prime})(1-f_{\beta}(\epsilon^{\prime}-\omega))}{\frac{\Gamma^2}{4}+(\epsilon^{\prime}-\omega-\epsilon_d)^2}\delta_{\alpha \alpha^{\prime}}\quad+ \left\{c.c.\quad\text{with} \quad f_{\gamma}\rightarrow \left(1-f_{\gamma}\right), ~n_d\rightarrow \left(1-n_d\right)\right\}.
 \end{align}
\end{widetext}
We verified that the above expression satisfies the well-known symmetry property $S^{\he}_{\alpha\alpha',ss}(\omega)=S^{\he}_{\alpha'\alpha,ss}(-\omega)$~\cite{Blanter2000}.
% It can be shown that the above expression satisfies the well-known symmetry property $S^{\he}_{\alpha\alpha',ss}(\omega)=S^{\he}_{\alpha'\alpha,ss}(-\omega)$~\cite{Blanter2000}.
The zero-frequency noise can be calculated from the above expression by taking $\omega = 0$. By recognizing the transmission probability $\mathcal{T}(\epsilon)$ of Eq. \eqref{eq:Transmission_function}, we can show that the shot noise coincides with the one obtained within the $\lb$ approach, as given in Eq. \eqref{eq:shot_noise_lb}, 
\begin{eqnarray}
\label{eq:shot_HE}
    S^{\he}_{\alpha \alpha',ss}(\omega=0) = S^{\lb}_{\alpha\alpha'}(\omega=0).
\end{eqnarray}
From the above result, we can verify that the zero-frequency cross and auto-correlation functions in a two-terminal setup satisfy $S_{RL} = S_{LR} = - S_{RR} = - S_{LL}$. We also note that, as expected, the zero-frequency components of the noise are real, the imaginary part cancels out.
We have also verified explicitly that Eq.~\eqref{eq:noise_HE_frequency} agrees with finite-frequency expression of the noise in the Landauer-Büttiker framework defined in Ref.~\cite{Blanter2000}. Overall, the results obtained in Eqs.~\eqref{eq:Iss_Jss_MEvs_LB} and \eqref{eq:shot_HE}, demonstrate the connection between the $\he$ and $\lb$ pictures, as shown in Fig. \ref{fig:setups} panels (a) to (c).

%%%%%%%%%%%%%%%%%%%%%%%%%%%%%%%%%%%%%%%%%%%%%%%%%%%%%%%%%%%%%%%%%%%%%%%%%%%%
%%%%%%%%%%%%%%%%%%%%%%%%%%%%%%%%%%%%%%%%%%%%%%%%%%%%%%%%%%%%%%%%%%%%%%%%%%%%

\section{Master equation for the QD}
\label{sec_ME_for_QD}

For a single-level QD in the ME formalism, it is convenient to express $\hat{H}_S =  \epsilon_d \hat{d^\dagger}\hat{d}$ in terms of the raising and lowering operators for a two-level system in the canonical basis $\{ \ket{0}, \ket{1}\}$, namely $\hat \sigma_+ = \ketbra{1}{0}$ and $\hat \sigma_- = \ketbra{0}{1}$ respectively. In this way, we can associate the operator $\hat{d}^\dagger$ ($\hat{d}$) to the operator $\hat \sigma_+$ ($\hat \sigma_-$). 
The Lindblad equation for this model is given by~\cite{Breuer2007},
\begin{eqnarray}
\label{eq:lind_ME_single_QD}
\dot{\rho}(t)=\mathcal{L} \, \rho(t) = \left( \mathcal L_0 + \sum_{ \alpha } (\mathcal L^+_{\alpha} +\mathcal L^-_{\alpha }) \right)\rho(t)\,,
\end{eqnarray}
and corresponds to the general form of Eq.~\eqref{eq:lind} with a single energy transition $\epsilon_j = \epsilon_d$, and jump operators $\hat L_{j\alpha}=\delta_{j1}\hat \sigma_-$, $\hat L_{j\alpha}^{\dagger}=\delta_{j1}\hat \sigma_+$ with $j=1$. 
% \sk{[I don't understand why we introduce two energies when we focus exclusively on degenerate qubits. It makes reading difficult, for example here.]}} 
Here, $\mathcal L_0$, defined in Eq.~\eqref{eq:L0}, takes the following form in terms of the non-hermitian commutator,
\begin{equation}
\label{eq:L0_QD}
    \mathcal L_0 \rho = -i\left[\epsilon_d \hat \sigma_+ \hat \sigma_-\!\! - \frac{i}{2}\sum_{\alpha} \left(\Gamma_\alpha^- \hat \sigma_+\hat \sigma_- +\Gamma_\alpha^+ \hat \sigma_-\hat \sigma_+\right), \rho \right]_\dagger
\end{equation}
and the Lindblad jump superoperators defined in Eqs.~\eqref{eq:jump_lind_op1} and \eqref{eq:jump_lind_op2}, are given by,
\begin{eqnarray}
    && \mathcal{L}_{\alpha}^+ \rho = \Gamma_\alpha^+ \hat \sigma_+ \rho \hat \sigma_-,\\
    && \mathcal{L}_{\alpha}^- \rho = \Gamma_\alpha^- \hat \sigma_- \rho \hat \sigma_+.
\end{eqnarray}

\subsection{Solution for $\rho(t)$}

In order to compute observables within the ME approach, the time-dependent solution for the reduced density operator satisfying Eq.~\eqref{eq:lind_ME_single_QD} is required. Although this is not a trivial task in general, analytical results can be obtained in the case of the single-level QD setup~\cite{Schaller2014}. In this minimal model, no quantum coherences are formed during the dynamics, and only the populations for the ground and excited states are non-zero.
Thus, assuming zero coherence at time $t_0$, the reduced density matrix of the system takes a diagonal form,
$\rho(t)=\ketbra{0}{0}p_0(t)+\ketbra{1}{1}p_1(t)$, and the Lindblad master equation \eqref{eq:lind_ME_single_QD} reduces to the following rate equation for the populations%~\gh{add refs Schaller, strasberg, brandes}
,
$\dot p_i=\sum_{j}W_{ij}p_j$, or explicitly,
\begin{align}
\label{Rate_Equation}
\begin{pmatrix}
\dot p_0\\
\dot p_1
\end{pmatrix} =\left(\begin{array}{cc}
-\Gamma_L^+-\Gamma_R^+ & \Gamma_L^- +\Gamma_R^- \\
\Gamma_L^+ +\Gamma_R^+ &  -\Gamma_L^- -\Gamma_R^- \\
\end{array}\right)  
\begin{pmatrix}
p_0\\
p_1
\end{pmatrix}.
\end{align}
Here, the elements $W_{ij}$ of the $2\times 2$ matrix $W$ are the transition rates from state $j$ to state $i$ (where $i,j=0,1$). The solution for $p_1(t)$ is given by,
\begin{align}
\label{eq:sol_rho11}
p_1(t) &= \sum_{\alpha}\frac{\Gamma_\alpha}{\Gamma}\left[f_\alpha  + (n_d-f_\alpha)e^{-\Gamma (t-t_0)}\right].
\end{align}

% \begin{align}
% \label{eq:sol_rho11}
% p_1(t) &= \sum_{\alpha}\frac{\Gamma_\alpha}{\Gamma}\left[f_\alpha  + (f_\alpha - n_d)e^{-\Gamma (t-t_0)}\right].
% \end{align}
with initial occupation probability $p_1(t_0)= n_d$ at time $t=t_0$. The solution for the ground state is $p_0(t) = 1 -p_1(t)$, due to the conservation of probability. For convenience, we have used the notation $f_\alpha\coloneqq f_\alpha(\epsilon_d)$, where the Fermi-Dirac distributions are evaluated at the energy of the dot $\epsilon_d$, a consequence of the weak-coupling assumption. Interestingly, this solution explicitly depends on the occupation imbalance between both reservoirs and the dot through the factor $\left(f_\alpha - n_d\right)$, which exponentially decays for long time.

\subsection{Particle and energy currents with ME}

Particle and energy current superoperators in reservoir $\alpha=L,R$ can be directly calculated from their definitions in Eqs.~\eqref{eq:super_current_I} and \eqref{eq:super_current_J}, using the explicit expressions of the Lindblad jump superoperators for a single-level QD given in Eq.~\eqref{eq:jump_lind_op2}. They take the following form,
\begin{align}
    \mathcal{I}_\alpha \rho &= \left(\mathcal{L}_\alpha^+ - \mathcal{L}_\alpha^-\right)\rho=\Gamma_\alpha^+ \hat \sigma_+ \rho \hat \sigma_--\Gamma_\alpha^- \hat \sigma_- \rho \hat \sigma_+, \\
     \mathcal{J}_\alpha \rho &= \epsilon_d\! \left(\mathcal{L}_\alpha^+ - \mathcal{L}_\alpha^-\right)\rho=\epsilon_d\!\left(\Gamma_\alpha^+ \hat \sigma_+ \rho \hat \sigma_-\!\!-\Gamma_\alpha^- \hat \sigma_- \rho \hat \sigma_+\right)\!.
\end{align}
These superoperators reflect the positive (negative) count of particle or energy entering (leaving) the system, proportional to the rate  $\Gamma_\alpha^+$ ($\Gamma_{\alpha}^-$). 
By inserting the above expressions of the currents superoperators into Eqs.~\eqref{eq:Currents_ME}, and using the time-dependent solution of the reduced density matrix of the system $\rho(t)$ of Eq.~\eqref{eq:sol_rho11}, we obtain the following analytical expressions for the particle and energy current,
\begin{align}
I_\alpha^\me(t) &= \Tr{\mathcal{I}_\alpha \rho(t)} \nonumber \\
&= \sum_{\beta}\frac{\Gamma_{\alpha}\Gamma_{\beta}}{\Gamma}\left[\left(f_{\alpha}-f_{\beta}\right)+\left(f_{\beta}-n_d \right)e^{-\Gamma (t-t_0)}\right], \label{eq:IME_QD}\\
J_\alpha^\me(t) &= \Tr{\mathcal{J}_\alpha \rho(t)} = \epsilon_d I_\alpha^\me(t). \label{eq:JME_QD}
\end{align}
The steady-state expressions can be calculated directly by taking the limit $t \rightarrow \infty$. In this case, the time-dependent part decays exponentially, and as expected the stationary part only depends on the difference of the Fermi-Dirac functions. In contrast, in the $\he$ results, we saw a stationary part, an exponentially decaying part, as well as an oscillating exponentially decaying term.

\subsection{Comment on the activity $\mathcal{A}$}

The definition of the activity superoperator $\mathcal{A}$ comes naturally when deriving the exact expression for the current correlation function as discussed in Sec.~\ref{GFCS}. 
The average activity $A_\alpha \coloneqq \Tr{\mathcal{A}_\alpha \rho}$ can be interpreted as the sum of all quantum jumps between reservoir $\alpha$ and the quantum system~\cite{Landi2024}. It provides information about the dynamics of the quantum system, and has recently been shown to be of utility to derive lower bounds for signal-to-noise ratio in the context of thermodynamic uncertainty relations \cite{Terlizzi2019,Vo2022,Prech2023}.
In the case of the single-level QD, the activity superoperators in reservoir $\alpha=L,R$, can be made explicit by using the expressions of the Lindblad jump superoperators given in Eq.~\eqref{eq:jump_lind_op2},
\begin{align}
\mathcal{A}_\alpha \rho &= \left(\mathcal{L}_\alpha^+ + \mathcal{L}_\alpha^-\right)\rho=\Gamma_\alpha^+ \hat \sigma_+ \rho \hat \sigma_-+\Gamma_\alpha^- \hat \sigma_- \rho \hat \sigma_+ \,. 
\end{align}
As done for the currents, its average value can be computed using the time-dependent solution of the reduced density matrix of the system $\rho(t)$ of Eq.~\eqref{eq:sol_rho11}, 
\begin{align}
\label{activity_from_superoperator}
 A_\alpha(t) &\coloneqq \Tr{\mathcal{A}_\alpha \rho(t)} \nonumber\\
 &=\sum_{\beta}\frac{\Gamma_{\alpha}\Gamma_{\beta}}{\Gamma}\left\{e^{-\Gamma (t-t_0)}\left(f_{\beta}-n_d\right)\left[f_{\alpha}-(1-f_{\alpha})\right]\right.\nonumber\\
&+\left. \left[f_{\beta}\left(1-f_{\alpha}\right)+f_{\alpha}\left(1-f_{\beta}\right)\right]\right\}.
\end{align}
A particularly intuitive interpretation of the above expression becomes apparent in the stationary limit as $t\to \infty$, where the steady-state activity takes the following form,
\begin{align}
  A_{\alpha,ss} &= \lim_{t\to\infty} A_\alpha(t)\nonumber\\
&= \sum_{\beta}\frac{\Gamma_{\alpha}\Gamma_{\beta}}{\Gamma}\left[f_{\beta}\left(1-f_{\alpha}\right)+f_{\alpha}\left(1-f_{\beta}\right)\right].
\end{align}
The components of this expression have a natural interpretation. First, there is an overall factor, $\Gamma_{\alpha}\Gamma_{\beta}/\Gamma$,  which is the probability of particle exchange between reservoirs $\alpha$ and $\beta$. The terms in the bracket, such as $f_{\alpha/\beta}\left(1-f_{\beta/\alpha}\right)$ for $\alpha,\beta=L,R$ 
can be interpreted in terms of the occupation or de-occupation of the dots. These terms represent the product of the probability that the incoming particle (toward the system) occupies the energy level of the dot in the lead $\alpha/\beta$ ($f_{\alpha/\beta}$) multiplied by the probability that the arriving energy level in the reservoir $\beta/\alpha$ is available (empty, i.e., $(1-f_{\beta/\alpha})$). They are reminiscent of the contributions to noise in the Landauer-Büttiker approach \cite{Blanter2000}. 

In addition to this formal derivation, it is important to note that the average activity $A_\alpha$, for a single-level QD in absence of quantum coherence can also be calculated directly using the rate matrix $W$ of Eq.~\eqref{Rate_Equation}~\cite{Prech2023},  
\begin{eqnarray}
\label{eq:activity_rate}
    A_\alpha(t) &=& \sum_{i \neq j} (W_\alpha)_{i j} p_j(t) \,.
\end{eqnarray}
Inserting the solution for $p_1(t)$ and $p_0(t)=1-p_1(t)$ from Eq.~\eqref{eq:sol_rho11}, in the above relation, it is easy to show that expressions in Eqs.~\eqref{activity_from_superoperator} and \eqref{eq:activity_rate} are equivalent.\\
% \gh{my comment?} \gb{GB: the reason why I think Eq. 95 is important is that it represents a more natural definition of the activity, namely product of a population (probability of occupation) times a transition probability.} 

%%%%%%%%%%%%%%%%%%%%%%%%%%%%%%%%%%%%%%%%%%%%%%%%%%%%%%%%%%%%%%%%%%%%%%%%%%%%%
%%%%%%%%%%%%%%%%%%%%%%%%%%%%%%%%%%%%%%%%%%%%%%%%%%%%%%%%%%%%%%%%%%%%%%%%%%%%%

\subsection{Current correlation function with ME}

The current correlation function for the single-level QD can be computed by inserting the expressions of the superoperators $\mathcal{I}_\alpha, \mathcal{A}_\alpha$ and the solution $\rho(t)$ into Eq.~\eqref{eq:S_ME},
\begin{widetext}
\begin{eqnarray}
\label{eq:SME}
S^\me_{\alpha \alpha^{\prime}}(t,t') &=& \delta_{\alpha\alpha^{\prime}} \delta(t-t')\sum_{\beta}\frac{\Gamma_{\alpha}\Gamma_{\beta}}{\Gamma}\left\{e^{-\Gamma t}\left(f_{\beta}-n_d\right)\left[f_{\alpha}-(1-f_{\alpha})\right]+\left[f_{\beta}\left(1-f_{\alpha}\right)+f_{\alpha}\left(1-f_{\beta}\right)\right]\right\} \nonumber \\
&& -\frac{e^{-\Gamma (t-t')}}{\Gamma^2}\sum_{\beta,\beta'}\Gamma_{\alpha}\Gamma_{\alpha'}\Gamma_{\beta}\Gamma_{\beta'}e^{-\Gamma t'}\left(f_{\beta}-n_d\right)\left\{\left[(f_{\alpha}-f_{\beta'})+(f_{\alpha'}-f_{\beta'})+(f_{\alpha'}-f_{\alpha})\right]+e^{-\Gamma t'}(f_{\beta'}-n_d)\right\}\nonumber \\
&&-\frac{e^{-\Gamma (t-t')}}{\Gamma^2}\sum_{\beta,\beta^{\prime}}\Gamma_{\alpha}\Gamma_{\alpha^{\prime}}\Gamma_{\beta}\Gamma_{\beta^{\prime}}\left[(1-f_{\beta})f_{\alpha^{\prime}}+f_{\beta}(f_{\beta^{\prime}}-f_{\alpha^{\prime}})\right]\,.
\end{eqnarray}
\end{widetext}
In the above, we have assumed $t>t'$. Similar to the currents, all Fermi-Dirac distributions are evaluated at the energy of the dot $\epsilon_d$. 
% \gh{discuss: i) different terms ii) real quantity. why? Intuition? and then reformulate what is below. iii) should we discuss and present the steady-state expression? maybe good to also connect to other works in the literature.} 
By using the general relation $S_{\alpha\alpha^{\prime}}(t,t')=S^*_{\alpha^{\prime}\alpha}(t',t)$~\cite{Zhang2014}, and exploiting the fact that the current correlation is real in the weak-coupling regime, Eq. \eqref{eq:SME} can be written for $t<t'$ by exchanging $\alpha\leftrightarrow\alpha^{\prime}$ and $t\leftrightarrow t'$.
In the steady state, we can obtain the analytical expression of the stationary noise in the weak-coupling regime, which only depends on the time difference $\tau=t-t'$,
\begin{widetext}
\begin{align}
\label{SME_stationary}
S^\me_{\alpha \alpha^{\prime},ss}(\tau)&=\lim_{t\to \infty} S^\me_{\alpha \alpha^{\prime}}(t,t+\tau)\nonumber\\
&=\frac{\delta (\tau)\delta_{\alpha\alpha^{\prime}}}{\Gamma}\sum_{\beta}\Gamma_{\alpha}\Gamma_{\beta}\left\{f_{\beta}(\epsilon_d)\left(1-f_{\alpha}(\epsilon_d)\right)+f_{\alpha}(\epsilon_d)\left(1-f_{\beta}(\epsilon_d)\right)\right\}\nonumber\\
&-\frac{e^{-\Gamma \abs{\tau}}}{\Gamma^2}\sum_{\beta,\beta^{\prime}}\Gamma_{\alpha}\Gamma_{\alpha^{\prime}}\Gamma_{\beta}\Gamma_{\beta^{\prime}}\left\{\left(1-f_{\beta}(\epsilon_d)\right)\left[f_{\alpha}(\epsilon_d)\Theta(-\tau)+f_{\alpha^{\prime}}(\epsilon_d)\Theta(\tau)\right]\right\}\nonumber\\
&-\frac{e^{-\Gamma \abs{\tau}}}{\Gamma^2}\sum_{\beta,\beta^{\prime}}\Gamma_{\alpha}\Gamma_{\alpha^{\prime}}\Gamma_{\beta}\Gamma_{\beta^{\prime}}\left\{f_{\beta}(\epsilon_d)\left[f_{\beta^{\prime}}(\epsilon_d)-f_{\alpha}(\epsilon_d)\Theta(-\tau)-f_{\alpha^{\prime}}(\epsilon_d)\Theta(\tau)\right]\right\}.
\end{align}
\end{widetext}

\begin{figure}[!htb]
\centering
\includegraphics[width=1\columnwidth]{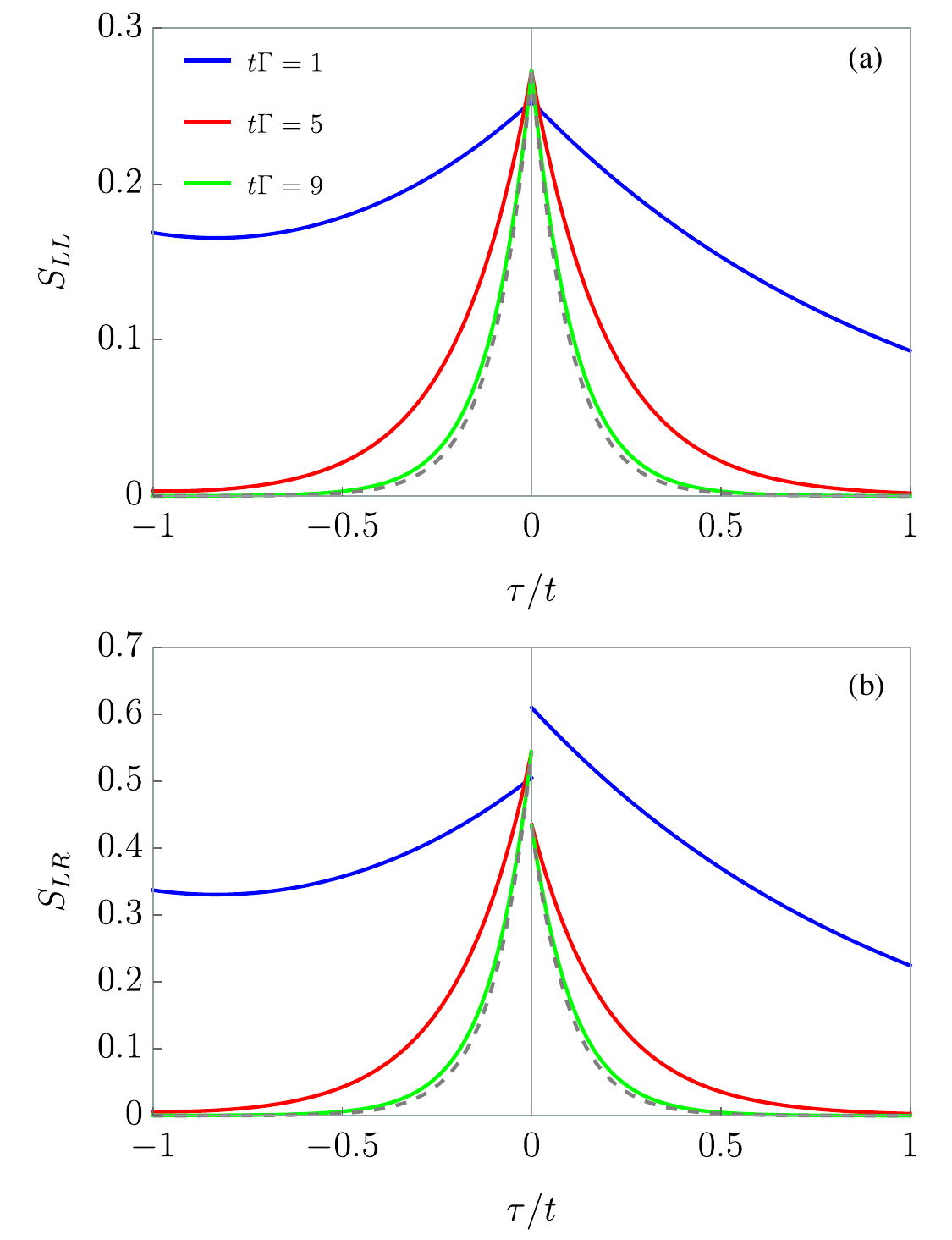}
\caption{Comparison between auto- and cross-current correlation functions in the weak-coupling regime, $S_{\alpha\alpha^{\prime}}^\me(t, t+\tau)$,  as a function of $\tau/t$. 
Since $\tau\in[-t, t]$, it is convenient to plot the functions in terms of the rescaled variable $\tau/t$ in the range $[-1,1]$. This choice allows easier comparison of the curves.
% \gb{Since $-t<\tau<t$, we have plotted the curves from between $-1<\tau/t<1$.} 
Different colors indicate different values of $t\Gamma=1,5,9$.  The gray dashed line represents the stationary limit $S^\me_{\alpha \alpha^{\prime},ss}(\tau)$. Notice the discontinuity in the cross-current correlation function at $\tau/t=0$. The other parameters are $\Gamma_L=1$, $\Gamma_R=2$, $\epsilon_d/\Gamma=1$, $\mu_L=-\mu_R=\Gamma/2$ and $T_L=T_R=\Gamma/100$ ($\hbar,k_B=1$). }
\label{SHE_SME_tau}
\end{figure}

Fig. \ref{SHE_SME_tau} shows the behavior of the auto- and cross-current correlations function in the $\me$ approach, $S^\me_{\alpha \alpha^{\prime}}(t,t+\tau)$, as functions of $\tau$, for different values of $t$ (different colors). The dashed curve corresponds to the stationary limit $S^\me_{\alpha \alpha^{\prime},ss}(\tau)$ of Eq.~\eqref{SME_stationary}. The numerical results approach the analytical stationary solution in the limit of large time. It is important to note that the correlation function features a discontinuity at $\tau=0$; this jump can be directly computed in the stationary limit from Eq. \eqref{SME_stationary},
\begin{align}
&\lim_{\tau \to 0^+}S^\me_{\alpha \alpha^{\prime},ss}(\tau)-\lim_{\tau \to 0^-}S^\me_{\alpha \alpha^{\prime},ss}(\tau)=\nonumber\\
&\sum_{\beta}\frac{\Gamma_{\alpha}\Gamma_{\alpha^{\prime}}\Gamma_{\beta}}{\Gamma}\left(f_{\alpha^{\prime}}-f_{\alpha}\right)\left[\left(1-f_{\beta}\right)-f_{\beta}\right].
\end{align}
As evident from the above expression, the discontinuity disappears for the auto-correlation noise ($\alpha=\alpha^{\prime}$), while it is finite for the cross-correlation noise ($\alpha=\alpha^{\prime}$). This can be clearly seen by comparing panels (a) and (b) of Fig. \ref{SHE_SME_tau}.

%%%%%%%%%%%%%%%%%%%%%%%%%%%%%%%%%%%%%%%%%%%%%%%%%%%%%%%%%%%%%%%%%%%%%%%%%%%%%%

\section{Connecting the different frameworks}
\label{sec:connection}

In Sec.~\ref{sec:HE_QD}, we have used the HE approach to derive the exact expressions for the particle and energy currents for a single-level QD in a two-terminal setup, as well as the auto- and cross-current correlation functions. Moreover, we have benchmarked the steady-state with the $\lb$ framework by taking the limit $t\to\infty$ of the exact solution. Moving forward to Section~\ref{sec_ME_for_QD}, we reproduced these expressions within a Master Equation (ME) framework. However, we have yet to address the crucial aspect of how the ME results can be derived from the HE and $\lb$ approaches by appropriate limiting procedures. This step is essential for a comprehensive understanding and completion of the overarching framework depicted in Fig.~\ref{fig:setups}. 
To pursue this, a weak-coupling limit must be considered to obtain meaningful results within the framework of $\me$. However, achieving this limit is generally not a straightforward task, see for instance Ref.~\cite{DAbbruzzo2023} discussing regularization procedures in the transient regime. While it might appear intuitive to take $\Gamma_L, \Gamma_R \rightarrow 0$ as the weak-coupling limit, this is not correct. A simple argument elucidates this point: currents in the HE approach are proportional to the coupling rates, as evident in Eqs.~\eqref{eq:IHE} and \eqref{eq:JHE}. Adopting this naive limit would invariably yield zero currents, essentially signifying a quantum dot completely decoupled from its reservoirs.
A similar argument holds for the current correlation function. This simple reasoning shows the necessity of a specific protocol to connect the three approaches. In the following, we discuss the steps to implement such a protocol and apply it to the currents and current correlations.

\subsection{Weak-coupling protocol}
\label{sec:weak_coupling_protocol}

There are two key aspects of the procedure to obtain $\me$ results from $\he$ approach. 

\noindent 1) The first crucial element is the limit of small system-reservoir coupling $\Gamma$, which is a fundamental characteristic of the $\me$ approach. 
In this respect, it is important to note that while the particle and energy currents are of the order $\mathcal O\left( \Gamma\right)$, the current correlation function is of the order $\mathcal O\left( \Gamma^2\right)$. Consequently, as mentioned above, directly taking the limit as $\Gamma\to 0$ would yield zero values for both currents and current correlation. To address this issue, we adopt a remedial approach. We first divide the respective expressions by $\Gamma$ and $\Gamma^2$, then take the limit $\Gamma\to 0$, and finally, multiply the obtained expressions by $\Gamma$ and $\Gamma^2$ again, giving us the contribution which is lowest order in the couplings. 
\\
\noindent 2) The second aspect concerns the time precision, $\Gamma^{-1}$, of any Markovian master equation, which is set by the system-reservoir coupling strength. Therefore, the transient description of the dynamics from the master equation is reliable only if the time interval $t-t_0$ is large enough to compensate for the small value of $\Gamma$. As a result, when taking the $\Gamma\to 0$ limit, it is crucial to treat the quantity $\Gamma \left(t-t_0\right)$ as a constant.
Mathematically, the weak-coupling protocol can be expressed in the following form,
\begin{align}
I_{\alpha}^\me(t)&=\Gamma\lim_{\Gamma\to 0}\frac{I_{\alpha}^\he(t)}{\Gamma}\\
J_{\alpha}^\me(t)&=\Gamma\lim_{\Gamma\to 0}\frac{J_{\alpha}^\he(t)}{\Gamma}\\
S_{\alpha\alpha'}^\me(t,t')&=\Gamma^2\lim_{\Gamma\to 0}\frac{S_{\alpha\alpha'}^\he(t,t')}{\Gamma^2},
\end{align}
with $\Gamma(t-t_0)$ kept constant. 
This is illustrated in Fig.~\ref{fig:setups}, panels (a) to (b).
We refer to the combined application of these steps as the \textit{weak-coupling protocol} of the exact solution. This approach allows us to properly obtain the desired $\me$ results from the $\he$ results.

\subsection{Weak-coupling limit of currents}

We now explain how the above weak-coupling protocol can be applied to the particle and energy currents in the single-level QD model. As explained above, we begin by dividing the exact current solutions given in Eqs. \eqref{eq:IHE} and \eqref{eq:JHE} by $\Gamma$, and then take the limit as $\Gamma\to 0$. This limit corresponds to taking the limit of the functions presented in Eqs. \eqref{eq:A_coeff}-\eqref{eq:C_coeff},
\begin{align}
\lim_{\Gamma \to 0}A_{\gamma}^m &= \epsilon_d^m f_\gamma(\epsilon_d) \\
\lim_{\Gamma \to 0} B_{\gamma}^m &= \epsilon_d^m e^{-\frac{\Gamma (t-t_0)}{2}} f_\gamma(\epsilon_d) \\
\lim_{\Gamma \to 0}C_{\gamma}^m &= \frac{\epsilon_d^m e^{-\frac{\Gamma (t-t_0)}{2}}}{2} f_\gamma(\epsilon_d)
\end{align}
To compute the above expressions, we have performed a change of variable $\epsilon\to \Gamma \epsilon$, which enables us to extract the Fermi-Dirac function evaluated at the energy of the dot outside the integral. Crucially, while taking the limit, we kept the factor $\Gamma(t-t_0)$ constant, respecting the temporal-resolution of $\me$ dynamics.

\begin{figure}[h]
\centering
\includegraphics[width=1\columnwidth]{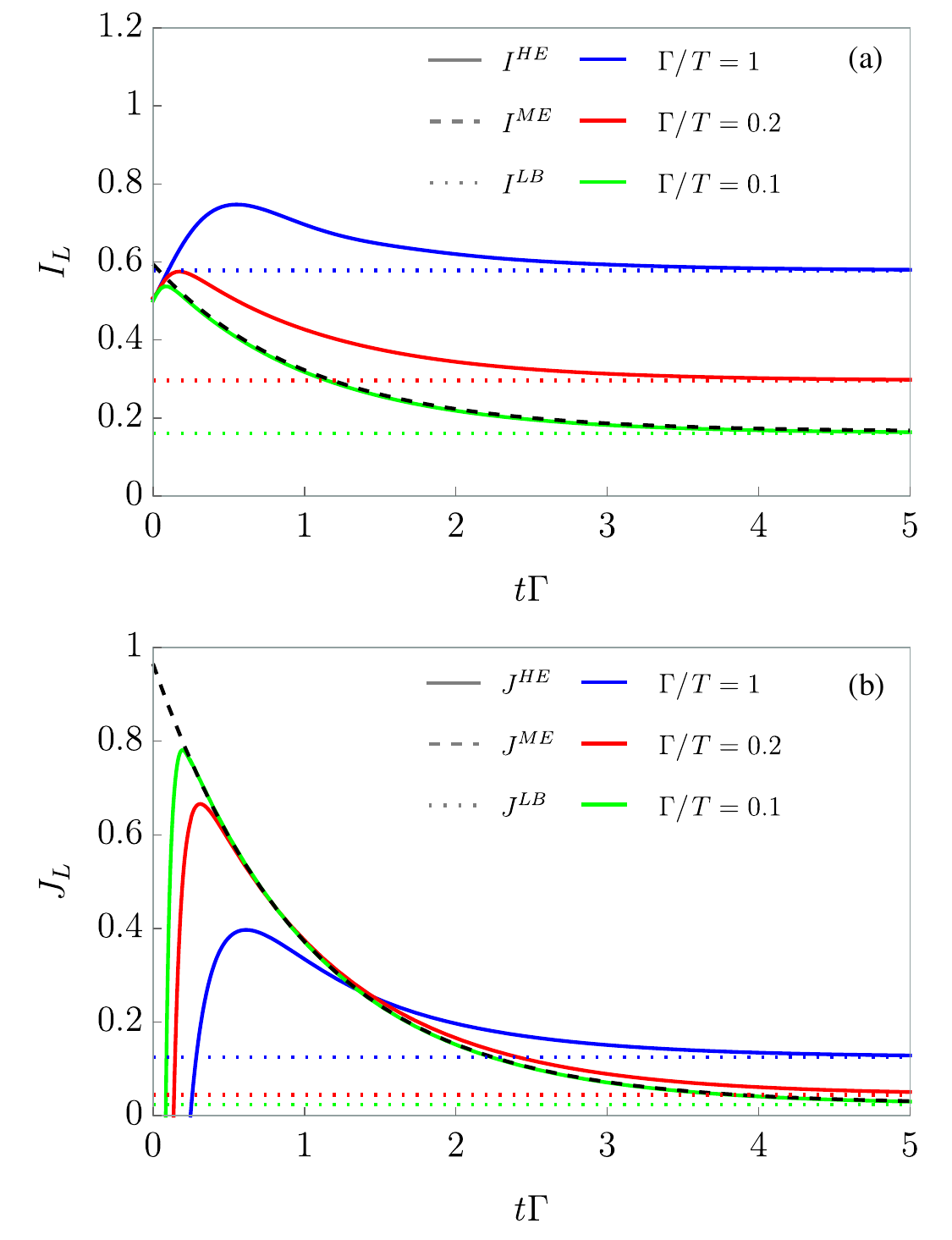}
\caption{(a) Particle current and (b) energy current as a function of $t\Gamma$, for different values of $\Gamma/T=1,0.2,0.1$. Solid curves correspond to the result obtained with the $\he$ (i.e., the exact solution), dashed black curves correspond to $\me$ (i.e., the weak-coupling limit), calculated with $\Gamma/ T=0.1$, and dotted lines to the $\lb$ (i.e., the stationary limit). The ME and HE results agree with each other for the weak-coupling case $\Gamma/ T=0.1$. The other parameters are $\Gamma_L=\Gamma_R=1$ and $\mu_L=-\mu_R=\Gamma/2$ ($h,k_B=1$).}
\label{fig:currents}
\end{figure}

Furthermore, we employed the following exact expressions,
\begin{align}
&\int \frac{d \epsilon}{2 \pi} \frac{1}{\frac{1}{4} + \epsilon^2} = 1,\\
&\int \frac{d \epsilon}{2 \pi} \frac{\cos(\epsilon \Gamma t)}{\frac{1}{4} + \epsilon^2} = e^{- \Gamma t/2},\\
&\int \frac{d \epsilon}{2 \pi} \frac{\epsilon \sin(\epsilon \Gamma t)}{\frac{1}{4} + \epsilon^2} = \frac{e^{- \Gamma t/2}}{2}.
\end{align}
By substituting the above expressions into Eqs. \eqref{eq:IHE} and \eqref{eq:JHE}, and multiplying again by $\Gamma$, we exactly recover the same results obtained with the $\me$ approach given in Eqs. \eqref{eq:IME_QD} and \eqref{eq:JME_QD}. As a technical remark, it is important to stress that our weak-coupling limit procedure leads to the correct $\me$ results, which otherwise cannot be obtained by simply considering the Lorentzian under the integrals of Eqs. \eqref{eq:A_coeff}-\eqref{eq:C_coeff} to become a Dirac delta in the limit of small $\Gamma$.

In Fig. \ref{fig:currents} we show (a) the particle and (b) energy currents  as functions of $t\Gamma$, for different values of $\Gamma/T$ and for all three approaches. For both currents, in the weak-coupling regime, i.e., for small values of $\Gamma/T$ (green curves), the currents obtained with the HE coincide with those obtained using the ME (i.e., solid green curves overlap with the dashed black ones). In general, the ME results are not expected to agree with the LB ones in the stationary limit (for $t\Gamma\gg1$, which violates the weak-coupling assumption). However, in the weak-coupling regime (when $\Gamma/ T\ll 1$), the steady state of the ME result matches the LB result exactly. On the other hand, the HE approach always reproduces the correct stationary solution. 

Moreover, Fig. \ref{fig:currents} demonstrates that although the currents are well described in the WBL at intermediate  and long time, there are peculiarities at short times. Specifically, we see that the particle current exhibits a jump at short times. On the other hand, the energy current shows a divergence at small time. This pathological behavior of the currents can be understood intuitively by noting that short times implies a wide spread in energy~\cite{Covito2018}, which invalidates the use of the WBL.
A possible way to solve this problem can be to consider an energy-dependent tunneling rate function $\Gamma(\epsilon)$ with an energy cut-off or decay. Possible choices include Lorentzian- and a boxcar-shaped function~\cite{Zhang2014}. 
% In this case however, no explicit analytical solutions can be obtained by using the $\he$ approach. 

\subsection{Weak-coupling limit of the current correlation function}

\begin{figure*}
\centering
\includegraphics[width=1\textwidth]{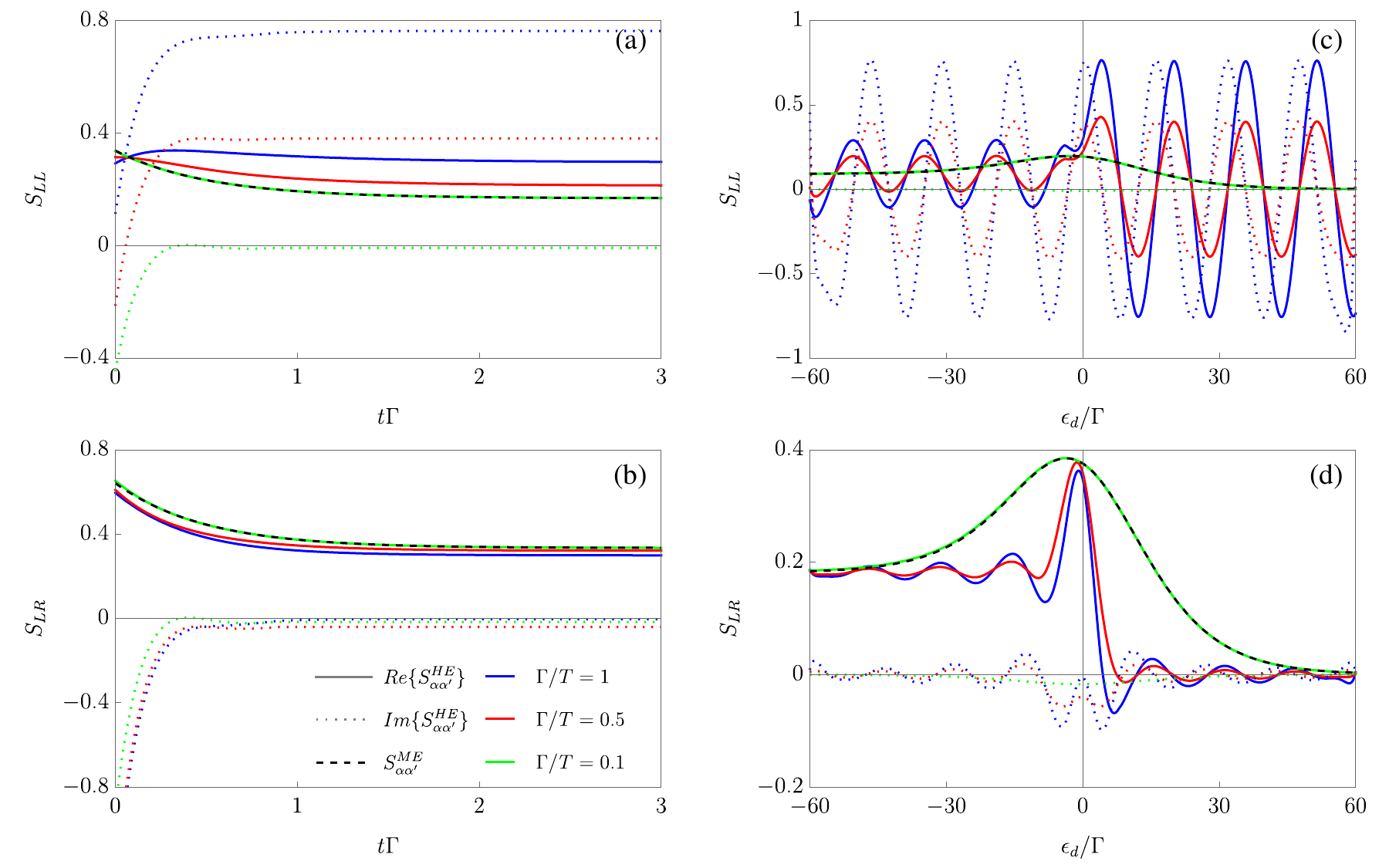}
\caption{Auto- ($S_{LL}$) and cross- ($S_{LR}$) current correlations as functions of time $t\Gamma$ (at fixed $\epsilon_d/\Gamma=1$, left panels, (a) and (b)) and as functions of the energy of the dot $\epsilon_d/\Gamma$ (at fixed $t\Gamma=1$, right panels, (c) and (d)). Similar results hold for $S_{RR}$ and $S_{RL}$. Solid curves correspond to $\text{Re}\{S_{\alpha\alpha^{\prime}}^{HE}\}$ and dotted curves correspond to $\text{Im}\{S_{\alpha\alpha^{\prime}}^{HE}\}$ obtained with the $\he$ approach, the dashed curves represents the $\me$ result (obtained with $\Gamma/T=0.1$).  Different colors indicate different values of $\Gamma/T =1,0.5,0.1$. The other parameters are $\Gamma_L=\Gamma_R=1$ and $\mu_L=-\mu_R=\Gamma/2$, $\tau\Gamma=0.4$.}
\label{fig:noise}
\end{figure*}

We now explain how the expression for the current correlation function of Eq. \eqref{eq:SME} obtained with the $\me$ approach can be computed starting from the exact solution of Eq. \eqref{SHE}, obtained within the $\he$ picture, by taking the weak-coupling limit. To this end, using the same weak-coupling protocol as for the currents, we take first the limit of $S_{\alpha\alpha'}^\he(t,t')/\Gamma^2$ as $\Gamma\to 0$. This is equivalent to taking the limit for small coupling of the $\Lambda$-functions defined in Eqs. \eqref{Lambda_HE},

\begin{widetext}
\begin{equation}
    \label{Lambda_ME}
\begin{aligned}
    &\lim_{\Gamma \to 0}	\Lambda_0(t,t^{\prime})= \frac{e^{i \epsilon_d (t-t^{\prime})}}{2}e^{- \frac{\Gamma}{2} (t+t^{\prime})}e^{\Gamma t_0}n_d,\\
    &\lim_{\Gamma \to 0}	\Lambda_{\gamma}(t,t^{\prime})= \frac{2e^{i \epsilon_d (t-t^{\prime})}}{\Gamma}f_{\gamma}(\epsilon_d)\delta (t-t'),\\
    &\lim_{\Gamma \to 0}	\Lambda^{(1)}_{\gamma}(t,t^{\prime})= e^{i \epsilon_d (t-t^{\prime})}\left[\Theta (t^{\prime}-t)e^{- \frac{\Gamma}{2} \abs{t-t^{\prime}}}-\Theta (t_0-t)e^{- \frac{\Gamma}{2} (t^{\prime}-t_0)}e^{- \frac{\Gamma}{2} \abs{t-t_0}}\right]f_{\gamma}(\epsilon_d),\\
    &\lim_{\Gamma \to 0}	\Lambda^{(2)}_{\gamma}(t,t^{\prime})= \frac{e^{i \epsilon_d (t-t^{\prime})}}{2}\left[e^{- \frac{\Gamma}{2} \abs{t-t^{\prime}}}+e^{- \frac{\Gamma}{2} (t+t^{\prime})}e^{\Gamma t_0}-e^{- \frac{\Gamma}{2} (t^{\prime}-t_0)}e^{- \frac{\Gamma}{2} \abs{t-t_0}}-e^{- \frac{\Gamma}{2} (t-t_0)}e^{- \frac{\Gamma}{2} \abs{t^{\prime}-t_0}}\right]f_{\gamma}(\epsilon_d).
\end{aligned}
\end{equation}
\end{widetext}
By substituting the above expressions back into Eq. \eqref{SHE}, it is possible to show analytically that the two approaches lead to the same result.

In Fig. \ref{fig:noise} we show the auto- and cross-current fluctuations as functions of time and as functions of the energy of the dot. As can be clearly seen, the result obtained with the $\he$ approaches the one obtained with the $\me$ approach in the limit of weak coupling, namely for small values of $\Gamma/ T$ (green curves). 
We emphasize that the $\me$ current correlation function is a real quantity as seen from Eq.~\eqref{eq:SME}.
This is further illustrated in Fig. \ref{fig:noise} where, for small coupling, the real part of the exact $\he$ result converges to the $\me$ one (black dashed curve), while the imaginary part vanishes. However, as the coupling strength increases, the $\me$ result deviates from the $\he$ one (red and blue curves), emphasizing that the $\me$ result remains valid only under weak-coupling conditions.\\
% In particular, we note that the imaginary part of the Heisenberg current fluctuations (dotted curves) converges to zero when $\Gamma/ T\to 0$ for all values of time (left panels) and energy of the dot (right panels), while the real part converges to the $\me$ result. 
%Moreover, two additional observations can be made \gb{in the case of large values of $\Gamma/T$ (red and blue curves)}. First, looking at the time-behavior of current fluctuations (left panels, (a) and (b)), we note that even if the $\me$ and the $\he$ results have the same behavior, the $\me$ approach fails to reproduce not only the transient regime, but also the stationary one. 
In panels (c) and (d), we look at the current correlation as a function of the energy of the dot. In this case, the behavior between the two results, from the $\he$ and $\me$ approaches, is completely different. In particular, the $\me$ approach fails to display the periodic oscillations which are present in the strong coupling regime using the exact $\he$ approach.
These oscillations arise from quantum coherent effects as we have already discussed in the case of the currents in Fig.~\ref{fig:currents1}.

As a final result, we illustrate the connection between shot noise in the $\lb$ framework and the $\me$ result. To achieve this, we must compute the integral with respect to $\tau$ from Eq.~\eqref{SME_stationary}. This corresponds to the Fourier transform at the frequency $\omega=0$, 
\begin{align}
	\int d\tau S^\me_{\alpha \alpha^{\prime},ss}(\tau)&=\pm\frac{\Gamma _L \Gamma_R}{\Gamma } \left[f_L \left(1-f_R\right)+f_R \left(1-f_L\right)\right]\nonumber\\
 &\mp2\frac{ \Gamma _L^2 \Gamma _R^2}{\Gamma^3} \left( f_L-f_R\right)^2,
\end{align}
where $\pm$ distinguishes between auto- ($\alpha=\alpha^{\prime}$) and cross-current ($\alpha\neq\alpha^{\prime}$) correlations. 
It is straightforward to show that, by following the weak-coupling protocol discussed previously in this section, the above result coincides with the $\lb$ shot noise for small $\Gamma$, i.e., $\Gamma^2\left(\lim_{\Gamma\to 0}S_{\alpha\alpha'}^\lb(\omega=0)/\Gamma^2\right)$.

%%%%%%%%%%%%%%%%%%%%%%%%%%%%%%%%%%%%%%%%%%%%%%%%%%%%%%%%%%%%%%%%%%%%%%%%%%%%%%%%%%%%%%%%%%%%%%%%%%%%%%

\section{Generalization: The double quantum dot system}
\label{sec:DQD}

\begin{figure}[!htb]%[h!]
\centering
\includegraphics[width=0.8\columnwidth]{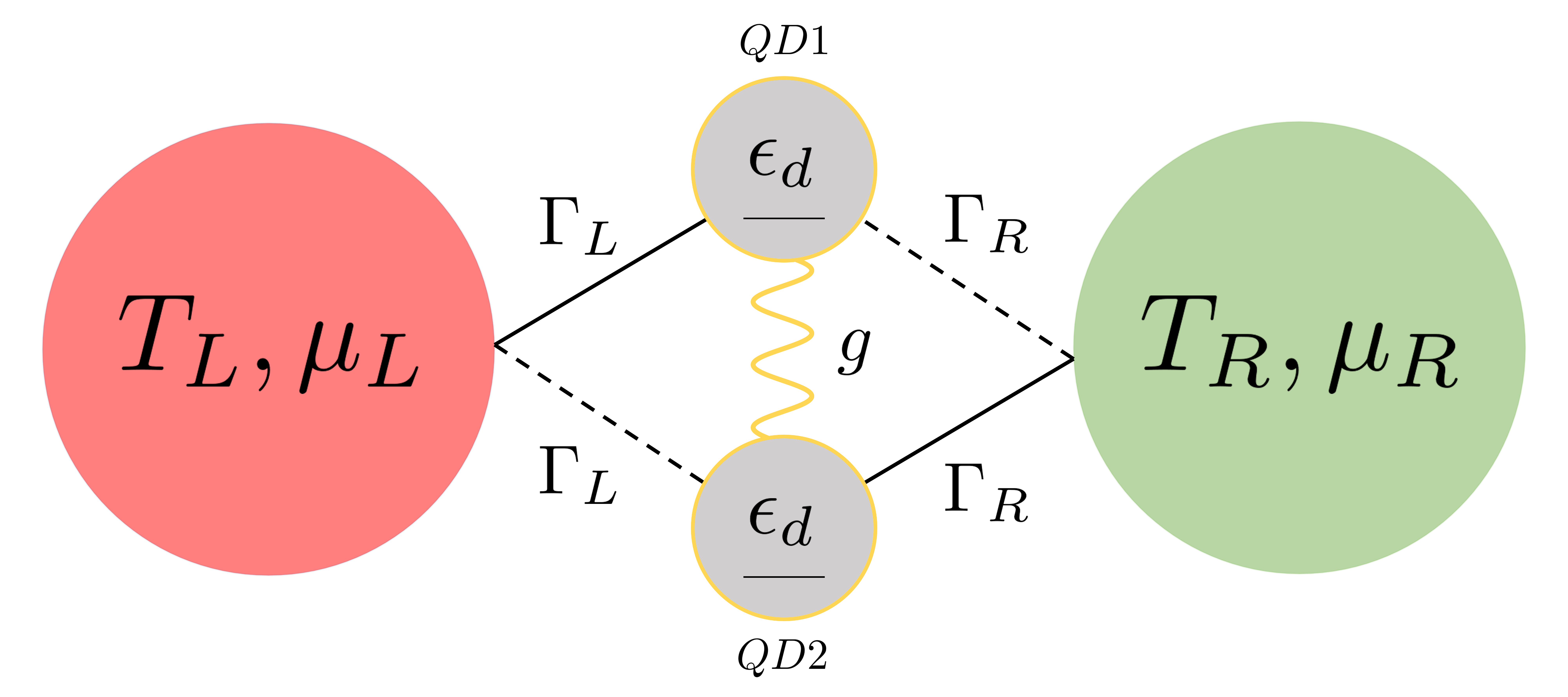}
\caption{Scheme of a double quantum-dot system: two degenerate quantum dots (QD1 and QD2) with energies $\epsilon_1=\epsilon_2=\epsilon_d$ and tunnel-coupling strength $g$, are attached to two reservoirs at temperature $T_\alpha$ and chemical potential $\mu_\alpha$, with $\alpha=L,R$. $\Gamma_L$ and $\Gamma_R$ represent the energy-independent bare tunneling rates (in the WBL) describing the coupling of the dots with the left and right reservoirs respectively. Solid lines correspond to a \textit{Series} configuration (each reservoir is attached to one dot only), while dashed lines correspond to a \textit{Parallel} configuration (both reservoirs are attached to both dots).} 
\label{fig:double_quantum_dot}
\end{figure}

In this section, we analyze the case of a double quantum-dot (DQD) coupled to two reservoirs as depicted in Fig.~\ref{fig:double_quantum_dot}. The Hamiltonian describing this system corresponds to the model introduced in Eq.~\eqref{eq:totalH} with $D=2$ and $N=2$, where the two dots have degenerate energies $\epsilon_1=\epsilon_2=\epsilon_d$, and interact with a tunneling strength $g$, 
\begin{align}
\label{eq:totalH_DQD}
		\hat{H}&=\sum_{n}\epsilon_{d} \hat{d}^{\dagger}_n\hat{d}_n + g\sum_{n\neq m}\hat{d}_n^{\dagger}\hat{d}_m+\sum_{k \alpha}\epsilon_{k\alpha}\hat{c}_{k\alpha}^{\dagger}\hat{c}_{k\alpha}\nonumber\\
  &+\sum_{n k \alpha }\left[t_{nk\alpha }^{*}\hat{c}_{k\alpha }^{\dagger}\hat{d}_n + t_{nk\alpha} \hat{d}^{\dagger}_n\hat{c}_{k\alpha}\right].
\end{align}
In the equations above and in the rest of the section, we adopt the notation $n,m=1,2$ to label the two dots, left and right reservoirs are labelled by $\alpha=L,R$. Each fermionic reservoir is characterized by its Fermi-Dirac distribution, $f_{\alpha}(\epsilon)=\{e^{(\epsilon-\mu_{\alpha})/ T_{\alpha}}+1\}^{-1}$, with $\mu_\alpha$ the chemical potential and $T_\alpha$ the temperature. In the following, we investigate two possible configurations for the two dots, the \textit{Parallel} and \textit{Series} configurations, defined as follows:
\begin{itemize}
    \item \textit{Parallel configuration ($P$)}: the two dots are attached to both reservoirs (solid and dashed lines in Fig.~\ref{fig:double_quantum_dot}). This configuration corresponds to a choice of the bare tunneling rate defined in Eq.~\eqref{eq:tunnel_rate} of the form,
    \begin{equation}
        \Gamma_{nm}^{\alpha}(\epsilon)=\Gamma_\alpha \delta_{nm}\,.
    \end{equation}
    \item \textit{Series configuration ($S$)}: each dot is attached to a single reservoir (solid lines in Fig.~\ref{fig:double_quantum_dot}). The corresponding form of the bare tunneling rate is
    \begin{equation}
        \Gamma_{nm}^{\alpha}(\epsilon)=\Gamma_\alpha \delta_{nm} \delta_{n\alpha}\,.
    \end{equation}
\end{itemize}

Let us note that, for the sake of clarity and given the complexity of the analytical results, we restrict our analysis to the evolution of the populations of the two quantum dots. These results are sufficient to demonstrate the validity and applicability of our weak-coupling procedure, as well as for bringing out the distinction between global and local master equations. In future works, it would be interesting to push these calculations to the coherences of the density operator, as well as to transport observables. However, we do not expect any technical difficulties in these directions, beyond the ones discussed below.

The section is organized as follows. In subsection~\ref{HE_DQD}, we compute the exact expressions of the state of the system using the $\he$ approach for the two configurations described above. Subsequently, in subsection~\ref{ME_DQD}, we present the results obtained from the $\me$ approach considering the \textit{local} and the \textit{global} Lindblad master equation respectively \cite{Hofer2017,Gonzalez2017}. Finally, in subsection~\ref{WC_DQD}, we demonstrate how our weak coupling protocol enables us to recover the local $\me$ state of the system from the exact series configuration and the global $\me$ result from the parallel configuration. These results demonstrate the validity of our connections and weak-coupling protocol beyond the minimal model of a single-level quantum dot. We provide all analytical details in the main text and Appendix \ref{app:DQD} to allow the interested reader to exploit our results considering other setups.

\vspace{0.5cm}
\subsection{HE exact solutions for the DQD}
\label{HE_DQD}

The Heisenberg equations for the DQD considering the Hamiltonian Eq.~\eqref{eq:totalH_DQD} take the form,
\begin{align}
	\label{diff_Eq_DQD_d_c}
	\frac{d}{d t} \hat{d}_n&= i[\hat{ H}, \hat{d}_n]=-i\epsilon_d \hat{d}_n-i g \sum_{m\neq n} \hat{d}_m  -i \sum_{k \alpha} t_{n k \alpha}\hat{c}_{k \alpha}, \nonumber \\
	&& \\
	\frac{d}{d t} \hat{c}_{k \alpha} &= i[\hat{ H}, \hat{c}_{k \alpha}]=-i \epsilon_{k \alpha} \hat{c}_{k \alpha}-i \sum_{m} t_{m k \alpha}^* \hat{d}_m\,.
\end{align}
%\begin{eqnarray}
%\label{diff_Eq_DQD_d_c}
%\frac{d}{d t} \hat{d}_n&=& i[\hat{ H}, \hat{d}_n]=-i\epsilon_d \hat{d}_n-i g \sum_{m\neq n} \hat{d}_m  -i \sum_{k \alpha} t_{n k \alpha}\hat{c}_{k \alpha}, \nonumber \\
%&& \\
%\frac{d}{d t} \hat{c}_{k \alpha} &=& i[\hat{ H}, \hat{c}_{k \alpha}]=-i \epsilon_{k \alpha} \hat{c}_{k \alpha}-i \sum_{m} t_{m k \alpha}^* \hat{d}_m\,.
%\end{eqnarray}
%where we have utilized the DQD Hamiltonian \eqref{eq:totalH_DQD} to simplify the expressions. 
In analogy with the single resonant-level model, these two equations can be manipulated to obtain an equation for the operator $\hat{d}_n$ of dot $n$,
\begin{equation}
   \frac{d}{d t} \hat{d}_n=-\left(\frac{\Gamma_{nn}}{2}+i\epsilon_d \right)\hat{d}_n-\sum_{m\neq n}\left(\frac{\Gamma_{nm}}{2}+i g\right) \hat{d}_m + \hat{\xi}_{n}(t)\,,
\end{equation}
with
\begin{eqnarray}
\Gamma_{nm} &=&\sum_{\alpha}\Gamma_{nm}^\alpha \,, \\
\hat{\xi}_{n}(t) &=& -i\sum_{k\alpha}t_{nk\alpha}e^{-i\epsilon_{k\alpha}(t-t_0)}\hat{c}_{k\alpha}(t_0)\,.
\end{eqnarray}
The evolution equations for the annihilation operators $\hat{d}_1, \hat{d}_2$ of the two dots can be written in a compact form in terms of the vectors $\vec{\hat{d}} = (\hat{d}_1, \hat{d}_2)^T$ and $\vec{\hat{\xi}} = (\hat{\xi}_1, \hat{\xi}_2)^T$,
\begin{equation}
\label{eq:matrix_diff_eq}
    \frac{d}{d t}\vec{\hat{d}}(t)=A^{\sigma}\vec{\hat{d}}(t)+\vec{\hat{\xi}}(t)
\end{equation}

The matrix $A^\sigma$ differs depending on the configuration. We provide below its explicit form for the \textit{Parallel} ($\sigma=P$) and \emph{Series} ($\sigma=S$) configurations, as well as its eigenvalues $\lambda^{P,S}_\pm$: 
%with $\sigma=\left\{P,S\right\}$ assumes different forms in different configurations, i.e. $\sigma=P$ for the \textit{Parallel} and $\sigma=S$ for the \textit{Series} configuration respectively.
\begin{itemize}
    \item \textit{Parallel}: 
    \begin{equation}
    \label{eq:matrix_A_Parallel}
        A^P=-\begin{pmatrix}
\frac{\Gamma}{2}+i \epsilon_d &  i g \\
i g & \frac{\Gamma}{2}+i \epsilon_d 
\end{pmatrix},
\quad \lambda^P_{\pm}=-\frac{\Gamma}{2}-i\left(\epsilon_d\pm g\right)\,,
    \end{equation}

\item \textit{Series}: 
    \begin{equation}
     \label{eq:matrix_A_Series}
        A^S=-\begin{pmatrix}
\frac{\Gamma_L}{2}+i \epsilon_d &  i g \\
i g & \frac{\Gamma_R}{2}+i \epsilon_d 
\end{pmatrix},
\quad \lambda^S_{\pm}=-\frac{\Gamma}{4}\pm\eta-i\epsilon_d \,,
    \end{equation}
\end{itemize}
with $\gamma\equiv \frac{\Gamma_L-\Gamma_R}{4}$ and $\eta\equiv \sqrt{\gamma^2-g^2}$. In this work, we restrict our calculations to the case $\eta$ being real, corresponding to $g\leq\abs{\gamma}$. This situation corresponds to the overdamped regime of the dynamics of this open quantum system, see Ref.~\cite{Khandelwal2021} for a discussion.
%real for $g\leq\abs{\gamma}$. In the following, we will always assume $\eta \in \mathbb{R}$. The case $\eta \in \mathbb{C}$ leads to oscillations in the dynamics as discussed in \cite{Khandelwal2021} for instance. Similar calculations can be conducted. Defining} 
Defining the matrix $D^{\sigma}(t)=e^{A^\sigma t}$, the solution of Eq.~\eqref{eq:matrix_diff_eq} can be expressed componentwise as follows
\begin{equation}
  \hat{d}_n(t)=\sum_m D^\sigma_{nm}\hat{d}_m(t_0) +\sum_m\int_{t_0}^{t}ds~D^\sigma_{nm}(t-s)\hat{\xi}_{m}(s).
\end{equation}
This allows us to express the average population of the dots for the configuration $\sigma=P,S$ as
% \begin{align}
% \label{exp_val_populations}
% &\expval{\hat{d}^{\dagger}_n(t)\hat{d}_{n'}(t)}_{\sigma}=\sum_{m m'} D^{\sigma}_{nm}(t)^*D^\sigma_{m'n'}(t)\expval{\hat{d}^{\dagger}_m(t_0)\hat{d}_{m'}(t_0)}\nonumber\\
% &+\sum_{m m'}\int_{t_0}^t ds~ds' D^{\sigma}_{nm}(t-s)^*D^\sigma_{m'n'}(t-s')\expval{\hat{\xi}_m^{\dagger}(s)\hat{\xi}_{m'}(s')}.
% \end{align}
% By using the WBL approximation, and the eigenvalue decomposition of the matrix $A^\sigma$, we can explicit the analytical solution of the above equation for the populations of the two quantum dots QD1 and QD2, in both configurations. For QD1 we obtain 
\begin{widetext}
\begin{equation}
\label{exp_val_populations}
\expval{\hat{d}^{\dagger}_n(t)\hat{d}_{n'}(t)}_{\sigma}=\sum_{m m'} D^{\sigma}_{nm}(t)^*D^\sigma_{m'n'}(t)\expval{\hat{d}^{\dagger}_m(t_0)\hat{d}_{m'}(t_0)}+\sum_{m m'}\int_{t_0}^t ds~ds' D^{\sigma}_{nm}(t-s)^*D^\sigma_{m'n'}(t-s')\expval{\hat{\xi}_m^{\dagger}(s)\hat{\xi}_{m'}(s')}.
\end{equation}
By using the WBL approximation, and the eigenvalue decomposition of the matrix $A^\sigma$, we can explicit the analytical solution of the above equation for the populations of the two quantum dots QD1 and QD2, in both configurations. For QD1 we obtain 
\begin{itemize}
    \item In the \textit{Parallel} configuration:
    \begin{equation}
    \label{eq:rho_P_exact}
    \rho_{\he,1}^P(t)\equiv \expval{\hat{d}^{\dagger}_{1} \left(t\right)\hat{d}_{1}(t)}_{P}=e^{-\Gamma (t-t_0)}\left[\bar{n}+\Delta n\cos{\left(2g(t-t_0)\right)}\right]+2\sum_{\alpha m}\frac{\Gamma_{\alpha}}{\Gamma}\left[M_{\alpha}^P(t)\right]_{mm},
\end{equation}

\item In the \textit{Series} configuration:
\begin{align}
\label{eq:rho_S_exact}
    \rho_{\he,1}^S(t)&\equiv \expval{\hat{d}^{\dagger}_1 (t)\hat{d}_{1}(t)}_{S}=\frac{e^{-\frac{\Gamma}{2} (t-t_0)}}{\eta^2}\left[-g^2\bar{n}+\left(\gamma^2 n_1-g^2\Delta n\right)\cosh{\left(2\eta(t-t_0)\right)}-n_1\gamma~\eta \sinh{\left(2\eta(t-t_0)\right)}\right]\nonumber\\
    &-\frac{g^2}{\eta^2}\sum_{\alpha m m'}\left(-1\right)^{\delta_{m m'}\delta_{\alpha R}}\frac{\Gamma_{\alpha}}{\Gamma}\left[M^{S}_{\alpha}(t)\right]_{m m'}+2\frac{\Gamma_{L}}{\Gamma}\frac{\gamma}{\eta^2}\left\{\left[M^{S}_{L}(t)\right]_{--}\left(\gamma+\eta\right)+\left[M^{S}_{L}(t)\right]_{++}\left(\gamma-\eta\right) \right\}.
\end{align}
\end{itemize}
\end{widetext}
where $\bar{n}=\frac{n_1+n_2}{2}$, $\Delta n=\frac{n_1-n_2}{2}$ and $n_1, n_2$ the initial populations of the two dots. The time-dependent population of QD2 in both configurations, $ \rho_{\he,2}^P(t)$ and $\rho_{\he,2}^S(t)$, are obtained by exchanging $L\leftrightarrow R$ and $1\leftrightarrow 2$ in the above equations. In Eqs.~\eqref{eq:rho_P_exact} and \eqref{eq:rho_S_exact}, the elements of the matrix $M^{\sigma}_{\alpha}$ have an analytical form, and depend on the eigenvalues of the matrix $A^\sigma$,
\begin{align}
\label{eq:M_general_main}
  \left[M^{\sigma}_{\alpha}(t)\right]_{m m'} &=\Gamma \int \frac{d\epsilon}{2\pi} \frac{e^{\lambda_m^{\sigma*}\frac{t-t_0}{2}}\sinh{\left[\frac{t-t_0}{2}\left(\lambda_m^{\sigma*}-i\epsilon\right)\right]}}{\lambda_m^{\sigma*}-i\epsilon}\nonumber\\
  &\times\frac{e^{\lambda_{m'}^{\sigma}\frac{t-t_0}{2}}\sinh{\left[\frac{t-t_0}{2}\left(\lambda_{m'}^{\sigma}+i\epsilon\right)\right]}}{\lambda_{m'}^{\sigma}+i\epsilon}f_{\alpha}(\epsilon).
\end{align}
All details of the derivation of the above equations are provided in Appendix~\ref{app:DQD}. To the best of our knowledge, Eqs.~\eqref{eq:rho_P_exact} and \eqref{eq:rho_S_exact} provide for the first time analytical expressions for the populations of the double-quantum dot setup in a parallel or series configurations. These results allow us to discuss connections with previous results obtained within a master equation approach, that we recall in the next subsection.

\subsection{Master Equation for the DQD}
\label{ME_DQD}

Unlike the single-resonant-level model, the Lindblad master equation for a two-terminal double quantum dot setup  is not uniquely defined. Indeed, the interaction between the dots brings a new energy scale in the system. The strength of this interaction determines whether the secular approximation holds and whether the resulting master equation treats dissipation \textit{locally} or \textit{globally} \cite{Hofer2017, Gonzalez2017, DeChiara2018, Dann2018, Mitchison2018, Cattaneo2019, Potts2021, Dann2023}. In the following, we use the upper index $\mu= \ell, \text{\textit{g}}$ to label the \emph{local} and \emph{global} configurations respectively. 
%The index $j$ runs over the relevant energy states for each description. 
The Lindblad equation takes the following form
% We introduce below a slightly different general notation as compared to Eq.~\eqref{eq:Lindblad}, convenient to express our results: we add the upper index $\mu= \ell, \text{\textit{g}}$ to highlight explicitly \emph{local} versus \emph{global}. The index $j$ runs over the relevant energy states for each description, in agreement with the rest of the manuscript. It reads:
%Here, we indicate with $\mu=\ell$ the \textit{local} and $\mu=\text{\textit{g}}$ the \textit{global} Lindblad master equations respectively, which take the following form for $\alpha=L,R$ and $\nu=\pm$ \sk{SK: I checked this notation.} \gh{If I am correct, we do not need the index $\nu$.}
% \begin{align}
%      \dot\rho_{\me}^\mu(t)
% = &-i[\hat H_S,\rho_{\me}^\mu(t)]+ \sum_{\alpha}\sum_{\nu=\pm}\Gamma^{\mu\nu}_{\alpha}  \mathcal D\left[\hat L_\alpha^{\mu\nu} \right]\rho_{\me}^\mu(t).  
%  \end{align}
\begin{align}
\label{eq:local_global_Lindblad}
     \dot\rho_{\me}^\mu(t)
= &-i[\hat H_S,\rho_{\me}^\mu(t)]\nonumber \\
&+ \sum_{j \alpha}\left(\Gamma^{\mu-}_{j \alpha}  \mathcal D\left[\hat L_{j \alpha}^{\mu} \right]+ \Gamma^{\mu+}_{j \alpha}  \mathcal D\left[\hat L_{j \alpha}^{\mu\dagger} \right]\right)\rho_{\me}^\mu(t).  
 \end{align}
The expressions of the rates and jump operators depend on whether the \emph{local} or \emph{global} description is used:

\begin{itemize}
    % \item \textit{Local} master equation ($\mu=\ell$): we consider energy-degenerate dots, i.e., $\epsilon_j = \epsilon_d$, with $j=1,2$. In this case, dot 1 (2) only couples to the left (right) reservoir, as shown in Fig.~\ref{fig:double_quantum_dot}. Thus, the rates and jump operators are given by:
    \item \textit{Local} master equation ($\mu=\ell$): In this case, dot 1 (2) only couples to the left (right) reservoir, as shown in Fig.~\ref{fig:double_quantum_dot}, and the jump operators act on the charge states of the dots with energies $\epsilon_1, \epsilon_2$. Here, we consider energy-degenerate dots, i.e., $\epsilon_j = \epsilon_d$, with $j=1,2$. Therefore, the rates and jump operators take the following form:
\begin{eqnarray}
\Gamma_{j \alpha}^{\ell +}=\Gamma_\alpha f_{\alpha}(\epsilon_d);\quad \Gamma_{j \alpha}^{\ell -}=\Gamma_\alpha\left(1- f_{\alpha}(\epsilon_d)\right)\,,
\end{eqnarray}
and
\begin{eqnarray}
&& \hat L_{j L}^{\ell}=\delta_{1j}\hat \sigma_{-}^{(j)};\quad \hat L_{j R}^{\ell}= \delta_{2 j}\hat \sigma_{-}^{(j)}\,.
% \\
% && \hat L_{j L}^{\ell \dagger}=\delta_{1j}\hat \sigma_{+}^{(j)};\quad \hat L_{j R}^{\ell \dagger}= \delta_{2 j}\hat \sigma_{+}^{(j)}\,.
\end{eqnarray}
with $\hat \sigma_{\pm}^{(1)}\equiv \left(\hat \sigma_{\pm}\otimes \mathds{1}\right)$ and $\hat \sigma_{\pm}^{(2)}\equiv \left(\mathds{1}\otimes\hat \sigma_{\pm}\right)$.
\end{itemize} 
%It is important to notice that $L_{L\pm}^{\ell}$ removes particles from the system and adds them to the left reservoir, acting only on the first dot while leaving the second dot unchanged. Similarly, $L_{R\pm}^{\ell}$ removes particles from the system and adds them to the right reservoir, acting only on the second dot while leaving the first dot unchanged. This reflects the local nature of the Lindblad master equation.

\begin{itemize}
   % \item \textit{Global} master equation  ($\mu=\text{\textit{g}}$): the relevant energies entering the rates are the eigenenergies $\epsilon_\pm$ of $\hat{H}_S$. The rates are therefore evaluated at the transition energies between energy eigenstates, and the index $j$ runs over $+,-$. For the \textit{global} master equation, we have 
   \item \textit{Global} master equation ($\mu=\text{\textit{g}}$): In this case, the relevant energies are the eigenenergies of $\hat H_S$: $0$, $\epsilon_\pm = \epsilon_d \pm g$, and $2\epsilon_d$, with the corresponding eigenstates denoted by $\ket{0}$, $\ket{\epsilon_\pm}$, and $\ket{2}$, respectively.
    The rates are evaluated at the transition energies $\epsilon_j = \epsilon_\pm$, with the index $j$ running over $+,-$. The only non-zero jump operators correspond to transitions with energies $\epsilon_+$ and $\epsilon_-$. Thus, the rates and jump operators are given by:
\begin{equation}
\label{eq:Gamma_global_ME}
\Gamma_{\pm \alpha}^{\text{\textit{g}} +}=2 \Gamma_\alpha f_{\alpha}(\epsilon_\pm);\quad \Gamma_{\pm \alpha}^{\text{\textit{g}} -}=2\Gamma_\alpha\left(1- f_{\alpha}(\epsilon_\pm)\right),\\
\end{equation}
and
\begin{align}
&& \hat L^{\text{\textit{g}}}_{\pm L} = \ketbra{0}{0}\hat \sigma_{-}^{(1)}\ketbra{\epsilon_\pm}{\epsilon_\pm} + \ketbra{\epsilon_\mp}{\epsilon_\mp}\hat \sigma_{-}^{(1)}\ketbra{2}{2} \,,\nonumber\\
&&\hat L^{\text{\textit{g}}}_{\pm R} = \ketbra{0}{0}\hat \sigma_{-}^{(2)}\ketbra{\epsilon_\pm}{\epsilon_\pm} + \ketbra{\epsilon_\mp}{\epsilon_\mp}\hat \sigma_{-}^{(2)}\ketbra{2}{2}\,.
\end{align}
\end{itemize}
%In this case  the eigenenergies of $\hat H_S$ are $0$, $\epsilon_\pm =\epsilon_d \pm g$, and $2\epsilon_d$, and the corresponding eigenstates are denoted by $\ket{0}$, $\ket{\epsilon_\pm}$, and $\ket{2}$, respectively.

Let us comment on the factor 2 that appears in the rates Eq.~\eqref{eq:Gamma_global_ME}. Our objective in this section is to demonstrate the applicability of our weak coupling protocol for the case of two quantum dots, \textit{i.e.} the possibility to derive the \emph{global} and \emph{local} master equations from the exact solutions obtained from $\he$.  Given the structure of the jump operators in the \emph{global} ME, it is intuitive that the \emph{global} master equation must be derived from the \emph{Parallel} configuration from the $\he$ approach. However, in the $\me$ approach, only one channel couples each reservoir to the system, as opposed to the \emph{Parallel} configuration, where clearly two channels are present. As we show below with our analytical results, the solutions from the $\he$ and $\me$ approaches match when accounting for this factor 2, as done in Eq.~\eqref{eq:Gamma_global_ME}. %Alternatively, one could have divided each rate by 2 in the \emph{Parallel} configuration to match with the \emph{global}. 
Hence, this factor 2 is not derived from the microscopic description, but has to be accounted for when connecting the different theoretical approaches from the above simple argument. \\

To solve the Lindblad equation \eqref{eq:local_global_Lindblad} for both the \textit{local} and \textit{global} case, we utilize the vectorization technique outlined in Refs.~\cite{Khandelwal2020,Bourgeois2024}. The resulting analytical expressions for the time-dependent population of QD1 can be written as: 
% \sk{I don't like this notation. Logically these symbols should represent the reduced density matrix and not the population.}
\begin{widetext}
\begin{itemize}
\item For the \textit{local} master equation:
    \begin{align}
\label{eq:rho_L_ME}
    \rho_{\me,1}^\ell(t)&\equiv \Tr{\hat \sigma_{+}^{(1)}\hat \sigma_{-}^{(1)}\rho_{\me}^{\ell}(t)}=\frac{e^{-\frac{\Gamma}{2} (t-t_0)}}{\eta^2}\left[-g^2\bar{n}+\left(\gamma^2 n_1-g^2\Delta n\right)\cosh{\left(2\eta(t-t_0)\right)}-n_1\gamma~\eta \sinh{\left(2\eta(t-t_0)\right)}\right]\nonumber\\
        &-\frac{g^2}{\eta^2}\sum_{\alpha}\sum_{m m'=\pm}\left(-1\right)^{\delta_{m m'}\delta_{\alpha R}}\frac{\Gamma_{\alpha}}{\Gamma}f_{\alpha}(\epsilon_d)\frac{e^{-(t-t_0)\left(\frac{\Gamma}{4}+\eta \frac{m+m'}{2}\right)}}{4}\frac{\sinh{\left[(t-t_0)\Gamma\left(\frac{1}{4}+ \frac{m+m'}{2}\frac{\eta}{\Gamma}\right)\right]}}{\frac{1}{4}+ \frac{m+m'}{2}\frac{\eta}{\Gamma}}\nonumber\\
            &+\frac{\Gamma_L}{\Gamma}f_L(\epsilon_d)\frac{\gamma}{\eta^2}\sum_{m=\pm} \frac{e^{-(t-t_0)\left(\frac{\Gamma}{4}+\eta m\right)}}{2}\frac{\sinh{\left[(t-t_0)\Gamma\left(\frac{1}{4}+ m\frac{\eta}{\Gamma}\right)\right]}}{\frac{1}{4}+ m\frac{\eta}{\Gamma}}\left(\gamma+ m \eta\right).
    \end{align}
        \item For the \textit{global} master equation:
        \begin{equation}
        \label{eq:rho_G_ME}
        \rho_{\me,1}^{\text{\textit{g}}}(t)\equiv \Tr{\hat \sigma_{+}^{(1)}\hat \sigma_{-}^{(1)}\rho_{\me}^{\text{\textit{g}}}(t)}=e^{-\Gamma (t-t_0)}\left[\bar{n}+\Delta n\cos{\left(2g(t-t_0)\right)}\right]+\sum_{\alpha}\frac{\Gamma_\alpha}{\Gamma}e^{-\Gamma \frac{t-t_0}{2}}\sinh{\left(\Gamma \frac{t-t_0}{2}\right)}\left(f_\alpha\left(\epsilon_{+}\right)+f_\alpha\left(\epsilon_{-}\right)\right).
        \end{equation}
\end{itemize}
\end{widetext}
The populations of QD2, $\rho_{\me,2}^\ell(t)$ and $ \rho_{\me,2}^{\text{\textit{g}}}(t)$, are obtained by exchanging $L\leftrightarrow R$ and $1\leftrightarrow 2$.\\
%Both local and global solutions allow for interesting remarks. 
We can make two consistency checks at this point. First, considering the solution of the local master equation, Eq.~\eqref{eq:rho_L_ME}, the populations of each dot converge to the Fermi-Dirac distribution of their respective reservoirs in the limit of decoupled dots ($g=0$) and infinite time (stationary regime, $t \rightarrow \infty$):
\begin{equation}
    \rho_{\me,1}^\ell=f_L(\epsilon_d), \quad \rho_{\me,2}^\ell=f_R(\epsilon_d).
\end{equation}

%it can be shown that by taking the stationary limit as $t\rightarrow \infty$ of Eq.~\eqref{eq:rho_L_ME} and setting $g=0$, the populations of the dots converge to the Fermi-Dirac distribution of the respective leads to which they are coupled, i.e., $\rho_{\me,1}^\ell=f_L(\epsilon_d)$ and $\rho_{\me,2}^\ell=f_R(\epsilon_d)$.

Second, in the same limit of decoupled quantum dots ($g=0$), the solution of the global master equation, Eq.~\eqref{eq:rho_G_ME}, corresponds to the one of the single resonant-level, see Eq.~\eqref{eq:sol_rho11} with $n_1=n_2=n_d$. %when dots are decoupled, i.~e.~$g=0$, then we recover the same solution of the single dot given in Eq.~\eqref{eq:sol_rho11} with $n_1=n_2=n_d$.}

As a final remark, it is interesting to note that the first term in the solutions for the local and global master equations that depend on the initial populations $n_1, n_2$ exactly correspond to the one in the exact solutions for the series and parallel configuration of the $\he$ approach, without further approximation, see Eqs.~\eqref{eq:rho_P_exact} and \eqref{eq:rho_S_exact}. In particular, these terms being equal without the necessity of a Markovian assumption, implies that potential non-Markovian effects present in the exact solutions are independent of the initial populations of the dots. It also implies that the weak coupling protocol that we discuss below effectively acts only on the second terms in the exact solutions.

%to notice that in both cases, \textit{local} and \textit{global}, the part of the state that only depends on the initial population of the dots coincides with the exact solution obtained with the $\he$ approach using the \textit{Series} and \textit{Parallel} configuration respectively, see Eqs.~\eqref{eq:rho_S_exact} and \eqref{eq:rho_P_exact}. This underscores the master equation's ability to accurately capture the internal dynamics of the system. Moreover, it reveals that the non-Markovian aspect of the exact solution remains independent of the initial populations of the dots.

\subsection{Weak coupling protocol for the DQD}
\label{WC_DQD}

To demonstrate the general applicability of our weak-coupling protocol outlined in Sec.~\ref{sec:weak_coupling_protocol} for a single resonant level, we now apply it to the case of the double quantum dot. Importantly, we show how to obtain the dots population from the \textit{global} and \textit{local} Lindblad master equations, properly taking the weak-coupling limit of the exact results in the \textit{Parallel} and \textit{Series} configurations respectively.\\

We follow the same steps as described in subsection \ref{sec:weak_coupling_protocol} when introducing our weak coupling protocol. We take the weak-coupling limit $\Gamma\to 0$ of the matrix elements $\left[M^{\sigma}_{\alpha}(t)\right]_{m m'}$,
% . From our analytical expressions, it can be shown that this limit leads to terms of order $\mathcal O(1)$ in the coupling strength $\Gamma$
see Appendix~\ref{app:WCP_DQD} for the details). We obtain for both configurations $P$ and $S$,
%Since our focus lies on the dot populations, which remain of order $\mathcal O(1)$ in the coupling strength $g$, there is no necessity to first divide by $\Gamma$.
%Therefore, as detailed in Appendix~\ref{app:WCP_DQD}, we obtain:
\begin{itemize}
    \item For the \textit{Parallel} configuration:
    \begin{equation}
    \label{eq:Mmm'_WCP_main_text_Parallel}
    \lim_{\Gamma\to 0}\left[M^{P}_{\alpha}(t)\right]_{\pm\pm}=\frac{e^{-\Gamma\frac{t-t_0}{2}}}{2}f_{\alpha}(\epsilon_\pm) \sinh{\left[\frac{t-t_0}{2}\Gamma\right]},
    \end{equation}
    
    \item For the \textit{Series} configuration:
    \begin{align}
    \label{eq:Mmm_WCP_main_text_Series}
    \lim_{\Gamma\to 0}\left[M^{S}_{\alpha}(t)\right]_{m m'} &=\frac{e^{-\frac{\Gamma}{2}}}{2}e^{(m+m')\eta\frac{t-t_0}{2}} f_{\alpha}(\epsilon_d)\nonumber\\
        &\times\frac{\sinh{\left[\frac{t-t_0}{2}\Gamma\left(-\frac{1}{2}+\left(m+m'\right)\frac{\eta}{\Gamma} \right)\right]}}{-\frac{1}{2}+\left(m+m'\right)\frac{\eta}{\Gamma}}.
    \end{align}
\end{itemize}
Similar to the case of a single quantum dot, these calculations also assume $\Gamma \left(t-t_0\right)$ to be constant as $\Gamma\to 0$. Additionally, for the \textit{Series} configuration (Eq.~\eqref{eq:Mmm_WCP_main_text_Series}), we also considered $\lim_{\Gamma\to 0} \eta/\Gamma\sim \text{constant}$, which corresponds to the assumption underlying the energy scales of the local master equation, i.e., $g\lesssim\Gamma$. Finally, substituting Eqs.~\eqref{eq:Mmm'_WCP_main_text_Parallel} and \eqref{eq:Mmm_WCP_main_text_Series} into Eqs.~\eqref{eq:rho_P_exact} and \eqref{eq:rho_S_exact}, we verify that
\begin{align}
\lim_{\Gamma\to 0}\rho_{\he,n}^P(t)&=\rho_{\me,n}^{\textit{g}}(t) \label{eq:limit_global}\\
\lim_{\Gamma\to 0}\rho_{\he,n}^S(t)&=\rho_{\me,n}^{\ell}(t), \label{eq:limit_local}
\end{align}
for both dots, i.e., $n=1,2$. 
Remarkably, the two limits Eqs.~\eqref{eq:limit_global} and \eqref{eq:limit_local} provide a novel insight onto the global and local master equations, allowing for exact derivations considering either the parallel or series configurations with the $\he$ approach.
Notably, our analytical results reveal a critical insight: the series configuration within the HE framework fails to recover the global ME results. This peculiarity arises due to distinct $A^P$ and $A^S$ matrices (Eqs. \eqref{eq:matrix_A_Parallel} and \eqref{eq:matrix_A_Series}), governing the exact evolution of the dot operators in parallel and series configurations, respectively. Merely introducing a finite inter-dot coupling in the series setup does not suffice to accurately recover populations in the transient regime. On the other hand, starting from a parallel configuration, where each dot connects to both reservoirs, in the weak-coupling limit, allows the recovery of correct populations corresponding to the global master equation. This underscores that the DQD system's global description, particularly under strong inter-dot coupling, does not treat the dots as separate entities tied to specific reservoirs as assumed in the series configuration.
%\gb{Notably, our analytical results reveal a critical insight: the global master equation fails to derive the dot populations from a series configuration within the HE framework. This discrepancy arises due to distinct $A^P$ and $A^S$ matrices of Eqs. \eqref{eq:matrix_A_Parallel} and \eqref{eq:matrix_A_Series}, governing the dot operators' exact evolution in parallel versus series configurations. Merely introducing a finite dot-dot coupling from the series setup does not suffice to accurately recover populations in the transient regime. Conversely, starting from a parallel configuration, where each dot connects to both reservoirs, allows recovery of correct populations using the global master equation in the weak-coupling regime. This underscores that the DQD system's global description, particularly under strong coupling (g), cannot adequately treat the dots as separate entities tied to specific reservoirs, as assumed in the series configuration.}\\

Before, we conclude, we would like to explain how our weak coupling protocol captures the typical assumptions made when using the $\me$ approach: weak system-bath coupling corresponding to the Born approximation, the Markov assumption that implies the reservoir's correlation time to be much shorter than the system's timescales, leading to a time-local master equation (neglecting memory effects), and the secular assumption (not always necessary) which averages out rapidly oscillating terms in the interaction picture, capturing the system's slow dynamics. 

The first step of our weak-coupling protocol consists in taking the limit $\Gamma\to 0$. It therefore corresponds to the weak system-bath coupling of the Born approximation that ensures the validity of the perturbative expansion of the solution. Then, keeping  $\Gamma t$ constant provides an appropriate description of the dynamics at all times. It reflects the Markovian approximation as it selects timescales over which reservoir correlations decay rapidly compared to the system's typical timescale, making memory effects negligible. Finally, in the specific case of two quantum dots in a \emph{Series} configuration, we assume $g\lesssim\Gamma$, where $g$ is the internal coupling strength between the dots. 
This ensures that the internal dynamics (timescale $g^{-1}$) is slower than the relaxation dynamics (timescale $\Gamma^{-1}$), a situation where the secular approximation can be used.

% \st{This ensures that the internal dynamics (timescale $g^{-1}$) does not exceed the relaxation dynamics (timescale $\Gamma^{-1}$).} 
% It aligns with the secular approximation by eliminating fast oscillating terms.
% }

%%%%%%%%%%%%%%%%%%%%%%%%%%%%%%%%%%%%%%%%%%%%%%%%%%%%%%%%%%%%%%%%%%%%%%%%%%%%%%%%%%%%%%%%%%%%%%%%%%%%%%

\section{Conclusion}
\label{sec:conc}

Exploiting two paradigmatic models for open quantum systems, our work establishes connections between three distinct theoretical frameworks for quantum transport and thermodynamics: Heisenberg equations of motion, master equations and Landauer-B\"uttiker approach. These connections take the form of a weak-coupling protocol for recovering master-equation results from the exact solutions obtained from the Heisenberg equations, the Landauer-B\"uttiker approach serving as benchmark of our results in the steady-state regime. This protocol not only provides a formal procedure to connect these frameworks but also reveals critical insights into master equations. It highlights the temporal resolution determined by the system-environment coupling strength ($\Gamma^{-1}$) and the interplay of the relevant energy scales in open quantum systems.%\gh{the role of all energy scales involved in the dynamics of open quantum systems.}

For a single resonant level, we derived analytical expressions for the currents and current correlation functions at all times, first from the Heisenberg equations, and then from a generalized approach to full counting statistics within a master equation approach. Importantly, this generalization accounts for the presence of multiple reservoirs, non-instantaneous quantum jumps, and multiple times in the transient dynamics. We believe this will be a significant addition to the master equation toolbox.
The calculation of currents and current correlations are crucial for topics such as fluctuation-dissipation relations ~\cite{Altaner2016,Barker2022}, thermodynamic uncertainty relations ~\cite{Guarnieri2019,Miller2021,Potts2019,Lopez2023} and quantum metrology ~\cite{Cavina2018,Salvia2023,Rodriguez2023}, which are important problems in open quantum systems, as well as for experiments in these directions.

In the case of two interacting quantum dots, we obtained analytical expressions for the populations of the quantum system. Notably, we recovered existing results within the master equation approach, corresponding to the local and global master equation descriptions, using exact results from the Heisenberg equation for series and parallel configurations of the dots.

Focusing on integrable systems characterized by quadratic tunneling Hamiltonians was essential to rigorously establish the weak-coupling protocol, which is the key to highlight fundamental differences among different frameworks. Beyond analytically solvable models, with only numerical solutions at hand, we expect that establishing a weak-coupling procedure will be challenging.\\
%\gh{believe it is an open question to determine how to establish the implementation of a weak-coupling protocol.}
Overall, our analysis provides important insights into the consistency of different theoretical frameworks for quantum transport and thermodynamics, with their respective strengths and limitations. We believe that this work will serve as a comprehensive resource for researchers in the field of open quantum systems, especially for those seeking guidance when utilizing multiple techniques or when delving into less familiar methodologies to address specific problems.

\section*{Acknowledgements}
We are thankful for fruitful discussions with Gabriel Landi, Rosa Lopez, Patrick Potts and Alberto Rolandi, and to Liliana Arrachea, Michael Moskalets and David Sánchez for useful feedback. G.B. and G.H. acknowledge support from the Swiss National Science Foundation through the NCCR SwissMAP, and
G.H. additionally thanks the SNSF for support through a starting Grant PRIMA PR00P2\textunderscore
179748. S.K. acknowledges support from the Swiss National Science Foundation grant P500PT\textunderscore222265 and the Knut and Alice Wallenberg Foundation through the Wallenberg Center for Quantum Technology (WACQT). 

 \bibliography{References}
\newpage

%%%%%%%%%%%%%%%%%%%%%%%%%%%%%%%%%%%%%%%%%%%%%%%%%%%%%
% \newpage

\clearpage
\widetext

\appendix

% \begin{center}
% \textbf{\large SUPPLEMENTARY MATERIAL\vspace{0.3cm}\\Exact finite-time correlation functions for multi-terminal setups: \\ Connecting theoretical frameworks for quantum transport and thermodynamics}
% \end{center}
% %%%%%%%%%%%%%%%%%%%%%%%%%%%%%%%%%%%%%%%%%%%%%%%%%%%%%
	
% 	%%%% Prefix a "S" to all equations, figures, tables and reset the counter %%%%%
% 	\setcounter{equation}{0}
% 	\setcounter{section}{0}
% 	\setcounter{figure}{0}
% 	\setcounter{table}{0}
% 	%\setcounter{page}{1}
% 	\makeatletter
% 	\renewcommand{\theequation}{S\arabic{equation}}
% 	\renewcommand{\thefigure}{S\arabic{figure}}
% 	% \renewcommand{\bibnumfmt}[1]{[S#1]}
% 	% \renewcommand{\citenumfont}[1]{S#1}
% %%%%%%%%%%%%%%%%%%%%%%%%%%%%%%%%%%%%%%%%%%%%%%%%%%%%%

\section{Multi-time multi-terminal joint probability with FCS}
\label{appendix:multi-time_joint_probability}
In this section, we derive the multi-time expression for the joint probability $P^>\left(\vec{x}^{(1)},t_{1} ; \vec{x}^{(2)},t_{2}; \ldots; \vec{x}^{(N)},t_{N}\right)$, defined as the probability of having a net amount of \(\vec{x}^{(1)}\) quanta transferred within time \(t_1-t_0\), \(\vec{x}^{(2)}\) within time \(t_2-t_0\), and so on. Eq.~\eqref{eq.x-joint_probability} of the main text corresponds to the case of $N=2$.
The full joint probability distribution can be derived from the marginal probability $P\left(\vec{x}_N ; t_N\right)$. Using the Chapman–Kolmogorov property for Markovian evolution, the marginal probability can be expressed in the following form,
\begin{align}
P\left(\vec{x}_N ; t_N\right)= \sum_{\vec{x}_1, \ldots, \vec{x}_{N-1}} \Tr{\Omega\left(\vec{x}^{(N)}-\vec{x}^{(N-1)} ; t_N-t_{N-1}\right) \Omega\left(\vec{x}^{(N-1)}-\vec{x}^{(N-2)} ; t_{N-1}-t_{N-2}\right) \ldots \Omega\left(\vec{x}^{(1)} ; t_1\right) \rho_0}.
\end{align}
This corresponds to Eq.~(25) in Ref.~\cite{Marcos2010}, generalized to the case of multi-terminals, and for any quanta \(\vec{x}\) of particles, charges, or energies exchanged between the system and the reservoirs.
Then, since we have $P\left(n_N ; t_N\right)=\sum_{n_1, \ldots, n_{N-1}} P^{>}\left(n_1, t_1, \ldots, n_N, t_N\right)$, by direct inspection, we find,
\begin{align}
 P^>\left(\vec{x}^{(1)},t_{1} ; \vec{x}^{(2)},t_{2}; \ldots; \vec{x}^{(N)},t_{N}\right) =\Tr{\left[\prod_{k=1}^{N}\Omega\left(\vec{x}^{(N-k+1)}-\vec{x}^{(N-k)} ; t_{N-k+1}-t_{N-k}\right)\right]\rho_0}, 
\end{align}
which in the $\vec{\chi}$-space takes the following form,
\begin{align}
 P^>\left(\vec{\chi}^{(1)},t_{1} ; \vec{\chi}^{(2)},t_{2}; \ldots; \vec{\chi}^{(N)},t_{N}\right) =\Tr{\left[\prod_{k=1}^{N}\Omega\left(\sum_{i=N}^{N-k+1}\vec{\chi}^{(i)} ; t_{N-k+1}-t_{N-k}\right)\right]\rho_0}.
\end{align}
The above result may be alternatively derived using Bayes' theorem for the conditional density operator and using von Neumann’s projection postulate, as presented in Refs.~\cite{Marcos2007,Marcos2010}.

%%%%%%%%%%%%%%%%%%%%%%%%%%%%%%%%%%%%%%%%%%%%%%%%%%%%%%%%%%
 \section{Calculation of observables with FCS}
 \label{appendix:observables_GFCS}

In this section, we show how to compute expectation values of observables using the FCS method introduced in Sec.~\ref{GFCS}. First, we consider the average number of tunneled particles, charges, or energy quanta exchanged with reservoir $\alpha$ within time $t'-t_0$. Using the marginal property of the joint probability distribution, it takes the following form,
\begin{align}
\label{eq:appendix_general_number}
\expval{x'_\alpha(t')}_{\me} &=\int_{t_0}^{t^{\prime}}ds^\prime F(t^{\prime}-s^\prime) \sum_{\vec{x}'} x'_\alpha  P({\vec x}',s')=\int_{t_0}^{t^{\prime}}ds^\prime F(t^{\prime}-s^\prime)\sum_{\vec{x}\vec{x}'} x'_\alpha  P^>(\vec{x},s;\vec {x}',s')\nonumber\\
&= \int_{t_0}^{t^{\prime}}ds^\prime F(t^{\prime}-s^\prime) \partial_{i\chi'_\alpha} \sum_{\vec{x}\vec{x}'} P^>(\vec{x},s;\vec {x}',s')e^{i\vec{x}\cdot\vec{\chi}}e^{i\vec{x}'\cdot\vec{\chi}'}\big\rvert_{\vec{\chi},\vec{\chi}'=0}\nonumber\\
&=\int_{t_0}^{t^{\prime}}ds^\prime F(t^{\prime}-s^\prime) \partial_{i\chi'_\alpha} P^>(\vec{\chi},s;\vec{\chi}',s')\big\rvert_{\vec{\chi},\vec{\chi}'=0},
\end{align}
where we have introduced the detection response function $F$, and the joint probability in the $\vec{\chi}$-space,
\begin{align}
    P^>(\vec{\chi},s;\vec{\chi}',s')&=\sum_{\vec{x}\vec{x}'} P^>(\vec{x},s;\vec {x}',s')e^{i\vec{x}\cdot\vec{\chi}}e^{i\vec{x}'\cdot\vec{\chi}'}\nonumber\\
    &=\Tr{\Omega(\vec{\chi}',t'-t)\Omega(\vec{\chi}+\vec{\chi}',t-t_0)\rho_0}\nonumber\\
    &=\Tr{e^{\mathcal{L} (\vec{\chi}')(t'-t)}e^{\mathcal{L} (\vec{\chi}+\vec{\chi}')(t-t_0)}\rho_0}.
\end{align}
In the last line we have rewritten the propagator, $\Omega(\vec{\chi},t)=e^{\mathcal{L} (\vec{\chi})t}$.

\subsection*{Currents with FCS}

The current associated to the counting variable $x_\alpha$ can be obtained taking the time derivative of Eq.~\eqref{eq:appendix_general_number},
\begin{align}
\label{eq:appendix_general_current}
\partial_{t'}\expval{x_\alpha(t')}_{\me}&=\int_{t_0}^{t^{\prime}}ds^\prime F(t^{\prime}-s^\prime) \partial_{i\chi'_\alpha}\partial_{s'} P^>(\vec{\chi},s;\vec{\chi}',s')\big\rvert_{\vec{\chi},\vec{\chi}'=0}\nonumber\\
&= \int_{t_0}^{t}ds'\,F\left(t'-s'\right)\partial_{i\chi'_\alpha}\partial_{s'}\Tr{ e^{\mathcal L(\vec{\chi}') \left(s'-t_0\right)}\rho_0 }\big\rvert_{\vec{\chi}'=0}\nonumber\\
&= \int_{t_0}^{t}ds'\,F\left(t-s'\right)\Tr{ \partial_{i\chi'_\alpha}\mathcal L(\vec{\chi}')\rvert_{\vec{\chi}'=0} e^{\mathcal L \left(s'-t_0\right)}\rho_0 }.
\end{align}
Here in the first line, we used the property of the derivative of the convolution and the boundary conditions of the detection response function, $F(0)=F(t-t_0)=0$, while in the third line we have used the trace preserving property of the Lindbladian $\Tr{\mathcal L(0)\sigma}=0$ (valid for any density matrix $\sigma$).
In the last line, we recognize the generalized current superoperator at reservoir $\alpha$,
\begin{equation}
    \partial_{i\chi_\alpha}\mathcal L(\vec{\chi})\big\rvert_{\vec\chi=0}= \sum_j \nu_{j\alpha} \left( \mathcal L_{j \alpha}^+ - \mathcal L_{j \alpha}^- \right).
\end{equation}
The above expression corresponds to the particle current superoperator $\mathcal{I}_\alpha$ for $\nu_{j\alpha}=1$, and to the energy current superoperator $\mathcal{J}_\alpha$ for $\nu_{j\alpha}=\epsilon_j$ of Eqs.~\eqref{eq:super_current_I} and \eqref{eq:super_current_J} respectively.

\subsection*{Two-time current correlation with FCS}
\label{seb_sec:appendix_currents}
 
As mentioned in Eq.~\eqref{eq:two_time_current_correlation_main} of the main text, the two-time current correlation can be written in the following manner,
 \begin{align}
 \label{eq:two-current_correlation_appendix}
\left\langle I_\alpha(t)I_{\alpha^\prime}(t^\prime)\right\rangle_{\me} &=\int_{t_0}^{t}ds\int_{t_0}^{t^{\prime}}ds^\prime F(t-s)F(t^{\prime}-s^\prime)~\left[\partial_{i\chi_\alpha}\partial_{i\chi'_{\alpha'}} \partial_s\partial_{s'}P(\vec{\chi},s;\vec{\chi}',s')\big\rvert_{\vec{\chi},\vec{\chi}'=0}\right],
\end{align}
where we have again used the property of the derivative of the convolution and the boundary conditions of the detection response function to move the time derivatives inside the integral.
Here, we recall that the time-ordered joint probability takes the following form,
\begin{align}
\label{eq:total_joint_probability_appendix}
P\left(\vec{\chi},s;\vec{\chi}',s'\right)&=
\Theta (s'-s)P^>\left(\vec{\chi},s;\vec{\chi}',s'\right)+\Theta (s-s')P^<\left(\vec{\chi},s;\vec{\chi}',s'\right),
\end{align}
with $P^<\left(\vec{\chi},s;\vec{\chi}',s'\right)=P^>\left(\vec{\chi}',s';\vec{\chi},s\right)$.
The term in square parenthesis in Eq.~\eqref{eq:two-current_correlation_appendix}, can be computed by performing a change of variable. In particular by substituting $\tau=s'-s$ and $t=s$, and using the property of the  derivative of the Dirac delta function $-\delta'(\tau)f(\tau)=\delta(\tau)f'(0)$, we can write it in the following form,
\begin{align}
\label{eq:derivative_joint_probability}
\partial_{i\chi_\alpha}\partial_{i\chi'_{\alpha'}} \partial_s\partial_{s'}P(\vec{\chi},s;\vec{\chi}',s')\big\rvert_{\vec{\chi},\vec{\chi}'=0}&=
\partial_{i\chi_\alpha} \partial_{i\chi_{\alpha'}'}\delta(\tau)\left(\partial_t -\partial_\tau\right)\left[  P^>(\vec{\chi},t;\vec{\chi}',t+\tau)-  P^<(\vec{\chi},t;\vec{\chi}',t+\tau)\right]\big\rvert_{\vec \chi, \vec\chi^\prime=0}\nonumber\\
&+\partial_{i\chi_\alpha} \partial_{i\chi_{\alpha'}'}\Theta(\tau)(-\partial_{\tau}^2+\partial_\tau\partial_t)\mathcal P^>(\vec{\chi},t;\vec{\chi}',t+\tau)\big\rvert_{\vec \chi, \vec\chi^\prime=0}\nonumber\\
&+\partial_{i\chi_\alpha} \partial_{i\chi_{\alpha'}'}\Theta(-\tau)(-\partial_{\tau}^2+\partial_\tau\partial_t)\mathcal P^<(\vec{\chi},t;\vec{\chi}',t+\tau)\big\rvert_{\vec \chi, \vec\chi^\prime=0}
\end{align}
We can now calculate the above three terms. The first term can be simplified as,
\begin{align*}
% \label{eq:(i)}
\text{(i)}\rightarrow\quad&\partial_{i\chi_\alpha} \partial_{i\chi_{\alpha'}'}\delta(\tau)\left(\partial_t -\partial_\tau\right)\left[  P^>(\vec{\chi},t;\vec{\chi}',t+\tau)-  P^<(\vec{\chi},t;\vec{\chi}',t+\tau)\right]\Big\rvert_{\vec \chi, \vec\chi^\prime=0}\nonumber\\
&~~~~=\partial_{i\chi_\alpha} \partial_{i\chi_{\alpha'}'}\delta(\tau)\left(\partial_t-\partial_\tau\right)\Tr{e^{\mathcal{L}(\vec{\chi}')\tau}e^{\mathcal{L}(\vec{\chi}+\vec{\chi}')t}\rho_0-e^{\mathcal{-L(\vec{\chi})\tau}}e^{\mathcal{L}(\vec{\chi}+\vec{\chi}')(t+\tau)}\rho_0}\Big\rvert_{\vec \chi, \vec\chi^\prime=0}\nonumber\\
&~~~~=\partial_{i\chi_\alpha} \partial_{i\chi_{\alpha'}'}\delta(\tau)\Tr{\mathcal{L}(\vec{\chi}'+\vec{\chi})\rho(\vec{\chi}'+\vec{\chi},t)-\mathcal{L}(\vec{\chi}')\rho(\vec{\chi}'+\vec{\chi},t)-\mathcal{L}(\vec{\chi})\rho(\vec{\chi}'+\vec{\chi},t)}\big\rvert_{\vec \chi, \vec\chi^\prime=0}\nonumber\\
&~~~~=\delta(\tau)\Tr{\partial_{i\chi_\alpha} \partial_{i\chi_{\alpha'}'}\mathcal{L}(\vec{\chi}'+\vec{\chi})\big\rvert_{\vec \chi, \vec\chi^\prime=0}\rho(t)}\nonumber\\
&~~~~=\delta(\tau)\Tr{\partial_{i\chi_\alpha} \partial_{i\chi_{\alpha'}'}\left[\mathcal L_0 +\sum_{j\beta} \left(e^{i(\chi_\beta+\chi'_\beta) \nu_{j\beta}}\mathcal L_{j\beta}^+ +e^{-i(\chi_\beta+\chi'_\beta) \nu_{j\beta}}\mathcal L_{j\beta}^-\right)\right]\Big\rvert_{\vec \chi, \vec\chi^\prime=0}~\rho(t)}\nonumber\\
&~~~~=\delta(\tau)\Tr{\partial_{i\chi_\alpha} \left[\sum_{j}\nu_{j\alpha'} \left(e^{i\chi_{\alpha'} \nu_{j\alpha'}}\mathcal L_{j\alpha'}^+ -e^{-i\chi_{\alpha'} \nu_{j\alpha'}}\mathcal L_{j\alpha'}^-\right)\right]\Huge\rvert_{\vec \chi \vec=0}~\rho(t)}\nonumber\\
&~~~~=\delta(\tau)\delta_{\alpha\alpha'}\Tr{ \sum_{j}\nu_{j\alpha}^2 \left(\mathcal L_{j\alpha}^+ +\mathcal L_{j\alpha}^-\right)\rho(t)}\nonumber\\
&~~~~=\delta(\tau)\delta_{\alpha\alpha'}\Tr{ \mathcal{A}_{\alpha}e^{\mathcal{L}(t-t_0)}\rho_0}.
\end{align*}
Here, in the third line we have used the property of the Dirac delta function, $\delta(\tau)f(\tau)=\delta(\tau)f(0)$, and the trace preserving property of the Lindbladian $\Tr{\mathcal L(0)\sigma}=0$ for any density matrix $\sigma$. In the fourth line, we have substituted the expression of $\mathcal{L}(\vec{\chi}'+\vec{\chi})$ as presented in Eq.~\eqref{eq:Lindbladian_fourier_transformed}. Finally, in the last line, we have identified the activity, $\mathcal A_\alpha$, which counts the total number of quantum jumps in or out of the system.
The second term of Eq.~\eqref{eq:derivative_joint_probability} takes the following form,
\begin{align*}
% \label{eq:(ii)}
\text{(ii)}\rightarrow\quad &\partial_{i\chi_\alpha} \partial_{i\chi_{\alpha'}'}\Theta(\tau)(-\partial_{\tau}^2+\partial_\tau\partial_t)P^>(\vec{\chi},t;\vec{\chi}',t+\tau)\big\rvert_{\vec \chi, \vec\chi^\prime=0}\nonumber\\
&~~~~= \partial_{i\chi_\alpha} \partial_{i\chi_{\alpha'}'}\Theta(\tau)(-\partial_{\tau}^2+\partial_\tau\partial_t)\Tr{e^{\mathcal{L}(\vec{\chi}')\tau}e^{\mathcal{L}(\vec{\chi}+\vec{\chi}')t}\rho_0}\big\rvert_{\vec \chi, \vec\chi^\prime=0}\nonumber\\
&~~~~=\Theta(\tau)\partial_{i\chi_\alpha} \partial_{i\chi_{\alpha'}'}\Tr{\mathcal{L}(\vec{\chi}')\left(e^{\mathcal{L}(\vec{\chi}')\tau}\mathcal{L}(\vec{\chi}+\vec{\chi}')-\mathcal{L}(\vec{\chi}')e^{\mathcal{L}(\vec{\chi}')\tau}
\right)\rho(\vec{\chi}+\vec{\chi}',t)}\big\rvert_{\vec \chi, \vec\chi^\prime=0}\nonumber\\
&~~~~=\Theta(\tau)\partial_{i\chi_\alpha} \Tr{\left[\partial_{i\chi_{\alpha'}'}\mathcal{L}(\vec{\chi}')\right]_{\vec\chi^\prime=0} \left(e^{\mathcal{L}(0)\tau}\mathcal{L}(\vec{\chi})-\mathcal{L}(0)e^{\mathcal{L}(0)\tau}
\right)\rho(\vec{\chi},t)}\big\rvert_{\vec \chi=0}\nonumber\\
&~~~~=\Theta(\tau)\partial_{i\chi_\alpha} \Tr{\left[\partial_{i\chi_{\alpha'}'}\mathcal{L}(\vec{\chi}')\right]_{\vec\chi^\prime=0} e^{\mathcal{L}(0)\tau}\left[\partial_{i\chi_{\alpha}}\mathcal{L}(\vec{\chi})\right]_{\vec\chi=0}
\rho(t)}\nonumber\\
&~~~~=\Theta(\tau) \Tr{\mathcal{I}_{\alpha'} e^{\mathcal{L}\tau}\mathcal{I}_{\alpha}
e^{\mathcal{L}(t-t_0)}\rho_0},
\end{align*}
and similarly, the last term of Eq.~\eqref{eq:derivative_joint_probability} can be written as,
\begin{align*}
% \label{eq:(iii)}
\text{(iii)}\rightarrow\quad \partial_{i\chi_\alpha} \partial_{i\chi_{\alpha'}'}\Theta(-\tau)(-\partial_{\tau}^2+\partial_\tau\partial_t)P^<(\vec{\chi},t;\vec{\chi}',t+\tau)\big\rvert_{\vec \chi, \vec\chi^\prime=0}=\Theta(-\tau) \Tr{\mathcal{I}_{\alpha} e^{-\mathcal{L}\tau}\mathcal{I}_{\alpha'}e^{\mathcal{L}(t+\tau-t_0)}
\rho_0}.
\end{align*}Substituting expressions (i)-(iii) in Eq. \eqref{eq:derivative_joint_probability}, and
going back to the time variables $s'=t+\tau$ and $s=t$, the two-time current correlation takes the final form given in the main text, which we rewrite here for reference,
\begin{align}
\left\langle I_\alpha(t)I_{\alpha^\prime}(t^\prime)\right\rangle_{\me} &= 
\delta_{\alpha\alpha^\prime}\int_{t_0}^{t}ds\,F\left(t-s \right)F\left(t^{\prime}-s\right)\text{Tr}\{\mathcal A_{\alpha} \, e^{\mathcal L\left(t^\prime-t_0 \right)} \rho_0\} \nonumber \\
& + \int_{t_0}^{t}ds\int_{t_0}^{t^{\prime}}ds^\prime F(t-s)F(t^{\prime}-s^\prime) \Theta\left(s^\prime-s\right)\text{Tr}\{\mathcal I_{\alpha'} e^{\mathcal L \left(s'-s\right)}\mathcal I_{\alpha}e^{\mathcal{L}(s-t_0)}\rho_0\}\nonumber\\
& + \int_{t_0}^{t}ds\int_{t_0}^{t^{\prime}}ds^\prime F(t-s)F(t^{\prime}-s^\prime) \Theta\left(s-s'\right)\text{Tr}\{\mathcal I_\alpha e^{\mathcal L \left(s-s'\right)}\mathcal I_{\alpha'}e^{\mathcal{L}(s'-t_0)}\rho_0\}.
\end{align}

%%%%%%%%%%%%%%%%%%%%%%%%%%%%%%%%%%%%%%%%%%%%%%%%%%%%%%%%%%%%%%%%%

\section{Single-level QD transmission probability for the LB approach}
\label{app:trans}
	
As discussed in Section~\ref{sub_sec_LB}, the Landauer-B\"uttiker approach expresses observables in terms of the energy-dependent transmission function $\mathcal{T}(\epsilon)$ \cite{Landauer1957,Buttiker1986}. In the subsequent discussion, we elucidate the methodology for computing the transmission function of a single-resonant level quantum dot. To achieve this, we model the quantum dot system as a double-well junction~\cite{Stone1985,Buttiker1988,Datta1997}. The barriers to the left and right of the junction are defined by $2\times 2$ scattering matrices $\mathcal{S}_L$ and $\mathcal{S}_R$, which govern the relationship between outgoing and incident stationary states. These scattering matrices are characterized by the reflection amplitudes $r_i$ and $r_i^{\prime}$, as well as the transmission amplitudes $t_i$ and $t_i^{\prime}$ for the left ($i=L$) and right ($i=R$) barriers, respectively. Here, we assume symmetric scattering matrices, i.e., $t_i = t_i^{\prime}$.
By combining the scattering matrices, $\mathcal{S}=\mathcal{S}_L\circ \mathcal{S}_R$~\cite{Datta1997,Blasi2023_a,Blasi2023_b}, we can calculate the transmission function by squaring its off-diagonal entries as follows,
\begin{equation}
\mathcal{T}(\epsilon)=\abs{\mathcal{S}_{LR}}^2=\abs{\mathcal{S}_{RL}}^2=\frac{T_LT_R}{1+R_LR_R-2\sqrt{R_LR_R}\cos{\theta(\epsilon)}}
\end{equation}
Here, $T_i=\abs{t_i}^2=\abs{t_i^{\prime}}^2$ and $R_i=\abs{r_i}^2=\abs{r_i^{\prime}}^2$ represent the transmission and reflection probabilities for the left ($i=L$) and right ($i=R$) barriers, respectively, such that $T_i+R_i=1$ due to unitarity. Furthermore, $\theta(\epsilon)=\arg{r_L^{\prime}}+\arg{r_R}$ denotes the phase shift accumulated during one complete round-trip between the scatterers. Under the assumption that $R_L$ and $R_R$ are approximately equal to 1, the transmission function exhibits pronounced resonances when the denominator approaches zero. This situation occurs when the round-trip phase $\theta(\epsilon_d)$ is a multiple of $2\pi$ at the resonance energy $\epsilon=\epsilon_d$ (representing the energy of the quantum dot). In the vicinity of these resonance values, we can expand the cosine function, obtaining the so-called Breit-Wigner formula for the transmission function~\cite{Buttiker1988,Blanter2000},
\begin{align}
\label{Transmission_function}
\mathcal T\left(\epsilon \right) \approx \frac{T_LT_R}{\left(\frac{T_L+T_R}{2}\right)^2+2\left(1-\cos{\theta(\epsilon)}\right)}\approx \frac{\Gamma_L\Gamma_R}{\frac{\Gamma^2}{4}+(\epsilon-\epsilon_d)^2}.
\end{align}
Here, we used the fact that close to the resonance $2\left(1-\cos{\theta(\epsilon)}\right)\approx \left(d\theta/d\epsilon\right)^2\left(\epsilon-\epsilon_d\right)^2$, and we defined $\Gamma_{L,R}=\frac{d\epsilon}{dk}T_{L,R}$. Physically, $\Gamma_L$ and $\Gamma_R$ (divided by $\hbar$) represent the rates at which an electron placed between the barriers would escape into the left and right leads, respectively. To elucidate this, we note that if we express the round-trip phase shift as $\theta\approx 2kW$, where $W$ is the effective width of the well, then $\frac{d\epsilon}{d\theta}=\frac{\hbar}{\tau}$, where $\tau=2W/v$ represents the time it takes for the electron to travel from one barrier to another and back, with $v=\frac{1}{\hbar}\frac{d\epsilon}{dk}$ being velocity of the particle. 
Substituting in their definition, we obtain,
\begin{equation}
    \Gamma_L=\hbar\frac{T_L}{\tau} \quad \text{and} \quad \Gamma_R=\hbar\frac{T_R}{\tau}.
\end{equation}
A fraction $T_L$ of the attempts on the left barrier is successful, while a fraction $T_R$ of the attempts on the right barrier is successful. Hence, $\Gamma_L/\hbar$ and $\Gamma_R/\hbar$ tell us the number of times per second that an electron succeeds in escaping through the left and right barriers, respectively.

The expression for the transmission function provided in Eq. \eqref{Transmission_function}, fully determines all transport quantities, including currents and current correlation functions, as specified in the last column of Table~\ref{Table1}.

%%%%%%%%%%%%%%%%%%%%%%%%%%%%%%%%%%%%%%%%%%%%%%%%%%

\section{Heisenberg equation for the single-level QD}
\label{appendix_A}
	
	  In this section we provide the details of how to compute Eqs. \eqref{d_op_sol} and \eqref{c_op_sol}. In order to do so, we start by first writing the formal solution of the second line of Eq. \eqref{diff_Eq_d_c} as,
	\begin{equation}
		\hat{c}_{k \alpha}(t)=e^{-i \epsilon_{k \alpha} (t-t_0)} \hat{c}_{k\alpha}(t_0)-i \int_{t_0}^{t} ds~ e^{-i \epsilon_{k \alpha}(t-s)} t_{k \alpha}^* \hat{d}(s).
	\end{equation}
 % \begin{equation}
	% 	\hat{c}_{k \alpha}^{\dagger}(t)=e^{i \epsilon_{k \alpha} (t-t_0)} \hat{c}_{k\alpha}^{\dagger}(t_0)+i \int_{t_0}^{t} ds~ e^{i \epsilon_{k \alpha}(t-s)} t_{k \alpha} \hat{d}^{\dagger}(s).
	% \end{equation}
	By substituting the above expression into the first line of of Eq. \eqref{diff_Eq_d_c} we obtain,
	\begin{align}
		\label{d_intermediate}
		\frac{d}{d t} \hat{d} &=-i\epsilon_d \hat{d} -i \sum_{k, \alpha} t_{k \alpha} e^{-i \epsilon_{k \alpha} (t-t_0)} \hat{c}_{k\alpha}(t_0)-\int_{t_0}^{t} ds~\sum_{k, \alpha} \abs{t_{k \alpha}}^2e^{-i \epsilon_{k \alpha} (t-s)}\hat{d}(s).
	\end{align}
 % \begin{align}
	% 	\label{d_intermediate}
	% 	\frac{d}{d t} \hat{d}^{\dagger} &=i\epsilon_d \hat{d}^{\dagger} +i \sum_{k, \alpha} t_{k \alpha}^{*} e^{i \epsilon_{k \alpha} (t-t_0)} \hat{c}_{k\alpha}^{\dagger}(t_0)-\int_{t_0}^{t} ds~\sum_{k, \alpha} \abs{t_{k \alpha}}^2e^{i \epsilon_{k \alpha} (t-s)}\hat{d}^{\dagger}(s).
	% \end{align}
	The last term in the above expression can be written in the following manner,
	\begin{align}
		\int_{t_0}^{t} ds~\sum_{k, \alpha} \abs{t_{k \alpha}}^2e^{-i \epsilon_{k \alpha} (t-s)}\hat{d}(s)&=\int_{t_0}^{t} ds~\int \frac{d\epsilon}{2\pi}~\overbrace{2\pi\sum_{k, \alpha}\delta(\epsilon-\epsilon_{k\alpha}) \abs{t_{k \alpha}}^2}^{\Gamma}e^{-i \epsilon (t-s)}\hat{d}(s)\nonumber\\
		&=\Gamma\int_{t_0}^{t} ds~\overbrace{\int \frac{d\epsilon}{2\pi}~e^{-i \epsilon (t-s)}}^{\delta(t-s)}\hat{d}(s)=\frac{\Gamma}{2}\hat{d}(t)
	\end{align}
 % \begin{align}
	% 	\int_{t_0}^{t} ds~\sum_{k, \alpha} \abs{t_{k \alpha}}^2e^{i \epsilon_{k \alpha} (t-s)}\hat{d}^{\dagger}(s)&=\int_{t_0}^{t} ds~\int \frac{d\epsilon}{2\pi}~\overbrace{2\pi\sum_{k, \alpha}\delta(\epsilon-\epsilon_{k\alpha}) \abs{t_{k \alpha}}^2}^{\Gamma}e^{i \epsilon (t-s)}\hat{d}^{\dagger}(s)\nonumber\\
	% 	&=\Gamma\int_{t_0}^{t} ds~\overbrace{\int \frac{d\epsilon}{2\pi}~e^{i \epsilon (t-s)}}^{\delta(t-s)}\hat{d}^{\dagger}(s)=\frac{\Gamma}{2}\hat{d}^{\dagger}(t)
	% \end{align}
where we used the property, $\int_{t_0}^{t} ds~\delta(t-s)\hat{d}(s)=\hat{d}(t)/2$, and the definition of the bare tunneling rate function of Eq. \eqref{eq:tunnel_rate}. It is important to recall that in our case, since we are considering the wide-band approximation, $\Gamma$ is a constant and can thus be taken out of the integral. Then, by using the definition of $\hat{\xi}(t)$ given in Eq. \eqref{xi}, Eq. \eqref{d_intermediate} takes the following form,
	\begin{equation}
		\frac{d}{d t} \hat{d}= \hat{\xi}(t)+\left(-i\epsilon_d-\frac{\Gamma}{2}\right)\hat{d}(t),
	\end{equation}
 % \begin{equation}
	% 	\frac{d}{d t} \hat{d}^{\dagger} = \hat{\xi}^{\dagger}(t)+\left(\epsilon_d-\frac{\Gamma}{2}\right)\hat{d}^{\dagger}(t),
	% \end{equation}
	which can be integrated, leading to the final expression given in Eq. \eqref{d_op_sol}.
A similar expression was used in Ref.~\cite{Rodriguez2023}.

%%%%%%%%%%%%%%%%%%%%%%%%%%%%%%%%%%%%%%%%%%%%%%%%%

\section{Single-level QD exact finite-time correlation functions}
\label{app_Noise}
Using Eq. \eqref{eq:def_HE_current_correlation}, the expression of the current correlation function takes the following form,
\begin{equation}
\label{Noise_1}
S^\he_{\alpha\alpha^{\prime}}(t,t^{\prime})=\expval{\hat I_{\alpha}(t)\hat I_{\alpha^{\prime}}(t^{\prime})}-\expval{\hat I_{\alpha}(t)}\expval{\hat I_{\alpha^{\prime}}(t^{\prime})}.
\end{equation}
Furthermore, by inserting the Hamiltonian \eqref{eq:totalH} into the current operator for a single quantum dot \eqref{def_current}, we obtain
\begin{align}
    \hat I_\alpha(t) &= - \frac{d}{dt} \hat{N}_\alpha  = - i \left[ \hat H, \hat{N}_\alpha \right]=i\sum_{k}\left[t_{k\alpha}^{*}\hat{c}_{k\alpha}^{\dagger}(t)\hat{d}(t)-t_{k\alpha} \hat{d}^{\dagger}(t)\hat{c}_{k\alpha}(t)\right].
\end{align}
Using the above expression, the first term of Eq.~\eqref{Noise_1} becomes,
	\begin{align}
		\expval{\hat I_{\alpha}(t)\hat I_{\alpha^{\prime}}(t^{\prime})}&=\expval{\left[i\sum_{k}\left(t_{k\alpha}^{*}\hat{c}_{k\alpha}^{\dagger}(t)\hat{d}(t)-t_{k\alpha} \hat{d}^{\dagger}(t)\hat{c}_{k\alpha}(t)\right)\right]\left[i\sum_{k^{\prime}}\left(t_{k^{\prime}\alpha^{\prime}}^{*}\hat{c}_{k^{\prime}\alpha^{\prime}}^{\dagger}(t^{\prime})\hat{d}(t^{\prime})-t_{k^{\prime}\alpha^{\prime}} \hat{d}^{\dagger}(t^{\prime})\hat{c}_{k^{\prime}\alpha^{\prime}}(t^{\prime})\right)\right]}\nonumber\\
		&=-\sum_{k k^{\prime}}\left\{t_{k\alpha}^{*}t_{k^{\prime}\alpha^{\prime}}^{*}\expval{\hat{c}_{k\alpha}^{\dagger}(t)\hat{d}(t)\hat{c}_{k^{\prime}\alpha^{\prime}}^{\dagger}(t^{\prime})\hat{d}(t^{\prime})}-t_{k\alpha}^{*}t_{k^{\prime}\alpha^{\prime}}\expval{\hat{c}_{k\alpha}^{\dagger}(t)\hat{d}(t) \hat{d}^{\dagger}(t^{\prime})\hat{c}_{k^{\prime}\alpha^{\prime}}(t^{\prime})}\right.\nonumber\\
		&\left.~~~-t_{k\alpha}t_{k^{\prime}\alpha^{\prime}}^{*}\expval{ \hat{d}^{\dagger}(t)\hat{c}_{k\alpha}(t)\hat{c}_{k^{\prime}\alpha^{\prime}}^{\dagger}(t^{\prime})\hat{d}(t^{\prime})}+t_{k\alpha}t_{k^{\prime}\alpha^{\prime}}\expval{ \hat{d}^{\dagger}(t)\hat{c}_{k\alpha}(t) \hat{d}^{\dagger}(t^{\prime})\hat{c}_{k^{\prime}\alpha^{\prime}}(t^{\prime})}\right\}.
	\end{align}
Now, using Wick's theorem, we can decompose the four-operator expectation values into the sum of products of two-operator expectation values by performing contractions,
	\begin{align}
		\expval{\hat{c}_{k\alpha}^{\dagger}(t)\hat{d}(t)\hat{c}_{k^{\prime}\alpha^{\prime}}^{\dagger}(t^{\prime})\hat{d}(t^{\prime})}&=\expval{\hat{c}_{k\alpha}^{\dagger}(t)\hat{d}(t)}\expval{\hat{c}_{k^{\prime}\alpha^{\prime}}^{\dagger}(t^{\prime})\hat{d}(t^{\prime})}+\expval{\hat{c}_{k\alpha}^{\dagger}(t)\hat{d}(t^{\prime})}\expval{\hat{d}(t)\hat{c}_{k^{\prime}\alpha^{\prime}}^{\dagger}(t^{\prime})}\nonumber\\
		\expval{\hat{c}_{k\alpha}^{\dagger}(t)\hat{d}(t) \hat{d}^{\dagger}(t^{\prime})\hat{c}_{k^{\prime}\alpha^{\prime}}(t^{\prime})}&=\expval{\hat{c}_{k\alpha}^{\dagger}(t)\hat{d}(t)}\expval{ \hat{d}^{\dagger}(t^{\prime}) \hat{c}_{k^{\prime}\alpha^{\prime}}(t^{\prime})}+\expval{\hat{c}_{k\alpha}^{\dagger}(t)\hat{c}_{k^{\prime}\alpha^{\prime}}(t^{\prime})}\expval{\hat{d}(t) \hat{d}^{\dagger}(t^{\prime})}\nonumber\\
		\expval{ \hat{d}^{\dagger}(t)\hat{c}_{k\alpha}(t)\hat{c}_{k^{\prime}\alpha^{\prime}}^{\dagger}(t^{\prime})\hat{d}(t^{\prime})}&=\expval{ \hat{d}^{\dagger}(t) \hat{c}_{k\alpha}(t)}\expval{\hat{c}_{k^{\prime}\alpha^{\prime}}^{\dagger}(t^{\prime})\hat{d}(t^{\prime})}+\expval{ \hat{d}^{\dagger}(t)\hat{d}(t^{\prime})}\expval{\hat{c}_{k\alpha}(t)\hat{c}_{k^{\prime}\alpha^{\prime}}^{\dagger}(t^{\prime})}\nonumber\\
		\expval{ \hat{d}^{\dagger}(t)\hat{c}_{k\alpha}(t) \hat{d}^{\dagger}(t^{\prime})\hat{c}_{k^{\prime}\alpha^{\prime}}(t^{\prime})}&=\expval{ \hat{d}^{\dagger}(t)\hat{c}_{k\alpha}(t)}\expval{ \hat{d}^{\dagger}(t^{\prime})\hat{c}_{k^{\prime}\alpha^{\prime}}(t^{\prime})}+\expval{ \hat{d}^{\dagger}(t)\hat{c}_{k^{\prime}\alpha^{\prime}}(t^{\prime})}\expval{\hat{c}_{k\alpha}(t) \hat{d}^{\dagger}(t^{\prime})}.
	\end{align}
	The calculation yields
	\begin{align}
		\expval{I_{\alpha}(t)I_{\alpha^{\prime}}(t^{\prime})}=-\sum_{k k^{\prime}}&\left\{t_{k\alpha}^{*}t_{k^{\prime}\alpha^{\prime}}^{*}\left[\expval{\hat{c}_{k\alpha}^{\dagger}(t)\hat{d}(t)}\expval{\hat{c}_{k^{\prime}\alpha^{\prime}}^{\dagger}(t^{\prime})\hat{d}(t^{\prime})}+\expval{\hat{c}_{k\alpha}^{\dagger}(t)\hat{d}(t^{\prime})}\expval{\hat{d}(t)\hat{c}_{k^{\prime}\alpha^{\prime}}^{\dagger}(t^{\prime})}\right]\right.\nonumber\\
		-&\left.t_{k\alpha}^{*}t_{k^{\prime}\alpha^{\prime}}\left[\expval{\hat{c}_{k\alpha}^{\dagger}(t)\hat{d}(t)}\expval{ \hat{d}^{\dagger}(t^{\prime}) \hat{c}_{k^{\prime}\alpha^{\prime}}(t^{\prime})}+\expval{\hat{c}_{k\alpha}^{\dagger}(t)\hat{c}_{k^{\prime}\alpha^{\prime}}(t^{\prime})}\expval{\hat{d}(t) \hat{d}^{\dagger}(t^{\prime})}\right]\right.\nonumber\\
		-&\left.t_{k\alpha}t_{k^{\prime}\alpha^{\prime}}^{*}\left[\expval{ \hat{d}^{\dagger}(t) \hat{c}_{k\alpha}(t)}\expval{\hat{c}_{k^{\prime}\alpha^{\prime}}^{\dagger}(t^{\prime})\hat{d}(t^{\prime})}+\expval{ \hat{d}^{\dagger}(t)\hat{d}(t^{\prime})}\expval{\hat{c}_{k\alpha}(t)\hat{c}_{k^{\prime}\alpha^{\prime}}^{\dagger}(t^{\prime})}\right]\right.\nonumber\\
		+&\left.t_{k\alpha}t_{k^{\prime}\alpha^{\prime}}\left[\expval{ \hat{d}^{\dagger}(t)\hat{c}_{k\alpha}(t)}\expval{ \hat{d}^{\dagger}(t^{\prime})\hat{c}_{k^{\prime}\alpha^{\prime}}(t^{\prime})}+\expval{ \hat{d}^{\dagger}(t)\hat{c}_{k^{\prime}\alpha^{\prime}}(t^{\prime})}\expval{\hat{c}_{k\alpha}(t) \hat{d}^{\dagger}(t^{\prime})}\right]\right\}.
	\end{align}
Putting together the first terms of all lines of the above expression, we find the product of the current expectation values. This cancels the corresponding term in the expression~\eqref{Noise_1}. Finally, the current correlation function takes the form given in Eq. \eqref{eq:noiseHE}, which we rewrite for reference,
 \begin{align}
S^\he_{\alpha\alpha^{\prime}}(t,t^{\prime})=-\sum_{k k^{\prime}}&\left\{t_{k\alpha}^{*}t_{k^{\prime}\alpha^{\prime}}^{*}\expval{\hat{c}_{k\alpha}^{\dagger}(t)\hat{d}(t^{\prime})}\expval{\hat{d}(t)\hat{c}_{k^{\prime}\alpha^{\prime}}^{\dagger}(t^{\prime})}-t_{k\alpha}^{*}t_{k^{\prime}\alpha^{\prime}}\expval{\hat{c}_{k\alpha}^{\dagger}(t)\hat{c}_{k^{\prime}\alpha^{\prime}}(t^{\prime})}\expval{\hat{d}(t) \hat{d}^{\dagger}(t^{\prime})}\right.\nonumber\\
&-\left.t_{k\alpha}t_{k^{\prime}\alpha^{\prime}}^{*}\expval{ \hat{d}^{\dagger}(t)\hat{d}(t^{\prime})}\expval{\hat{c}_{k\alpha}(t)\hat{c}_{k^{\prime}\alpha^{\prime}}^{\dagger}(t^{\prime})}+t_{k\alpha}t_{k^{\prime}\alpha^{\prime}}\expval{ \hat{d}^{\dagger}(t)\hat{c}_{k^{\prime}\alpha^{\prime}}(t^{\prime})}\expval{\hat{c}_{k\alpha}(t) \hat{d}^{\dagger}(t^{\prime})}\right\}.
\end{align}
It is convenient to rewrite the previous expression as follows,
\begin{align}
\label{S_CD_DC_CC_DD}
S^\he_{\alpha\alpha^{\prime}}(t,t^{\prime})=&-CD_{\alpha}(t,t^{\prime})\overline{DC}_{\alpha^{\prime}}(t,t^{\prime})+CC_{\alpha\alpha^{\prime}}(t,t^{\prime})\overline{DD}(t,t^{\prime})+\overline{CC}_{\alpha\alpha^{\prime}}(t,t^{\prime})DD(t,t^{\prime})-\overline{CD}_{\alpha}(t,t^{\prime})DC_{\alpha^{\prime}}.(t,t^{\prime})
\end{align}
where we have defined,
\begin{align}
\label{CD_DC_CC_DD}
CD_{\alpha}(t,t^{\prime})&=\sum_{k}t_{k\alpha}^{*}\expval{\hat{c}_{k\alpha}^{\dagger}(t)\hat{d}(t^{\prime})}& \overline{CD}_{\alpha}(t,t^{\prime})&=\sum_{k}t_{k\alpha}\expval{\hat{c}_{k\alpha}(t) \hat{d}^{\dagger}(t^{\prime})}\nonumber\\
 DC_{\alpha^{\prime}}(t,t^{\prime})&=\sum_{k^{\prime}}t_{k^{\prime}\alpha^{\prime}}\expval{ \hat{d}^{\dagger}(t)\hat{c}_{k^{\prime}\alpha^{\prime}}(t^{\prime})}&\overline{DC}_{\alpha^{\prime}}(t,t^{\prime})&=\sum_{k^{\prime}}t_{k^{\prime}\alpha^{\prime}}^{*}\expval{\hat{d}(t)\hat{c}_{k^{\prime}\alpha^{\prime}}^{\dagger}(t^{\prime})}\nonumber\\
CC_{\alpha\alpha^{\prime}}(t,t^{\prime})&=\sum_{k k^{\prime}}t_{k\alpha}^{*}t_{k^{\prime}\alpha^{\prime}}\expval{\hat{c}_{k\alpha}^{\dagger}(t)\hat{c}_{k^{\prime}\alpha^{\prime}}(t^{\prime})}& \overline{CC}_{\alpha\alpha^{\prime}}(t,t^{\prime})&=\sum_{k k^{\prime}}t_{k\alpha}t^{*}_{k^{\prime}\alpha^{\prime}}\expval{\hat{c}_{k\alpha}(t)c^{\dagger}_{k^{\prime}\alpha^{\prime}}(t^{\prime})}\nonumber\\
DD(t,t^{\prime})&=\expval{ \hat{d}^{\dagger}(t)\hat{d}(t^{\prime})}&\overline{DD}(t,t^{\prime})&=\expval{\hat{d}(t) \hat{d}^{\dagger}(t^{\prime})}.
\end{align}
We note that the above quantities can also be written in terms of the standard lesser and greater Green's functions,
% \begin{align}
% \label{Green_functions}
% G^{<}_{k\alpha, d}(t,t^{\prime})&=i\expval{\hat{c}_{k\alpha}^{\dagger}(t)\hat{d}(t^{\prime})}& G^{>}_{k\alpha, d}(t,t^{\prime})&=-i\expval{\hat{c}_{k\alpha}(t) \hat{d}^{\dagger}(t^{\prime})}\nonumber\\
%  G^{<}_{d,k^{\prime}\alpha^{\prime}}(t,t^{\prime})&=i\expval{ \hat{d}^{\dagger}(t)\hat{c}_{k^{\prime}\alpha^{\prime}}(t^{\prime})}&G^{>}_{d,\alpha^{\prime}k^{\prime}}(t,t^{\prime})&=-i\expval{\hat{d}(t)\hat{c}_{k^{\prime}\alpha^{\prime}}^{\dagger}(t^{\prime})}\nonumber\\
% G^{<}_{k\alpha,k^{\prime}\alpha^{\prime}}(t,t^{\prime})&=i\expval{\hat{c}_{k\alpha}^{\dagger}(t)\hat{c}_{k^{\prime}\alpha^{\prime}}(t^{\prime})}& G^{>}_{k\alpha,k^{\prime}\alpha^{\prime}}(t,t^{\prime})&=-i\expval{\hat{c}_{k\alpha}(t)c^{\dagger}_{k^{\prime}\alpha^{\prime}}(t^{\prime})}\nonumber\\
% G^{<}_{d,d}(t,t^{\prime})&=i\expval{ \hat{d}^{\dagger}(t)\hat{d}(t^{\prime})}&G^{>}_{d,d}(t,t^{\prime})&=-i\expval{\hat{d}(t) \hat{d}^{\dagger}(t^{\prime})}.
% \end{align}
\begin{align}
\label{Green_functions}
G^{<}_{d,k\alpha}(t^{\prime},t)&=i\expval{\hat{c}_{k\alpha}^{\dagger}(t)\hat{d}(t^{\prime})}& G^{>}_{k\alpha, d}(t,t^{\prime})&=-i\expval{\hat{c}_{k\alpha}(t) \hat{d}^{\dagger}(t^{\prime})}\nonumber\\
 G^{<}_{k^{\prime}\alpha^{\prime},d}(t^{\prime},t)&=i\expval{ \hat{d}^{\dagger}(t)\hat{c}_{k^{\prime}\alpha^{\prime}}(t^{\prime})}&G^{>}_{d,\alpha^{\prime}k^{\prime}}(t,t^{\prime})&=-i\expval{\hat{d}(t)\hat{c}_{k^{\prime}\alpha^{\prime}}^{\dagger}(t^{\prime})}\nonumber\\
G^{<}_{k^{\prime}\alpha^{\prime},k\alpha}(t^{\prime},t)&=i\expval{\hat{c}_{k\alpha}^{\dagger}(t)\hat{c}_{k^{\prime}\alpha^{\prime}}(t^{\prime})}& G^{>}_{k\alpha,k^{\prime}\alpha^{\prime}}(t,t^{\prime})&=-i\expval{\hat{c}_{k\alpha}(t)c^{\dagger}_{k^{\prime}\alpha^{\prime}}(t^{\prime})}\nonumber\\
G^{<}_{d,d}(t^{\prime},t)&=i\expval{ \hat{d}^{\dagger}(t)\hat{d}(t^{\prime})}&G^{>}_{d,d}(t,t^{\prime})&=-i\expval{\hat{d}(t) \hat{d}^{\dagger}(t^{\prime})}.
\end{align}

After substituting Eqs. \eqref{d_op_sol} and \eqref{c_op_sol} in Eqs. \eqref{CD_DC_CC_DD}, a long calculation yields,
\begin{align}
\label{CD}
CD_{\alpha}(t,t^{\prime}) &= i\sum_{\beta}\frac{\Gamma_{\alpha}\Gamma_{\beta}}{\Gamma}\left[\Lambda_0(t,t^{\prime})-\Lambda^{(1)}_{\alpha}(t,t^{\prime})+\Lambda^{(2)}_{\beta}(t,t^{\prime})\right]
\end{align}
\begin{align}
\label{DC}
DC_{\alpha^{\prime}}(t,t^{\prime}) &= -i\sum_{\beta}\frac{\Gamma_{\alpha^{\prime}}\Gamma_{\beta}}{\Gamma}\left[\Lambda_0(t^{\prime},t)-\Lambda^{(1)}_{\alpha^{\prime}}(t^{\prime},t)+\Lambda^{(2)}_{\beta}(t^{\prime},t)\right]^*
\end{align}
\begin{align}
\label{CC}
CC_{\alpha\alpha^{\prime}}(t,t^{\prime}) &=\frac{\Gamma_{\alpha}\Gamma}{2}\Lambda_{\alpha}(t,t')\delta_{\alpha\alpha^{\prime}}+\sum_{\beta}\frac{\Gamma_{\alpha}\Gamma_{\alpha^{\prime}}\Gamma_{\beta}}{2\Gamma}\left[\Lambda_0(t,t^{\prime})-\Lambda^{(1)}_{\alpha}(t,t^{\prime})-\Lambda^{(1)}_{\alpha^{\prime}}(t^{\prime},t)^*+\Lambda^{(2)}_{\beta}(t,t^{\prime})\right]
\end{align}
\begin{align}
\label{DD}
DD(t,t^{\prime}) &= 2\sum_{\beta}\frac{\Gamma_{\beta}}{\Gamma}\left[\Lambda_0(t^{\prime},t)+\Lambda^{(2)}_{\beta}(t^{\prime},t)\right]^*.
\end{align}
where the $\Lambda$-functions are defined in Eqs. \eqref{Lambda_HE}.
The barred quantities in Eqs. \eqref{CD_DC_CC_DD}, can be obtained from the above expressions by taking the complex conjugate and changing $\Lambda \leftrightarrow\overline{\Lambda}$; for example,
\begin{equation}
\overline{CD}_{\alpha}(t,t^{\prime})=-i\sum_{\beta}\frac{\Gamma_{\alpha}\Gamma_{\beta}}{\Gamma}\left[\overline{\Lambda}_0(t,t^{\prime})-\overline{\Lambda}^{(1)}_{\alpha}(t,t^{\prime})+\overline{\Lambda}^{(2)}_{\beta}(t,t^{\prime})\right]^*.
\end{equation}

\section{Contact energy current}
\label{appendix:contact_current}

The contact energy current between the system and the reservoir $\alpha$, $J_{c\alpha}^{\he}(t)=-\frac{d}{dt} \langle \hat{H}_{SR\alpha} \rangle=-i\left\langle [\hat{H},\hat  H_{SR_\alpha}] \right\rangle$, can be expressed using the $CD_{\alpha}(t,t)$ term introduced in Eq.~\eqref{CD_DC_CC_DD} in the following manner,
\begin{equation}
\begin{aligned}
\label{eq:contactcorr}
    J_{c\alpha}^{\he}(t) &= -2\text{Im}\left\{i\sum_{k\alpha}t^*_{k\alpha}\partial_t \left\langle \hat c^\dagger_{k\alpha}\left(t\right)d\left( t\right) \right\rangle\right\} =- 2\text{Im}\left\{i \partial_t CD_\alpha\left(t,t\right)\ \right\}\\
    &=2 \text{Im}\left\{\partial_t \sum_\beta \frac{\Gamma_\alpha\Gamma_\beta}{\Gamma}\left[\Lambda_0\left(t,t \right) -\Lambda_\alpha^{(1)}\left(t,t \right)+\Lambda_\alpha^{(2)}\left(t,t \right)\right] \right\}.
\end{aligned}
\end{equation}It is simple to see that the above expression does not contribute in the Landauer-Büttiker (long-time) or master equation (weak-coupling) limits. 

\begin{enumerate}
\item{In the long-time limit $t\to\infty$, we have from Eqs. \eqref{Lambda_long_time_1} that 
\begin{equation}
  \partial_t \left[\lim_{t\to \infty} \Lambda_0\left(t,t \right)\right]=\partial_t \left[\lim_{t\to \infty} \Lambda_\alpha^{(1)}\left(t,t \right)\right]=\partial_t \left[\lim_{t\to \infty} \Lambda_\alpha^{(2)}\left(t,t \right)\right] = 0  
\end{equation}
 This means the expression \eqref{eq:contactcorr} vanishes in the steady-state regime.}  
  
%   \item{
% In our weak-coupling procedure, we have to first take the limit $\Gamma\to0$ (with $\Gamma t$ set constant), using Eq. \eqref{Lambda_ME}, we have the following, 
%   \begin{equation}
%   \begin{aligned}
%  &\partial_t\left[\lim_{\Gamma\to0} \Lambda_0\left(t,t \right)\right] = \frac{1}{2}\partial_t e^{-\Gamma t}e^{\Gamma t_0}n_d = -\frac{\Gamma}{2}e^{-\Gamma\left(t-t_0 \right)}n_d\\
%  &\partial_t\left[\lim_{\Gamma\to0} \Lambda_\alpha^-\left(t,t \right)\right] =\partial_t\left\{\Theta\left(0 \right)-\Theta\left(t_0-t \right)e^{-\Gamma\left(t-t_0 \right)} \right\}f_\alpha\left(\epsilon_d \right) = \Gamma\Theta\left(t_0-t \right)e^{-\Gamma\left( t-t_0\right)}f_\alpha\left(\epsilon_d \right)\\
%  &\partial_t\left[\lim_{\Gamma\to0} \Lambda_\alpha^\pm\left(t,t \right)\right] =\frac{1}{2}\partial_t\left\{1-e^{-\Gamma\left(t-t_0 \right)} \right\}f_\alpha\left(\epsilon_d \right) = -\frac{\Gamma}{2}e^{-\Gamma\left(t-t_0 \right)}f_\alpha\left( \epsilon_d\right). 
%   \end{aligned}
%   \end{equation}
% To obtain the first-order coupling term of the current, we have to multiply the above expression by $\Gamma$. However, the above expression are already $\left(\Gamma\right)$ due to the time-derivative involved in the calculation. This means that the overall contact correction to the current is $J_{c\alpha}\left( t\right) = \mathcal O\left(\Gamma^2 \right)$, and can therefore be safely excluded from the total current.
% }

\item{
In the weak-coupling regime, we have first to consider the limit $\Gamma \to 0$ (with $\Gamma t$ held constant). Utilizing the set of Eqs. \eqref{Lambda_ME}, we obtain the following expressions:
\begin{equation}
\begin{aligned}
&\partial_t\left[\lim_{\Gamma\to0} \Lambda_0\left(t,t \right)\right] = \frac{1}{2}\partial_t e^{-\Gamma t}e^{\Gamma t_0}n_d = -\frac{\Gamma}{2}e^{-\Gamma\left(t-t_0 \right)}n_d\\
&\partial_t\left[\lim_{\Gamma\to0} \Lambda_\alpha^{(1)}\left(t,t \right)\right] =\partial_t\left\{\Theta\left(0 \right)-\Theta\left(t_0-t \right)e^{-\Gamma\left(t-t_0 \right)} \right\}f_\alpha\left(\epsilon_d \right) = \Gamma\Theta\left(t_0-t \right)e^{-\Gamma\left( t-t_0\right)}f_\alpha\left(\epsilon_d \right)\\
&\partial_t\left[\lim_{\Gamma\to0} \Lambda_\alpha^{(2)}\left(t,t \right)\right] =\frac{1}{2}\partial_t\left\{1-e^{-\Gamma\left(t-t_0 \right)} \right\}f_\alpha\left(\epsilon_d \right) = -\frac{\Gamma}{2}e^{-\Gamma\left(t-t_0 \right)}f_\alpha\left( \epsilon_d\right). 
\end{aligned}
\end{equation}
According to our weak coupling protocol outlined in Sec. \ref{sec:weak_coupling_protocol}, we substitute the above expressions into Eq. \eqref{eq:contactcorr} and multiply by $\Gamma$ to obtain the contribution at the lowest order in the couplings. This yields the overall contact energy current in the weak-coupling regime as $J_{c\alpha}\left( t\right) \sim \mathcal O\left(\Gamma^2 \right)$, allowing us to safely exclude it from the total current.
}
\end{enumerate}

\section{DQD exact finite-time dots populations}
\label{app:DQD}

In this Appendix, we detail how to compute the analytical expressions for the dots' population in both the \textit{Parallel} and the \textit{Series} configuration. In order to find the time-dependent  populations, we have to explicitly solve Eq.~\eqref{exp_val_populations} of the main text with $n=n'=1,2$, which we recall here for clarity, 
\begin{equation}
\label{exp_val_populations_appendix}
\expval{\hat{d}^{\dagger}_n(t)\hat{d}_{n}(t)}_{\sigma}=\sum_{m m'} D^{\sigma}_{nm}(t)^*D^\sigma_{m'n}(t)\expval{\hat{d}^{\dagger}_m(t_0)\hat{d}_{m'}(t_0)}+\sum_{m m'}\int_{t_0}^t ds~ds' D^{\sigma}_{nm}(t-s)^*D^\sigma_{m'n}(t-s')\expval{\hat{\xi}_m^{\dagger}(s)\hat{\xi}_{m'}(s')}.
\end{equation}
In order to do so, we first assume initial independent dots with no coherence, i.e., 
\begin{equation}
    \expval{\hat{d}^{\dagger}_m(t_0)\hat{d}_{m'}(t_0)}=n_m \delta_{m m'}
\end{equation}
and initial mean occupation of the $\alpha$ reservoir to be set by the Fermi-Dirac distribution,
\begin{equation}
    \expval{\hat{c}^{\dagger}_{k\alpha}(t_0)\hat{c}_{k'\alpha'}(t_0)}=\delta_{k\alpha,k'\alpha'}f_{\alpha}(\epsilon_{k\alpha}).
\end{equation}
Then, by computing the following term,
\begin{equation}
    \expval{\hat{\xi}_m^{\dagger}(s)\hat{\xi}_{m'}(s')}=\sum_{k k' \alpha\alpha'}t_{m k\alpha}^*t_{m' k'\alpha'}e^{i \epsilon_{k\alpha}(s-t_0)}e^{-i \epsilon_{k'\alpha'}(s'-t_0)}\expval{\hat{c}^{\dagger}_{k\alpha}(t_0)\hat{c}_{k'\alpha'}(t_0)}= \sum_{k \alpha}t_{m k\alpha}^*t_{m' k\alpha}e^{i \epsilon_{k\alpha}(s-s')}f_{\alpha}(\epsilon_{k\alpha}),
\end{equation}
and using the WBL approximation with $\Gamma_{ij}^{\alpha} = 2\pi \sum_{k}  t_{i k\alpha}^*t_{j k\alpha}  \delta\left(\epsilon - \epsilon_{k\alpha}\right)$, we can rewrite Eq.~\eqref{exp_val_populations} as follows:

\begin{equation}
\label{exp_val_populations_appendix_2}
\expval{\hat{d}^{\dagger}_n(t)\hat{d}_{n}(t)}_{\sigma}=\sum_{m} D^{\sigma}_{nm}(t)^*D^\sigma_{m n}(t)n_m+\sum_{m m'}\sum_{\alpha}\Gamma_{m m'}^{\alpha}\int \frac{d\epsilon}{2\pi} \tilde{D}^{\sigma}_{nm}(\epsilon)^*\tilde{D}^\sigma_{m'n}(\epsilon)f_{\alpha}(\epsilon),
\end{equation}
where 
\begin{equation}
\label{eq:Dtilde}
\tilde{D}^\sigma_{n n'}(\epsilon)=\int ds \left[\Theta\left(s+\frac{t-t_0}{2}\right)-\Theta\left(s-\frac{t-t_0}{2}\right)\right]D^\sigma_{n n'}\left(\frac{t-t_0}{2}-s\right)e^{-i\epsilon s}.
\end{equation}
In order to solve Eq.~\eqref{exp_val_populations_appendix_2}, it is convenient to write the matrix elements $D^\sigma_{m n}(t)$ using its eigen-decomposition form
\begin{equation}
\label{eq:eigendecomposition}
    D^\sigma_{n n'}(t)=\sum_{l}S^{\sigma}_{n l}~e^{\lambda_l^{\sigma}t}~S^{\sigma~-1}_{l n'},
\end{equation}
where $S^\sigma$ is the matrix whose columns are the eigenvectors of the matrix $A^\sigma$ introduced in the main text in Eqs.~\eqref{eq:matrix_A_Parallel} and \eqref{eq:matrix_A_Series}, and $\lambda^\sigma_\mp$ are the eigenvalues of $A^\sigma$.
Substituting Eq.~\eqref{eq:eigendecomposition} in Eqs.~\eqref{eq:Dtilde} we get
 \begin{equation}
\tilde{D}^\sigma_{n n'}(\epsilon)=2\sum_{l} S^{\sigma}_{n l}S^{\sigma~-1}_{l n'}\frac{e^{\lambda_{l}^{\sigma}\frac{t-t_0}{2}}\sinh{\left[\frac{t-t_0}{2}\left(\lambda_{l}^{\sigma}+i\epsilon\right)\right]}}{\lambda_{l}^{\sigma}+i\epsilon},
\end{equation}
which, in turn, allows us to rewrite Eq.~\eqref{exp_val_populations_appendix_2} in the following way,
% \begin{equation}
% \label{exp_val_populations_appendix_3}
% \expval{\hat{d}^{\dagger}_n(t)\hat{d}_{n}(t)}_{\sigma}=\sum_{m m'}\sum_{l l'} \left(S^{\sigma}_{n l}\right)^*\left(S^{\sigma~-1}_{l m}\right)^*S^{\sigma}_{m' l'}S^{\sigma~-1}_{l' n}\left[e^{\lambda_{l}^{\sigma*}t}e^{\lambda_{l'}^{\sigma}t}\gb{n_m}\delta_{m m'}+4\sum_{\alpha}\frac{\Gamma_{m m'}^{\alpha}}{\Gamma}\left[M_\alpha^\sigma\right]_{m m'}\right].
% \end{equation}
\begin{equation}
\label{exp_val_populations_appendix_3}
\expval{\hat{d}^{\dagger}_n(t)\hat{d}_{n}(t)}_{\sigma}=\sum_{m m'}\sum_{l l'} \left(S^{\sigma}_{n m}\right)^*\left(S^{\sigma~-1}_{m l}\right)^*S^{\sigma}_{l' m'}S^{\sigma~-1}_{m' n}\left[e^{\lambda_{m}^{\sigma*}t}e^{\lambda_{m'}^{\sigma}t}n_l\delta_{l l'}+4\sum_{\alpha}\frac{\Gamma_{l l'}^{\alpha}}{\Gamma}\left[M_\alpha^\sigma\right]_{m m'}\right].
\end{equation}
Here the matrix $M_\alpha^\sigma$ is the same as the one defined in the main text in Eq.~\eqref{eq:M_general_main} and that we report here for clarity,
\begin{equation}
\label{eq:M_general_appendix}
  \left[M^{\sigma}_{\alpha}(t)\right]_{m m'} = \Gamma \int \frac{d\epsilon}{2\pi} \frac{e^{\lambda_m^{\sigma*}\frac{t-t_0}{2}}\sinh{\left[\frac{t-t_0}{2}\left(\lambda_m^{\sigma*}-i\epsilon\right)\right]}}{\lambda_m^{\sigma*}-i\epsilon}\frac{e^{\lambda_{m'}^{\sigma}\frac{t-t_0}{2}}\sinh{\left[\frac{t-t_0}{2}\left(\lambda_{m'}^{\sigma}+i\epsilon\right)\right]}}{\lambda_{m'}^{\sigma}+i\epsilon}f_{\alpha}(\epsilon).
\end{equation}
Finally, in order to compute the populations of the dot given in Eqs.~\eqref{eq:rho_P_exact} and \eqref{eq:rho_S_exact} in the different configurations, for $\sigma=P,S$, it is sufficient to use the following relations:
\begin{itemize}
    \item in the \textit{Parallel} configuration 
    \begin{equation}
        \Gamma_{m m'}^{\alpha}(\epsilon)=\Gamma_\alpha \delta_{m m'}, \quad S^P=\frac{1}{\sqrt{2}}\begin{pmatrix}
1 &-1 \\
1 & 1
\end{pmatrix},
\quad \lambda^P_{\pm}=-\frac{\Gamma}{2}-i\left(\epsilon_d\pm g\right)
    \end{equation}

\item in the \textit{Series} configuration
\begin{equation}
        \Gamma_{m m'}^{\alpha}(\epsilon)=\Gamma_\alpha \delta_{m m'} \delta_{m \alpha}, \quad S^S=-\sqrt{\frac{\gamma}{2\eta}}\begin{pmatrix}
\sqrt{1+\eta/\gamma} &  i \sqrt{1-\eta/\gamma} \\
\frac{i g}{\sqrt{1+\eta/\gamma}} & \frac{-g}{\sqrt{1-\eta/\gamma}}
\end{pmatrix},
\quad \lambda^S_{\pm}=-\frac{\Gamma}{4}\pm\eta-i\epsilon_d
    \end{equation}
\end{itemize}
where $\eta\equiv \sqrt{\gamma^2-g^2}$ is real for $g\leq\abs{\gamma}$ with $\gamma\equiv \frac{\Gamma_L-\Gamma_R}{4}$.

\section{Weak-coupling protocol for DQD}
\label{app:WCP_DQD}

In this section we make explicit the application of our weak-coupling protocol to a double quantum dot (DQD) system. To retrieve the results of the \textit{local} and \textit{global} master equations, we apply the weak-coupling protocol to the matrix elements of Eq.~\eqref{eq:M_general_main} for both the \textit{Series} and the \textit{Parallel} configurations, respectively.

\subsection{From \textit{Parallel} configuration to \textit{global} master equation result}
\label{app:from_paraller_to_global}

\noindent For the \textit{Parallel} configuration, we only need to compute the weak-coupling limit of the diagonal matrix element $\left[M^{P}_{\alpha}(t)\right]_{m m}$ with $m=\pm$ using $\lambda^P_{\pm}=-\frac{\Gamma}{2}-i (\epsilon_d\pm g)$:

\begin{align}
    \lim_{\Gamma\to 0}\left[M^{P}_{\alpha}(t)\right]_{\pm\pm} &=\lim_{\Gamma\to 0} \Gamma\int \frac{d\epsilon}{2\pi} \frac{e^{\lambda_{\pm}^{P*}\frac{t-t_0}{2}}\sinh{\left[\frac{t-t_0}{2}\left(\lambda_{\pm}^{P*}-i\epsilon\right)\right]}}{\lambda_{\pm}^{P*}-i\epsilon}\frac{e^{\lambda_{\pm}^{P}\frac{t-t_0}{2}}\sinh{\left[\frac{t-t_0}{2}\left(\lambda_{\pm}^{P}+i\epsilon\right)\right]}}{\lambda_{\pm}^{P}+i\epsilon}f_{\alpha}(\epsilon)\nonumber\\
        &=\lim_{\Gamma\to 0}  e^{-\Gamma\frac{t-t_0}{2}} \Gamma \int \frac{d\epsilon}{2\pi} \frac{\sinh{\left[\frac{t-t_0}{2}\left(-\frac{\Gamma}{2}-i\left(\epsilon-\epsilon_d\mp g\right)\right)\right]}}{-\frac{\Gamma}{2}-i\left(\epsilon-\epsilon_d\mp g\right)}\frac{\sinh{\left[\frac{t-t_0}{2}\left(-\frac{\Gamma}{2}+i\left(\epsilon-\epsilon_d\mp g\right)\right)\right]}}{-\frac{\Gamma}{2}+i\left(\epsilon-\epsilon_d\mp g\right)}f_{\alpha}(\epsilon)\nonumber\\
            &=\lim_{\Gamma\to 0}  e^{-\Gamma\frac{t-t_0}{2}} \Gamma \int \frac{d\epsilon}{2\pi} \frac{\sinh{\left[\frac{t-t_0}{2}\left(-\frac{\Gamma}{2}-i\epsilon\right)\right]}}{-\frac{\Gamma}{2}-i\epsilon}\frac{\sinh{\left[\frac{t-t_0}{2}\left(-\frac{\Gamma}{2}+i\epsilon\right)\right]}}{-\frac{\Gamma}{2}+i\epsilon}f_{\alpha}(\epsilon+\epsilon_d\pm g)\nonumber\\
                &=e^{-\Gamma\frac{t-t_0}{2}}f_{\alpha}(\epsilon_\pm) \int \frac{d\epsilon}{2\pi} \frac{\sinh{\left[\frac{t-t_0}{2}\Gamma\left(-\frac{1}{2}-i\epsilon\right)\right]}}{-\frac{1}{2}-i\epsilon}\frac{\sinh{\left[\frac{t-t_0}{2}\Gamma\left(-\frac{1}{2}+i\epsilon\right)\right]}}{-\frac{1}{2}+i\epsilon}\nonumber\\
                    &=\frac{e^{-\Gamma\frac{t-t_0}{2}}}{2}f_{\alpha}(\epsilon_\pm) \sinh{\left[\frac{t-t_0}{2}\Gamma\right]}.
    \end{align}
From the second to the third line, we have shifted the energy $\epsilon\rightarrow \epsilon +\epsilon_\pm$ with $\epsilon_\pm=\epsilon_d \pm g$. From the third to the fourth line, we have performed the change of variable $\epsilon\rightarrow \Gamma \epsilon$  and taken the limit $\Gamma\to 0$, which allows us to extract the Fermi function evaluated at the energy $\epsilon_\pm$ outside the integral. Finally, as described in the main text, by substituting the above expression in Eq.~\eqref{eq:rho_P_exact} , we obtain the \textit{global} master equation solution for the population of each dot.

\subsection{From \textit{Series} configuration to \textit{local} master equation result}
\label{app:from_series_to_local}

For the \textit{Series} configuration, we compute the weak-coupling limit of the matrix elements $\left[M^{S}_{\alpha}(t)\right]_{m m'}$ with $m,m'=\pm$, using $\lambda^S_{m}=-\frac{\Gamma}{4}+m \eta-i\epsilon_d$:
\begin{align}
    &\lim_{\Gamma\to 0}\left[M^{S}_{\alpha}(t)\right]_{m m'} =\lim_{\Gamma\to 0} \Gamma\int \frac{d\epsilon}{2\pi} \frac{e^{\lambda_{m}^{S*}\frac{t-t_0}{2}}\sinh{\left[\frac{t-t_0}{2}\left(\lambda_{m}^{S*}-i\epsilon\right)\right]}}{\lambda_{m}^{S*}-i\epsilon}\frac{e^{\lambda_{m'}^{S}\frac{t-t_0}{2}}\sinh{\left[\frac{t-t_0}{2}\left(\lambda_{m'}^{S}+i\epsilon\right)\right]}}{\lambda_{m'}^{S}+i\epsilon}f_{\alpha}(\epsilon)\nonumber\\
        &=\lim_{\Gamma\to 0}  e^{\left(-\frac{\Gamma}{2}+(m+m')\eta\right)\frac{t-t_0}{2}} \Gamma \int \frac{d\epsilon}{2\pi} \frac{\sinh{\left[\frac{t-t_0}{2}\left(-\frac{\Gamma}{4}+m\eta-i\left(\epsilon-\epsilon_d\right)\right)\right]}}{-\frac{\Gamma}{4}+m\eta-i\left(\epsilon-\epsilon_d\right)}\frac{\sinh{\left[\frac{t-t_0}{2}\left(-\frac{\Gamma}{4}+m'\eta+i\left(\epsilon-\epsilon_d\right)\right)\right]}}{-\frac{\Gamma}{4}+m'\eta+i\left(\epsilon-\epsilon_d\right)}f_{\alpha}(\epsilon)\nonumber\\
            &=\lim_{\Gamma\to 0}  e^{\left(-\frac{\Gamma}{2}+(m+m')\eta\right)\frac{t-t_0}{2}} \Gamma \int \frac{d\epsilon}{2\pi} \frac{\sinh{\left[\frac{t-t_0}{2}\left(-\frac{\Gamma}{4}+m\eta-i\epsilon \right)\right]}}{-\frac{\Gamma}{4}+m\eta-i\epsilon}\frac{\sinh{\left[\frac{t-t_0}{2}\left(-\frac{\Gamma}{4}+m'\eta+i\epsilon\right)\right]}}{-\frac{\Gamma}{4}+m'\eta+i\epsilon}f_{\alpha}(\epsilon+\epsilon_d)\nonumber\\
                &=  e^{\left(-\frac{\Gamma}{2}+(m+m')\eta\right)\frac{t-t_0}{2}} f_{\alpha}(\epsilon_d)  \int \frac{d\epsilon}{2\pi} \frac{\sinh{\left[\frac{t-t_0}{2}\Gamma\left(-\frac{1}{4}+m\frac{\eta}{\Gamma}-i\epsilon \right)\right]}}{-\frac{1}{4}+m\frac{\eta}{\Gamma}-i\epsilon}\frac{\sinh{\left[\frac{t-t_0}{2}\Gamma\left(-\frac{1}{4}+m'\frac{\eta}{\Gamma}+i\epsilon\right)\right]}}{-\frac{1}{4}+m'\frac{\eta}{\Gamma}+i\epsilon}\nonumber\\
                     &=\frac{e^{-\Gamma \frac{t-t_0}{2}}}{2}e^{(m+m')\eta\frac{t-t_0}{2}} f_{\alpha}(\epsilon_d) \frac{\sinh{\left[\frac{t-t_0}{2}\Gamma\left(-\frac{1}{2}+\left(m+m'\right)\frac{\eta}{\Gamma} \right)\right]}}{-\frac{1}{2}+\left(m+m'\right)\frac{\eta}{\Gamma}}.
\end{align}
Similarly to the previous configuration, from the second to the third line, we have shifted the energy $\epsilon\rightarrow \epsilon +\epsilon_d$. From the third to the fourth line, we have performed the change of variable $\epsilon\rightarrow \Gamma \epsilon$ and taken the limit $\Gamma\to 0$, allowing us to extract the Fermi function evaluated at the energy $\epsilon_d$ outside the integral. When taking this limit, it is important to assume $\lim_{\Gamma\to 0}\eta/\Gamma\sim \text{constant}$. Intuitively, this corresponds to the assumption underlying the usage of the local master equation, i.~e.~ $g\lesssim\Gamma$. Finally, as described in the main text, substituting the above expression in Eq.~\eqref{eq:rho_S_exact} provides the \emph{local} master equation solution for the population of the two dots.

\end{document}